\documentclass[useAMS,usenatbib]{mn2e}
\usepackage{graphicx,natbib}
\usepackage{rotating}
\usepackage{lscape}
\usepackage[a4paper,margin=2cm]{geometry}

\usepackage{times}

\title[Extreme Infrared Variables from UKIDSS - I.]{Extreme Infrared Variables from UKIDSS - I. A Concentration in Star Forming Regions}
\author[C. Contreras Pe\~{n}a et al.: Extreme Infrared Variables from UKIDSS I.]{C. Contreras Pe\~{n}a,$^{1}$\thanks{E-mail:c.contreras@herts.ac.uk (CCP)} P. W. Lucas,$^{1}$ D. Froebrich,$^{2}$ M. S. N. Kumar,$^{3}$ J. Goldstein,$^{1}$ \newauthor J. E. Drew,$^{1}$ A. Adamson,$^{4}$ C. J. Davis,$^{5}$ G. Barentsen$^{1}$ and N. J. Wright$^{1}$ \\
$^{1}$Centre for Astrophysics Research, University of Hertfordshire, Hatfield, AL10 9AB, UK\\
$^{2}$Centre for Astrophysics and Planetary Science, University of Kent, Canterbury CT2 7NH, UK\\
$^{3}$Centro de Astrofísica da Universidade do Porto, Rua das Estrelas, 4150-762 Porto, Portugal\\
$^{4}$Joint Astronomy Centre, 660 North Aohoku Place, Hilo, HI 96720, USA\\
$^{5}$Astrophysics Research Institute, Liverpool John Moores University, Twelve Quays House, Egerton Wharf, Birkenhead CH41 1LD, UK}
\begin{document}

\date{\today}

\pagerange{\pageref{firstpage}--\pageref{lastpage}} \pubyear{2002}

\maketitle

\label{firstpage}

\begin{abstract}
We present initial results of the first panoramic search for high-amplitude near-infrared variability in the 
Galactic Plane. We analyse the widely separated two-epoch K-band photometry in the 5th and 7th data releases
of the UKIDSS Galactic Plane Survey. We find 45 stars with $\Delta K > 1$ mag, including 2 previously known 
OH/IR stars and a Nova. Even though the mid-plane is not yet included in the dataset, we find the
majority (66$\%$) of our sample to be within known star forming regions (SFRs), with two large concentrations 
in the Serpens OB2 association (11 stars) and the Cygnus-X complex (12 stars). Sources in SFRs show spectral 
energy distributions (SEDs) that support classification as Young Stellar Objects (YSOs). This indicates that 
YSOs dominate the Galactic population of high amplitude infrared variable stars at low luminosities and therefore likely dominate the total high amplitude population.
Spectroscopic follow up of the DR5 sample shows at least four stars with clear characteristics of eruptive 
pre-main-sequence variables, two of which are deeply embedded. Our results support the recent concept of eruptive variability comprising a 
continuum of outburst events with different timescales and luminosities, but triggered by a similar physical 
mechanism involving unsteady accretion. Also, we find what appears to be one of the most variable classical 
Be stars.
\end{abstract}


\section{Introduction}\label{sec:intro}

\nocite{*}

The Galactic Plane Survey \citep[GPS,][]{2008Lucas} is part of the
UKIRT Infrared Deep Sky Survey \citep[UKIDSS,][]{2007Lawrence}. The
GPS has covered the entire northern and equatorial Galactic plane that
is accessible to UKIRT, surveying 1868 deg$^{2}$ and providing $JHK$
photometry for $\sim$1 billion stars. The survey includes two epochs of K
band photometry, with the aims of (1) finding examples of brief
and rarely observed phases of stellar evolution via detection of high
amplitude variability and (2) providing a proper motion  catalogue. There has not previously been a panoramic survey for variability in the Galactic plane at near-infrared wavelengths, so even a 2 epoch survey represents an opportunity for new discoveries. This study provides a foretaste of the results that will come from the multi-epoch survey Vista Variables in the Via Lactea \citep[VVV,][]{2010Minniti} in the southern plane.

High amplitude variability in the near-infrared can be produced by several physical phenomena. Evolved giant and supergiant stars such as R Cor Bor stars, Mira variables and OH/IR stars are known to display large variability at infrared wavelengths \citep*[see e.g.][]{2004Tisserand,1991Whitelock,2006Jimenez}. Active Galactic Nuclei (AGN) can show up as point sources in the search through the GPS catalogues, and these objects are also variable in the near-infrared \citep{2002Enya,2013Cioni}. Other known classes of infrared variable stars include symbiotic stars \citep{2008Corradi}, Cataclysmic Variables/Novae \citep{2013Saito} and Classical Be stars \citep{1994Dougherty} \citep[see also][for a discussion on classes of near-infrared variable stars]{2013Catelan}.

Eruptive pre-main-sequence (PMS) are also among stars where high amplitude variability has been observed. This type of variable stars suffer episodic outbursts due to an abrupt increase in the mass accretion rate on to the central star, by a factor of up to 1000. The sudden rises in luminosity, of up to 6 magnitudes, can persist from months to $\sim$100 yr \citep{1996Hartmann}. Theoretical models of PMS evolution of sun-like stars generally assume that they accrete matter from their circumstellar discs in a continuous fashion \citep*[see discussion of some of the recent models in][]{2012Stamatellos}. They are assumed to gradually descend Hayashi tracks on the Hertzsprung-Russell (HR) diagram until they approach the main sequence. However, two problems with this simple picture have persisted for $\sim$20 years. The first is that HR diagrams of PMS clusters typically show a wide scatter about the best fitting isochrone. The second is the ``luminosity problem'' \citep[see e.g.][]{1990Kenyon,2009Evans,2012Caratti} which is that the typical luminosities of PMS stars in clusters are lower than expected ($\sim$1~L$_{\odot}$) for sun-like stars that should be above the main sequence. If eruptive variability is common amongst PMS stars it can resolve both problems: low average luminosity would be due to the great majority of the cycle being spent in the slowly accreting state, whilst the scatter would be due to the lingering effects of the more dramatic accretion events on the location of each star in the HR diagram \citep*{2009Baraffe,2012Baraffe}. If correct, this has serious consequences for the body of mass estimates of PMS stars in the literature. \citet{2012Baraffe} estimate that masses could be overestimated by $\sim 20\%$ to $\sim 40\%$ for brown dwarfs and low-mass-stars with $M>0.04 M_{\odot}$.

These eruptive YSOs have been classically divided into two sub-classes \citep[see e.g.][]{1989Herbig}: (1) FU Orionis variables (FUors), which show large increases in flux followed by a slow decay over $>$10 yrs. While in outburst they present strong CO absorption at 2.29 $\mu$m with H$_{2}$O absorption also present in their near-infrared spectrum. (2) EXor variables that have recurrent short-lived ($<$1~yr) outbursts and quiescent periods of 5-10 yrs. The IR spectrum is dominated by strong Br$\gamma$ in emission, with CO bands showing in emission while in outburst  and as absorption features while in quiescent states \citep[see e.g.][]{1996Hartmann,2007Fedele,2009Loren}. Both classes commonly power outflows seen as Herbig-Haro objects and near IR H$_2$ emission, although a direct connection between outbursts and the propagation of the jets is unclear. \citet{2013Magakian} show that V2494 Cyg would be the first FUor object to show such direct connection. These (optically defined) classifications struggle to include some intermediate objects and exclude younger protostars that have higher accretion rates but are too deeply embedded in circumstellar matter to be observed in the optical. Just three optically invisible protostars have been confidently identified as eruptive variables, with $K$ bandpass variability in excess of 2 magnitudes: OO Ser \citep{1996Hodapp,2007Kospal}, V2775~Ori (=[CTF93]216-2) \citep{2011Caratti} and GM Cha (=ISO-Cha I 192) \citep{2007Persi}. 

A recent study of protostars in Cygnus~OB7 found two more candidates that are currently under investigation,   \citep*{2012Rice}, whilst five more possible outbursting protostars are identified in \citet*{2013Scholz}. 

A few others have been proposed as embedded FUor-like objects based only on their spectral characteristics, e.g. PP13S \citep{2001Aspin} and AR6A+B \citep{2003Aspin}. At present only $\sim$10 FUors and 14-18 EXors are known \citep{2010Reip,2012Loren} but spectroscopic and H$_2$ studies of PMS stars suggest that eruptive variability may be common or universal \citep[see e.g][]{1996Hartmann,1997Reip,2010Conelley}.

Most PMS stars have been observed to show a low level near-infrared variability due to various physical processes \citep[such as rotation or hot spots, see e.g.][]{2004Lamm,2009Parihar}, but with a mean peak-to-trough variability of 0.15-0.17 mag in the K band \citep*[see e.g.][]{2001Carpenter,2008Oliveira,2012Scholz,2013Dorr}, with variations only very rarely exceeding 1 mag in K. YSOs are also known to vary at mid-infrared wavelengths \citep[see e.g][]{2009Vijh} and the same physical processes are not expected to produce variability larger than $0.6$~mag, whereas variations larger than $1$~mag are usually associated with eruptive variability \citep[see][and references therein]{2013Scholz}. A higher incidence of $K$ bandpass variations 
$>$1~mag ($\sim 13\% \pm 7\%$) has recently been reported in Class~I YSOs in the Braid 
nebula within Cygnus OB7 \citep{2012Rice}, which might be expected if fluctuations in the 
accretion rate are higher in these less evolved sources, which have a higher average 
accretion rate than the Class II and Class III YSOs that dominate the other studies. Two 
sources from \citet{2012Rice} and one source from \citet{2008Oliveira} 
were identified as possible eruptive variables, pending spectroscopic follow up. We note that variable extinction along the line of sight  can produce large change in the magnitudes of YSOs, such as the ones observed in UX Ori-like or AA Tau-like variables \citep[see e.g., ][]{2001Grinin,2013Bouvier}.  

\begin{figure*}
\centering

\includegraphics[width=2\columnwidth]{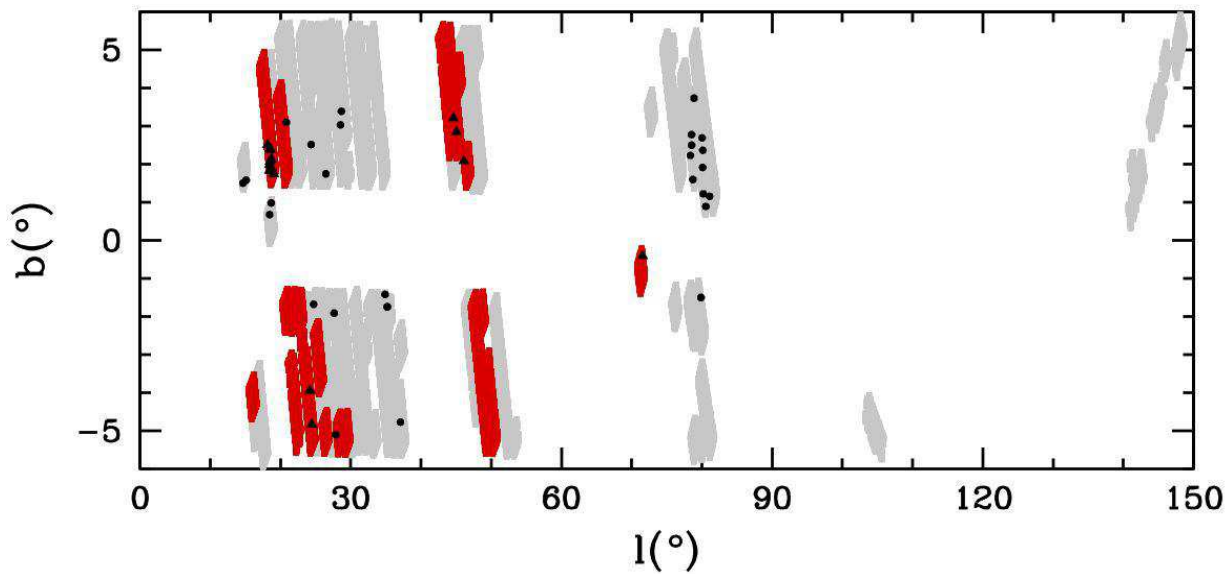}
\caption{UKIDSS GPS two epochs $K$ band photometry coverage up to DR7 ({\it grey}), with areas covered in DR5 shown in red. The locations of the 45 high-amplitude variables found in our searches are marked with black circles (DR7 search) and triangles (DR5 search). The higher concentration towards Serpens OB2 ($l\sim 18^{\circ}$) and the Cygnus-X complex ($l\sim 78^{\circ}-80^{\circ}$) is already apparent.}\label{fig:gps_cov}
\end{figure*}

This paper presents initial results of an ongoing project to perform an unbiased 
search for high amplitude near-infrared variables in the 2-epoch GPS data. The main focus of this study will be the analysis of YSOs, specifically likely young eruptive variable stars of the FUor/EXor class. The paper is divided as follows: Section 2 describes the methodology of the search for high amplitude variables in the 5th and 7th data releases (hereafter DR5 and DR7 respectively) of GPS and how these objects represent a very small fraction of the stars in the studied sample of the survey ($\sim$ 1 every 262000 stars). We then go on to describe the locations and properties of the objects found in our searches. In addition we discuss how YSOs are likely the dominant population of high-amplitude ($\Delta K>1$~mag) infrared variables. Starting from Section 3 we focus on the detailed analysis of objects arising from DR5 of GPS (the detailed analysis of DR7 objects will be the subject of a future publication). Therefore Section 3 describes follow up observations of DR5 objects ; in Section 4 we detail the results of follow up spectroscopy and photometry, in Section 5 we present a brief discussion and Section 6 presents a discussion on variable stars not associated with star forming regions.

\section{Candidate Selection from GPS}\label{target_sel}

\subsection{Search method and spatial bias}\label{sec:search}

We searched the GPS for high amplitude infrared variables identified via the two epochs 
of $K$ band photometry, which provide a minimium time baseline of 2 years. We used SQL queries 
of the WFCAM Science Archive \citep[WSA,][]{2008Hambly} to
select stars (objects with image profile classifier {\it mergedClass=-1}) that varied by $\Delta K \geq 1$~mag, having $K < 16$ mag in at least one epoch and for which no serious post-processing photometric errors are found ($K_{1,2}ppErrBits < 256$). The magnitude 
cut was designed to reduce the number of false positives, given that the faintest stars in 
crowded fields are the most likely to have incorrectly deblended fluxes. The cut used 
``apermag3'', the default 2$\arcsec$ diameter aperture. In order to further decrease the number of contaminant sources we also made additional cuts on the ellipticity of the detections ($K_{1,2}Ell < 0.3$) and in astrometric offsets between the observations (the detection in both epochs were required to be within $0.5\arcsec$ from each other). The queries returned a large number of candidates and in order to remove false positives, images of each candidate were inspected visually. Most false positives arose from bad pixels, diffraction spikes from bright stars or a small number of fields with bad astrometry. \footnote{These fields with bad astrometry were corrected for the 8th Data Release and the pipeline has now been made more robust against such errors.} 

Two epoch data were only available for a small proportion of the Galactic plane 
at the time of the searches, as indicated in Figure \ref{fig:gps_cov}. The figure shows that the area covered
is mostly located at Galactic latitudes $1^{\circ}<|b|<6^{\circ}$. In other words, two epoch 
data for the mid-plane were largely excluded from the available data releases. This is for historical
scheduling reasons. This spatial bias in the initial results reported here will be eliminated 
in future searches as the dataset becomes more complete. The fact that the mid-plane is not covered, implies that relatively few star formation regions or young stellar populations with similarly small scale heights will be included in our search. To be more specific, most star formation
regions at heliocentric distances, $d$\textgreater$3$~kpc will be excluded, assuming a scale height
of $\sim 67$~pc for molecular clouds located near $R$=$R_{\odot}$ \citep[e.g.][]{1987Knapp}. In addition, massive SFRs are also excluded \citep*[OB stars scale height of $\sim 30-50$~pc, e.g.][]{2000Reed,2006Elias}. 

\subsection{Search results}\label{sec:search_res}

Table \ref{table:par} presents the parameters for the variables selected in the searches to be described below. Column $1$ gives the source number sorted by ascending right ascension of the objects. Column $2$ presents the original designation given to the sources by the authors. This designation is maintained throughout the remaining of this work. Column $3$ corresponds to the full UKIDSS GPS designation for the source. Coordinates for the objects are given in Columns $4$ and $5$. The $K$ band magnitude of the source is presented in Column $6$, where $K_{1}$ represents the $K$ band epoch with contemporaneous $J$ and $H$ photometry\footnote{For the sources in this sample $K_{2}$ represents the earlier epoch, but this is not true for all GPS sources.}. The corresponding $J-K$ and $H-K$ colours are given in columns $7$ and $8$ respectively. Column $9$ gives the $K$ band magnitude difference between the 2 GPS epochs, $K_2-K_1$. Column $10$ presents $\Delta K_{all}$, the  absolute value of the peak-to-trough difference when using all of the available data from this work and the literature. Finally, the number of epochs used in the calculation of the latter, $N_{K}$, is given in column $11$, whilst the data release from which the objects arise is given in column $12$.

The initial search was performed in the first tranche of two epoch GPS variability data, which was released as part of DR5, on 28th of August 2009 and covered 31 deg$^{2}$ in the first Galactic quadrant. The search yielded 672 candidates. After inspection of the images to remove false positives, only 17 stars remained, with magnitude variations up to $\Delta K_{GPS} = 3.75$ mag (where $\Delta K_{GPS}$ is the absolute difference between the two GPS K band magnitudes). We conducted a second, similar search for extreme infrared variables in  DR7, which was released on the 10th of September 2010 and covered 155 deg$^{2}$, thus delivering an extra 124 deg$^{2}$ of two epoch data (the DR5 catalogue is embedded within the DR7 release). The query returned, after removal of candidates from DR5,  3365 candidates. The removal of false positives by visual inspection of the images yielded 28 real variables.

Adding the results from the DR5 and DR7 searches yields a total of 45 high amplitude infrared variables. We note that the amplitude of the variations are always observed to be much larger than individual errors. The amplitudes observed at 2.2 $\mu$m are also larger than the expected difference that arises from comparing different photometric passbands, e.g. UKIDSS $K$ vs 2MASS $K_{s}$, when ancillary data is available for the stars. Just to emphasize how rare these events are within the surveyed area of the Galactic plane, we note that when excluding the $\Delta K>1$~ mag condition in our searches, $\sim 11.8$ million stars are  selected from the DR7 catalogue (which includes the DR5 release). This implies that high amplitude variability occurs in $\sim$ 1 out of 262000 stars in GPS.

\begin{figure*}
\centering
\includegraphics[width=1\columnwidth]{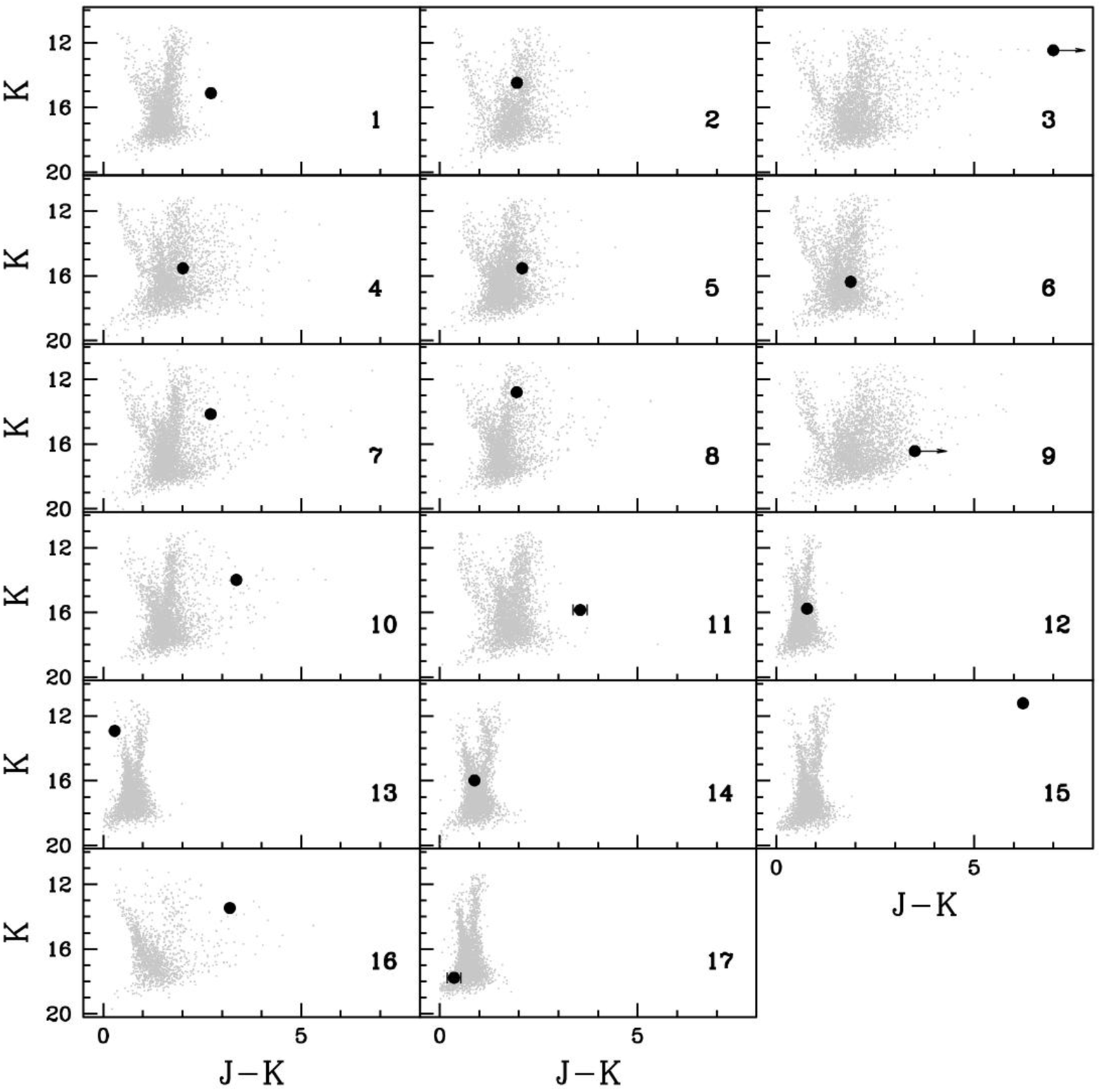}
\includegraphics[width=1\columnwidth]{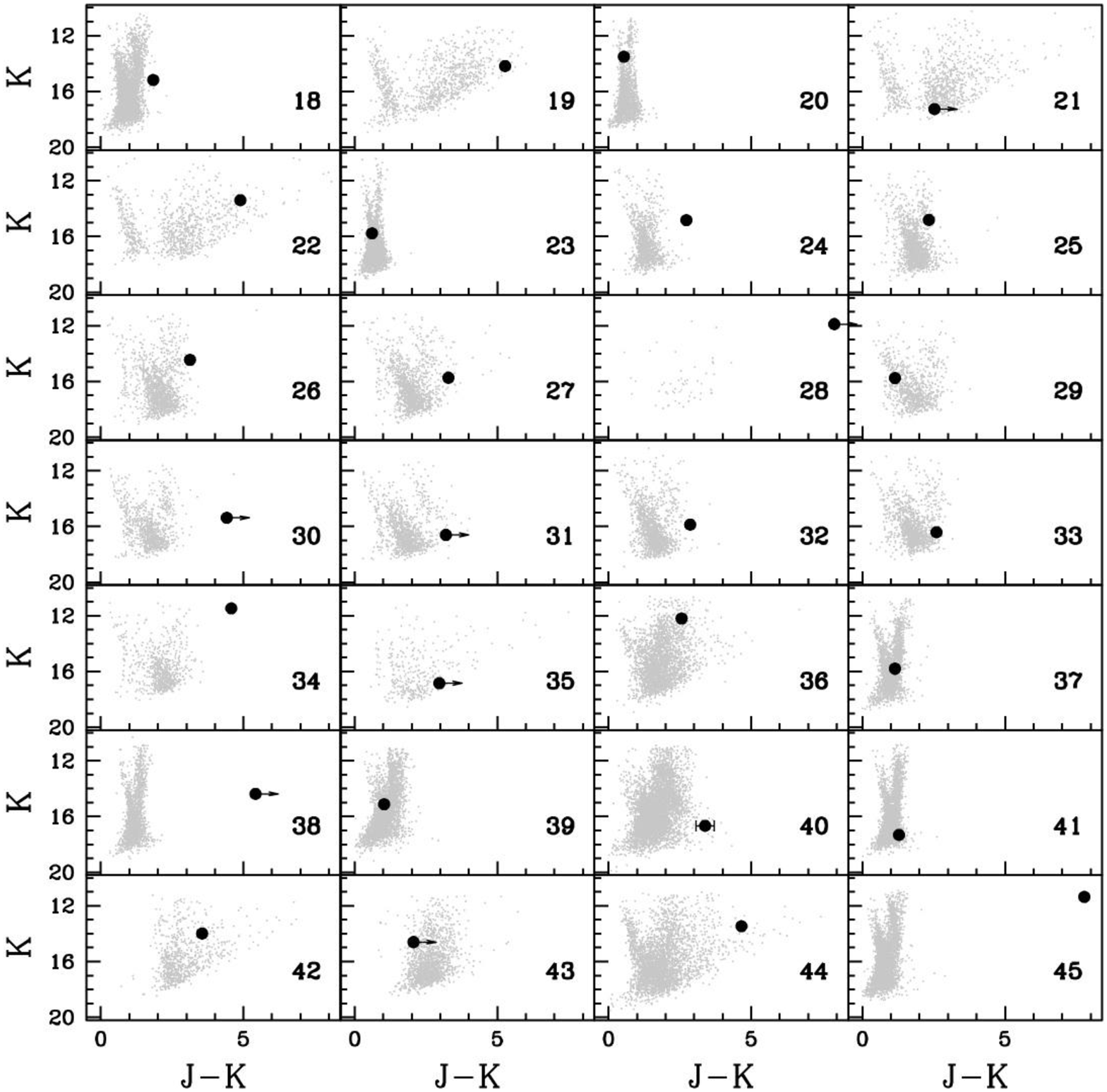}
\caption{Colour-magnitude diagrams of 6$^{'} \times$6$^{'}$ regions centered on each of the variable candidates. This is presented for candidates in DR5 ({\it left}) and DR7 ({\it right}). The arrow marks candidates for which $J$-$K$ colours represent lower limits. Numbers relate to the original designation of the objects given by the authors in column $2$ of table \ref{table:par}. Errors are plotted only for objects that present significant uncertainties on their measurements.}\label{dr57cmd}
\end{figure*}

\begin{figure*}
\centering
\includegraphics[width=1\columnwidth]{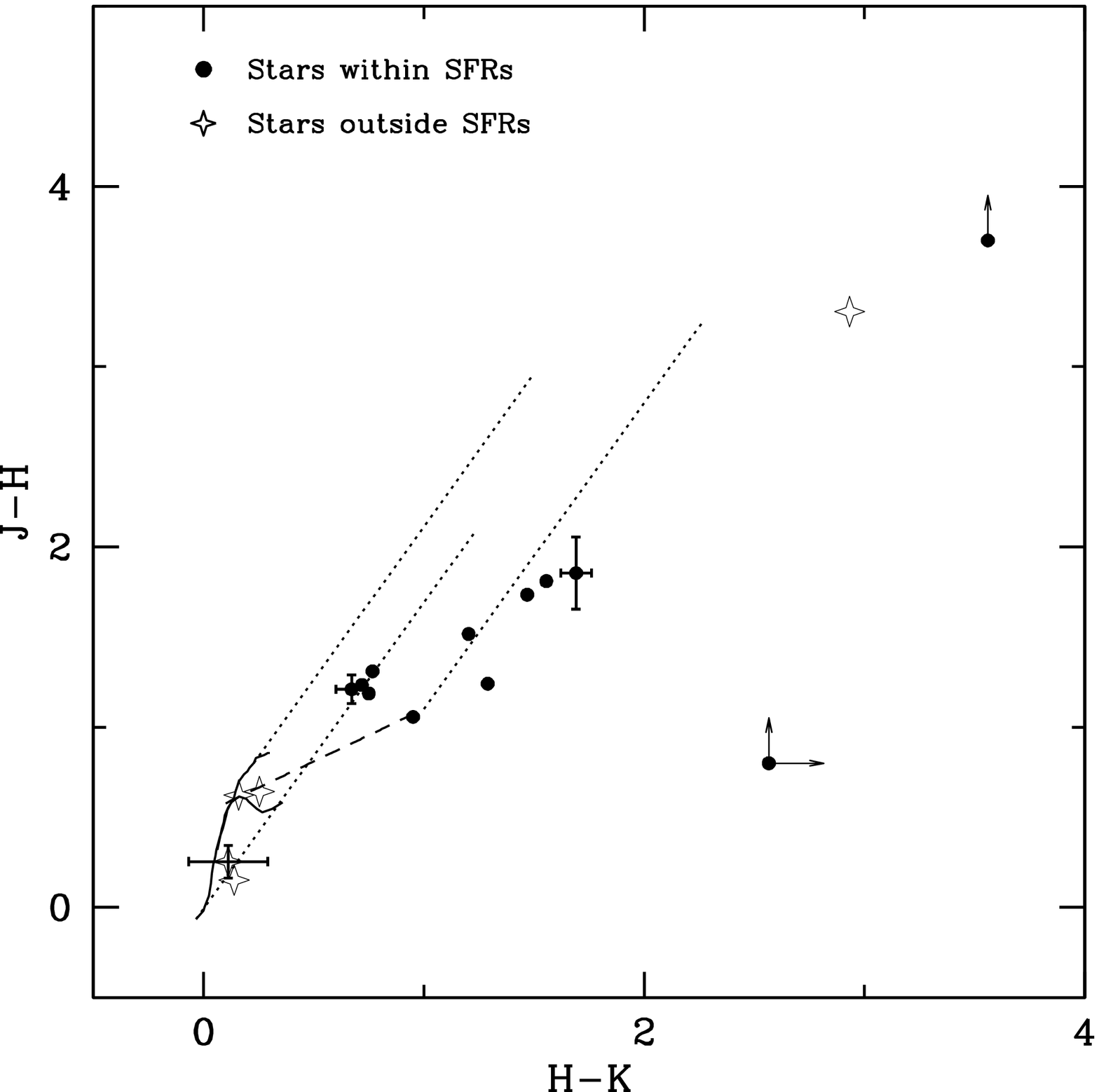}
\includegraphics[width=1\columnwidth]{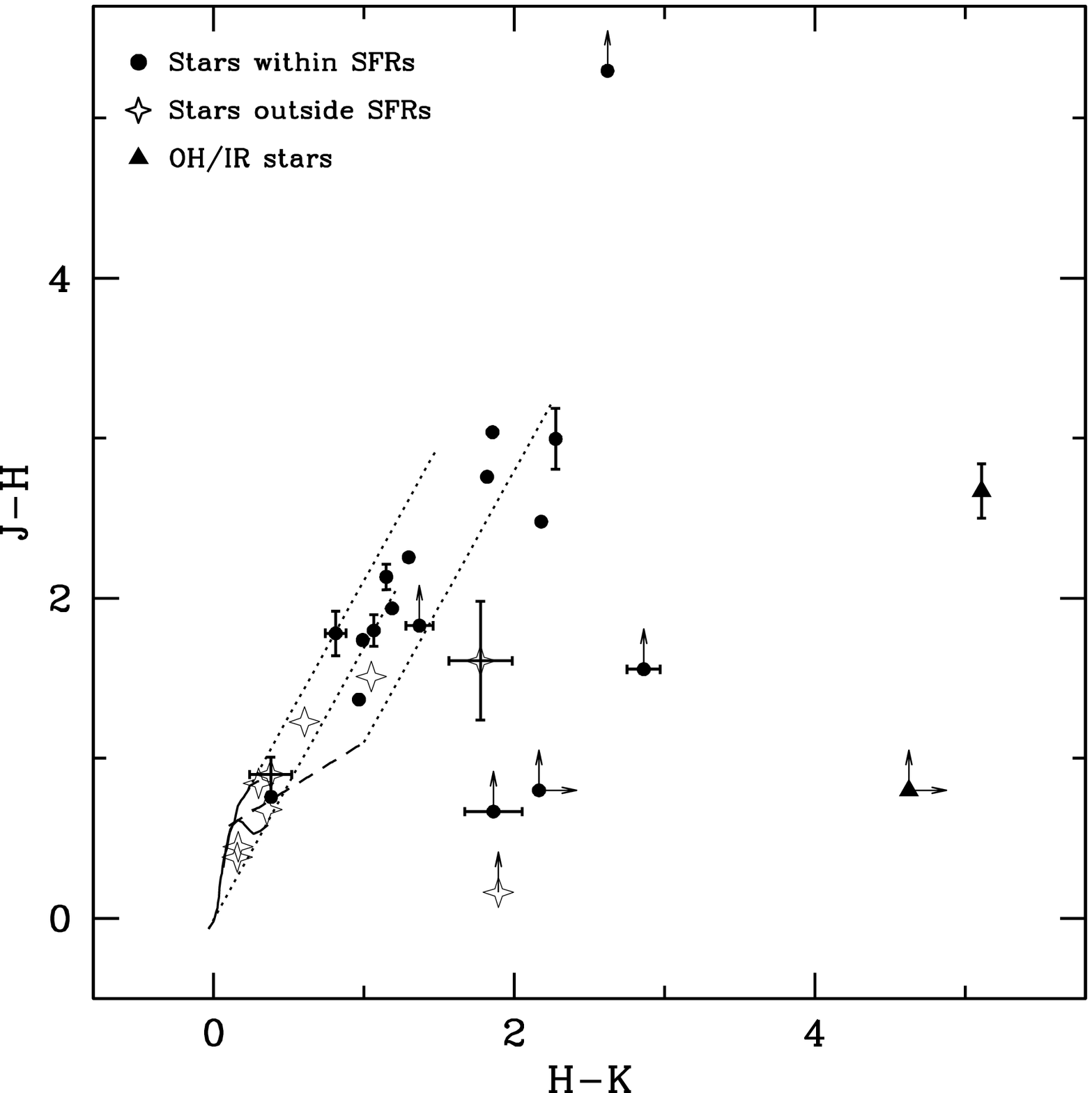}
\caption{Colour-colour diagram for GPS-selected candidates in DR5 ({\it left}) and DR7 ({\it right}). The objects are divided between those associated with SFRs as described in the text ({\it filled circles}) and objects not found within such areas ({\it open diamonds}). The two OH/IR stars found in the DR7 sample are marked with filled triangles. The classical T Tauri locus of \citet*{1997Meyer} is presented (\textit{long-dashed line}) along with intrinsic colours of dwarf and giants (solid lines) from \citet{1988Bessell}. Reddening vectors of $A_{V} = 20$~mag are shown as dotted lines. The arrows mark stars for which colours represent lower limits. Errors are plotted only for objects that present significant uncertainties on their measurements. }\label{nircdr57}
\end{figure*}

The colours and magnitudes of the 45 stars found in the analysis are shown in Figures \ref{dr57cmd} and \ref{nircdr57}, and listed in Table \ref{table:par}. Two of the 17 variables in DR5 show blue colours. One of them was identified via SIMBAD as 
Nova Sct 2003 \citep[][]{2003Nakano} corresponding to GPSV17 in Table \ref{table:par}.
The other 16 stars from DR5 are unknown in the literature, except GPSV13, which is listed in the \citet{2008Witham} catalogue of H$\alpha$ emission line stars. Also, some have previous 
detections in the Deep Near Infrared Survey of the Southern Sky (DENIS) or the 2 Micron All 
Sky Survey (2MASS). Two stars arising from DR7, GPSV38 and GPSV45, present exceptionally red $H-K$ colours and $\Delta K > 2$ mag. These correspond to the known OH/IR stars, GLMP755 \citep{2006Jimenez} and IRAS 18396-0807 \citep{1994Hashimoto}. OH/IR stars typically represent the most extreme oxygen-rich Asymptotic Giant Branch (AGB) stars, with mass loss rates greater than 10$^{-5}$ M$_{\odot}$yr$^{-1}$. They are thought to be the extension towards longer periods and higher luminosities of Mira variables \citep[][]{1991Whitelock}, being more evolved and in some cases perhaps in the process of becoming planetary nebulae. Given the higher mass loss rate, these objects are deeply embedded, being only detectable towards near-infrared wavelengths. Thus these stars will likely be a source of ``contamination'' in our search for eruptive PMS variables (see discussion below). However, these are still interesting objects. The mid-infrared colours of two sources detected here put them at the red end of the sequence of colours predicted for O-rich AGB stars with increasing mass loss \citep[][]{1987Bedijn} in the work of \citet{2006Jimenez}, implying that these stars are about to leave the sequence to become planetary nebulae. 

From the DR5 sample, eight of the stars appear very red in colour-magnitude diagrams (CMD, see left panel of Fig. 
\ref{dr57cmd}), i.e. redder than the giant branch plotted for stars in the vicinity.
Five of the remaining stars also appear quite red, being projected against the giant branch. These 13 ``red'' stars have colours indicating a $K$ band excess due to hot dust in our colour-colour diagrams, a common property of pre-main-sequence (PMS) stars with accretion discs. In fact, most of them are consistent with being reddened T Tauri stars (Fig. \ref{nircdr57}). We note that 4 of these are located in the boundary between reddened T Tauri and main sequence stars, however some of their properties favour the T Tauri interpretation. Three of them (GPSV9, GPSV3 and GPSV15) have larger K-band excesses, which could point to the objects being Class I YSOs. The results of the DR7 search are similar to those of DR5, in that most of the variables have redder ($J$-$K$) colours than normal dwarf branch or giant branch stars in a 6$\times$6 arcmin field centered on the variable objects (see Fig. \ref{dr57cmd}). There are some extreme cases: GPSV28 is extremely red and it may correspond to IRAS source 
IRAS 20226+4206. Gemini NIFS spectroscopy also shows that the object has a spatially extended 
H$_{2}$ outflow (Contreras Pe\~{n}a et al., in prep.). Figure \ref{nircdr57} shows the near-infrared colours of the selected candidates. Many of the objects fall in the region dominated by disc emission, i.e. between the reddened sequence of main sequence stars and the T Tauri locus of \citet{1997Meyer}. Some candidates show a larger excess, which would usually suggest classification as a Class I YSO, i.e. a disc $+$ envelope system. 

\begin{figure*}
\centering
\includegraphics[width=1.5\columnwidth]{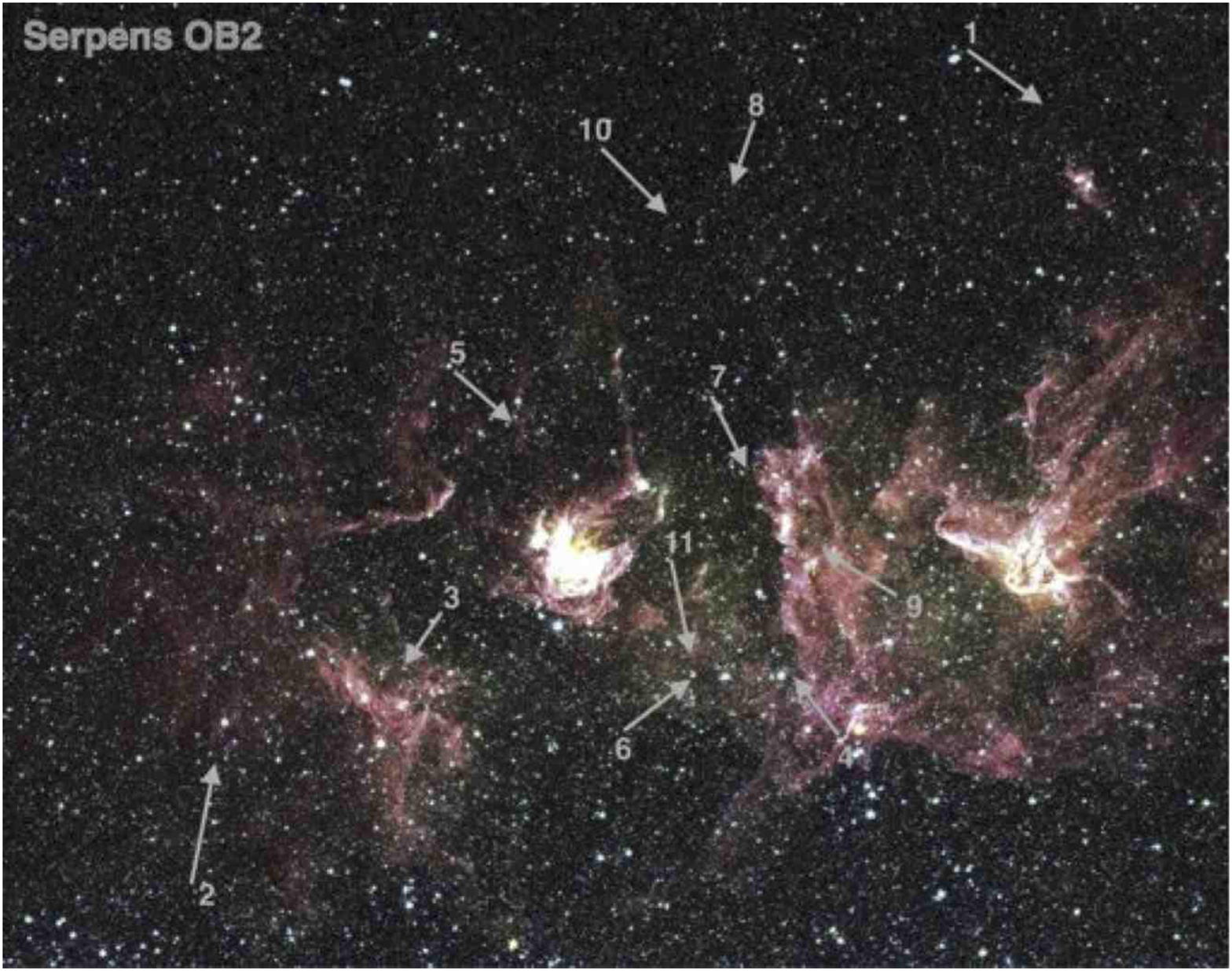}\\
\includegraphics[width=1\columnwidth]{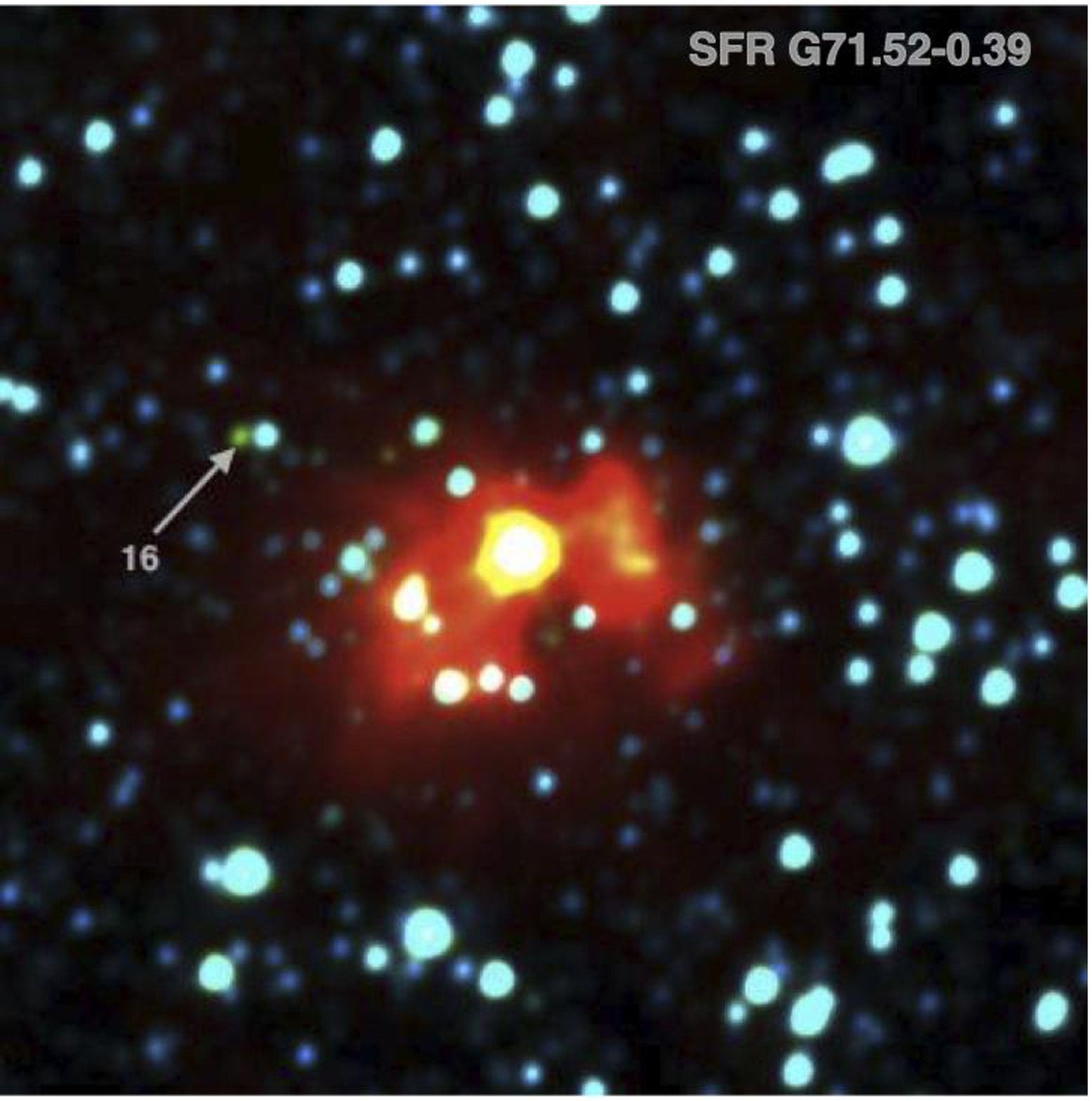}
\includegraphics[width=1\columnwidth]{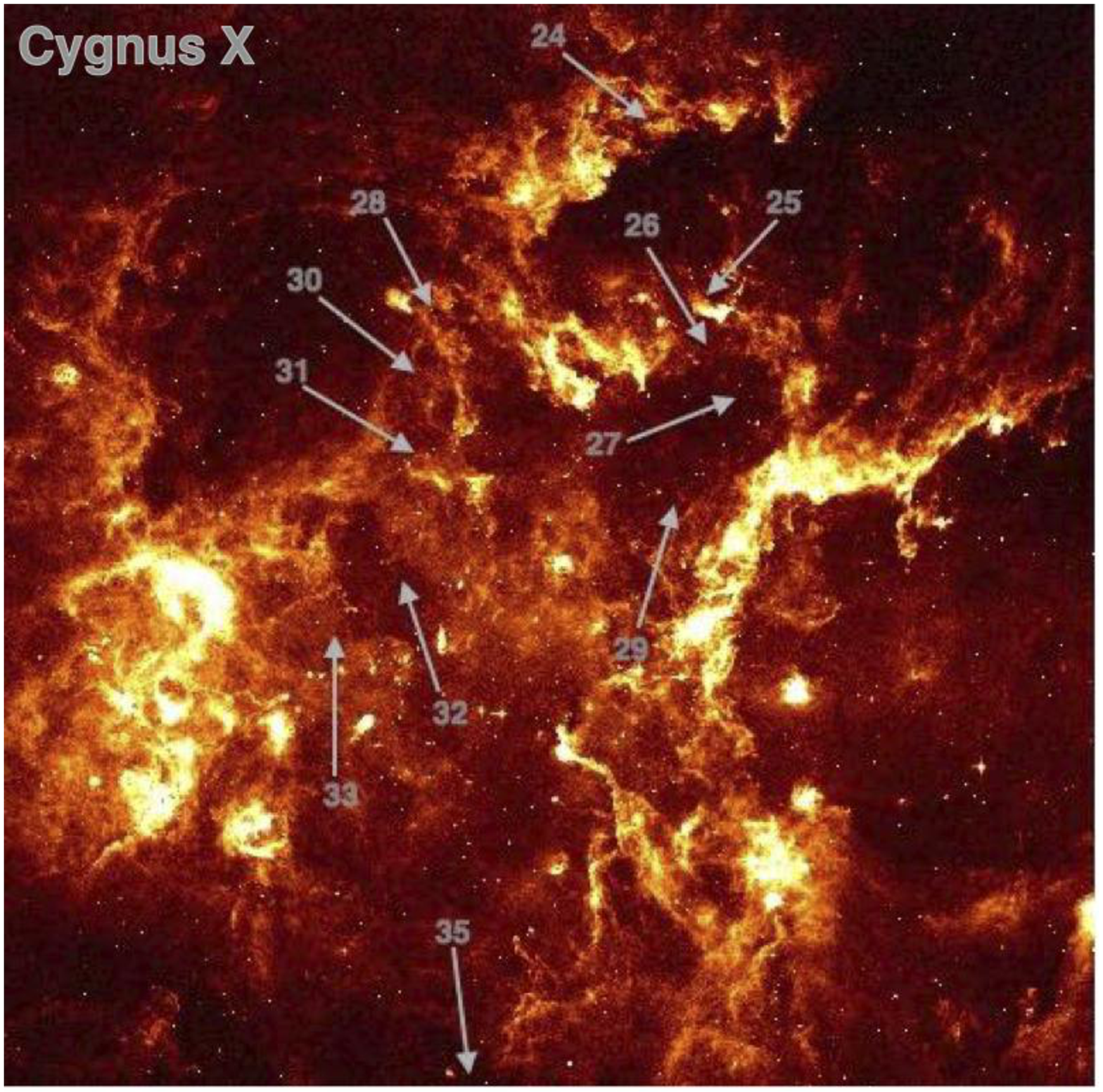}
\caption{({\it top}) False colour Spitzer GLIMPSE3D image (blue=3.6 $\mu$m, green=4.5$\mu$m, red=8.0$\mu$m) of the Serpens OB2 association. The projected location of 11 candidates found within this region are marked by the arrows with the corresponding source number from Table \ref{table:par}. ({\it bottom left}) False colour WISE image (blue=3.5 $\mu$m, green=4.6$\mu$m, red=12$\mu$m), the arrow marks the position of GPSV16; SFR G71.52-0.39 from \citet{2002Avedisova} catalog is clearly observed at the centre of the image.({\it bottom right}) MSX6C 8 $\mu$m image of the Cygnus-X star forming complex. Projected location of 12 sources found within this area are marked by arrows with their corresponding GPSV designations.}\label{fig:serp}
\end{figure*}

Strong evidence to support the YSO interpretation is given by the apparent high concentration of our sample towards SFRs. We performed the search for a possible association of the sources using SIMBAD and the \citet{2002Avedisova} catalog of SFRs. The main condition for the stars to be flagged as likely associated with a SFR was that evidence of star formation would show up in a 300$\arcsec$ radius search, centred on the star, in SIMBAD (i.e. PMS stars, dark clouds, HII regions) and the \citet{2002Avedisova} catalog. Four stars failed this main condition but are still flagged as likely associated with SFRs for the following reasons:

\begin{itemize}
\item GPSV5 and GPSV10 are within the large concentration of high amplitude variables in an area of 1 deg$^{2}$ that is coincident with the Serpens OB2 association (see below) and have colours consistent with a T Tauri interpretation (see Fig. 3). In addition GPSV5 is close to a bright 8 $\mu$m emission feature in Figure \ref{fig:serp}. GPSV10 (discussed in more detail in Section \ref{rc:sec}) also shows an SED that is consistent with a YSO classification.
\item GPSV26 is part of the sample that is within the Cygnus X star forming complex (see below). Its near infrared colours and SED are also consistent with a YSO interpretation (see figure \ref{app:ysos_seds1} in Appendix A).
\item GPSV19 is only 318$\arcsec$ from SFR G35.20-1.75 in the \citet{2002Avedisova} catalog (see below). Its colour is consistent with a YSO classification and it is an unusually red object compared to its local $6\times6\arcmin$ field. The lack of mid-infrared data for the object does not allow us to confirm this interpretation through its SED. However, the spectrum of the object shows H$_{2}$ emission at $2.12~\mu$m, probably 
associated with a molecular outflow, and Br$\gamma$ emission, both of which support a YSO classification (Contreras Pe\~{n}a et al, in prep.).  
\end{itemize}

Using our search criteria, it was found that 12 of the 17 DR5 variables present an apparent spatial association with known star forming regions. GPSV1 through 11 are located in an area of 1 deg$^{2}$ that is coincident with the Serpens OB2 association, a loose association of PMS stars located $\sim$2~kpc away \citep[][]{2000Forbes} and which contains several molecular clouds. The candidates do not appear to be located in the regions of most recent massive star formation that can be identified by the bright 8 $\mu$m emission in Figure \ref{fig:serp}. Two objects, GPSV3 and GPSV9, are actually within the 8 $\mu$m nebulosity and they have the reddest near-infrared colours of the Serpens OB2 objects. However, they do not seem to be Class I objects (see Section \ref{subsec:ev}). The variable object GPSV16 is found to be within $\sim 170\arcsec$ of the SFR listed as G71.52-0.39 \citep{1987Scoville} in the \citet{2002Avedisova} catalog (Fig.\ref{fig:serp}).
            
We find 17/28 DR7 variables likely associated with a known SFR. Three stars (GPSV19,GPSV21 and GPSV22) are likely associated with SFR G35.20-1.75 \citep{2009Zhang}, GPSV42 is in SFR G26.35+1.83 \citep{1990Hunter} and GPSV44 in SFR G28.96+3.54 \citep{1982Crutcher}. Twelve of our candidates are distributed along the Cygnus X star forming complex (Fig. \ref{fig:serp}), which is known to contain several rich and complex SFRs. The presence of FUor outbursts would not be new for this region \citep[see e.g.][]{2012Rice,2013Magakian}.

We note that 4 other stars fall in the YSO region in colour-colour plots but lack a spatial association with known SFRs. These are GPSV15, 36, 40 and 43. As discussed in section \ref{subsec:ev}, GPSV15 displays several characteristics associated with eruptive YSOs, and DR7 objects GPSV36 and 43 show SEDs similar to YSOs (see Appendix \ref{app:dr7}). GPSV40 is a faint object, and very red in its local CMD. However, mid-infrared data are not available to confirm its YSO nature.

In order to verify the significance of this spatial association with SFRs, we performed a Monte Carlo selection of many samples of 45 real GPS sources in the 2 epoch area, using the same cuts as the variable star selection, except for $\Delta K > 1$~mag. We find that only 10\% of randomly chosen sources would be within 300$\arcsec$ of
a known star forming region, i.e. 4-5 stars in our sample would be near SFRs by chance. However, as noted above, 4 stars in our sample have characteristics of YSOs but are not spatially associated with known SFRs. This indicates that chance associations have little effect on the proportion of variables (29/45) that would be classified as YSOs primarily on the basis of spatial association. In view of this, we believe that the strong concentration of high amplitude variables in SFRs found in our study is sufficient evidence on its own that most of 
the high amplitude variables are YSOs. Confirmation for the majority of the DR5 objects (based on SEDs 
and spectra) is discussed at length in section \ref{sec:res}. Supporting evidence for the DR7 objects, in the 
form of SEDs, is given in Appendix \ref{app:dr7}. Spectra for a subset of the DR7 objects will be included in 
a future paper focussing mainly on the Cygnus X region (Contreras Pe\~{n}a et al., in prep.). Preliminary 
inspection indicates that most of them show emission or absorption features that support a YSO classification.

Despite the strong evidence that the sample is dominated by YSOs, we still need to discuss other classes of 
high amplitude variable stars that could have the observed red colours and may account for those objects 
which are not in known SFRs.  

The faint K band magnitudes of the stars in our sample suggests that they are too faint to be R Cor Bor variables or luminous AGB stars, such as Mira-type variables. To verify this, we first derive the approximate 
extinction towards background objects in the local CMD of each of our sources. To do this, we noted the 
typical $J-K$ colour of the reddest sources in the local
field (excluding any extreme outliers) and then calculated an estimate of the maximum extinction that they 
could have by assuming that they have the intrinsic colours of $-$0.1, corresponding to early B type stars. 
We then derive the distance at which a typical AGB star \citep*[Assuming typical $M_{K}=-7.25$ mag, 
$(J-K)=1.25$ mag,][]{2008Whitelock} would have to be located in order to have the apparent magnitude of the 
individual variable source being considered (using the local $A_V$ estimated before). 
In each case a typical AGB star would have to be located well beyond the Galactic disc edge at Galactocentric
radius $R_G=14$~kpc \citep[see e.g.][]{2011Minniti}, indicating that there are no typical Mira variables in 
the sample.

AGB stars with high mass loss rates, such as OH/IR stars or extreme carbon stars 
\citep*[see e.g.][]{1992Volk}, are deeply embedded in circumstellar material and can suffer high extinction. These stars may only be observable at near-infrared wavelengths and appear  fainter than typical AGB stars \citep[see e.g.,][]{1997Vanloon}.
Therefore they can show up in our sample and in fact two of them are present in the DR7 sample, as discussed 
above. \citet*{2001Olivier} study a complete sample of nearby dust-obscured AGB stars, spread over $\sim 3/4$ of the sky,
that are undergoing heavy mass loss ($\dot{M} > 10^{-6} M_{\odot}$yr$^{-1}$). They find 30 stars within 1~kpc 
from the sun (the completeness limit of the study) with a scale height of 236~pc. This implies a space 
density of 9.35~kpc$^{-3}$. 
Most AGB stars are located at $R_G<$10~kpc (Ishihara et al.2011). For a typical inner disc sight line at 
$l=45^{\circ}$ this corresponds to a maximum heliocentric distance, $d\approx 14$~kpc. (Beyond this distance, 
the Galactic latitudes of our survey ($|b|\approx 1.5-5.5^{\circ}$) would place most sources a few scale 
heights above the plane, so very few AGB stars should be found). An area of 1~deg$^2$ along this $l=45^{\circ}$
sightline encompasses a conical volume of 1.1~kpc$^3$ out to d=14~kpc. After allowing for the effect of scale 
height on the source density this is equivalent to an effective volume of 0.12~kpc$^{3}$ in the mid-plane.
Then, 1.12 dust-enshrouded AGB stars deg$^{-2}$ should be contained in this volume. However, we 
expect to detect only a small fraction of them in a 2-epoch survey. In fact, we find that only $27\%$
should be detected, based on a Monte Carlo analysis using a large sample of randomly generated sinusoidal 
curves with different amplitudes and periods, similar to the sample of \citet{2001Olivier}, 
when randomly selecting two epochs separated by 3 years. In addition, only 2/30 stars in the
\citeauthor{2001Olivier} sample would have similar magnitudes to the brightest stars in our GPS sample, 
if they are located at $R_G<$10~kpc (the remaining stars would almost certainly saturate in UKIDSS even at 
these large distances). These two stars correspond to the first and third highest mass losing AGB stars in the 
\citeauthor{2001Olivier} sample ($\dot{M} > 10^{-4.8} M_{\odot}$yr$^{-1}$). AGB stars with this extremely
high mass loss rate, such as the OH/IR stars in the DR7 sample and the extreme carbon stars in 
\citet{1992Volk}, show a steep rise of their SEDs towards the mid-infrared, characterised 
by a very red $K-[12]$ colour, displaying $K-[12]>9$~mag.

When we apply the expected detection ratio (0.27) and the fraction of AGB stars with high mass loss rates
that would show up in our survey (2/30) we end up with 3.1 dust-enshrouded AGB stars in the area covered by 
our study. This number is remarkably close to our actual detection of two OH/IR stars in DR7. Only two 
objects from the remainder of the sample have $K-[12]$ colours similar to extreme AGB stars (GPSV3 and 
GPSV15) but we are confident that they are YSOs rather than AGB stars, see section 
\ref{sec:erup_3_15}. Despite this, it is reasonable to expect that we will find more very red AGB stars in 
future data releases, some of which may be previously unknown objects.

Symbiotic variable stars can be located in the same area as most of our variable stars in the JHK 
colour-colour diagram \citep[see Figure 2 in][]{2008Corradi}. The D-type symbiotic stars in particular have 
the fairly red {\it (H$-$K)} colours found among most of our sample. However, D-type symbiotic stars 
have AGB star companions, most of which can be expected to saturate in the GPS K band data (see the earlier discussion in this section).
\citet{2010Corradi} note that the space density of symbiotic stars is very uncertain but it is clear that 
they are a small Galactic population. As an illustration of their rarity, \citet{2008Corradi} selected a 
large number of possible symbiotic stars based on an H$\alpha$ and broad bandpass colour selection. 
Spectroscopic follow up \citep{2010Corradi} showed that most of their candidates were in fact T Tauri stars, 
despite their attempt to reduce contamination by avoiding spatially clustered candidates (Symbiotic 
variables, like AGB stars are expected to be found in isolation). It is unlikely therefore that there are 
many symbiotic stars in our sample but it is conceivable that some of the isolated objects with bluer 
{\it (H$-$K)} colours (after allowing for extinction) might be S-type symbiotic stars.

Active Galactic Nuclei (AGN) can also show up in our searches as point sources, with a wide range of 
near infrared colours. \citet{2008Maddox} finds an
expected density of $\sim$ 40 quasars down to $K=16$~mag in their study of 12.5 deg$^{2}$ of the early data 
release of the UKIDSS Large Area survey \citep[LAS,][]{2006Dye}, which implies $\sim$ 470 quasars
in the area covered in our study. However, the fraction of high-amplitude ($\Delta K>1$~mag) variable AGNs 
appears to be much lower than 1$\%$. For example, none of the 116 quasars in the sample of \citet{2013Cioni} 
displays $\Delta K>1$~mag. \citet{2012Kouzuma} searched for near-infrared counterparts of AGNs in the UKIDSS 
Large Area Survey (DR6, covering $\sim$ 1850 deg$^{2}$) and 2MASS catalogues. They identified 1920 AGN via 
$K$ band variability between the two epochs (UKIDSS and 2MASS), using several catalogue data quality cuts to 
remove most erroneous variables. Their sample included 15 candidate high amplitude AGNs ($\Delta K > 1$~mag) and they provided the 
coordinates to us at our request. Only 9 of these 15 have a point source profile-classification 
($mergedclass=-1$) in UKIDSS, and visual inspection reveals that only 6 of these 9 are real. The approximate 
completeness limit of their study only reaches to about $K=15.2$~mag (although some objects are still 
detected at fainter magnitudes). To compare with our GPS sample, we must first correct for the fact
that the \citet{2012Kouzuma} high amplitude variable sample will not include most AGN that were fainter at 
the less sensitive 2MASS epoch and then allow for our fainter magnitude limit. We find a factor of 1.6 for 
the first correction (using the proportion of highly variable AGN that were fainter in UKIDSS rather than in 
2MASS) and we conservatively estimate a factor of 4 more quasars arises from our $K=16$~mag limit, using Table 7 
of \citet{2008Maddox}. After applying these corrections and scaling to our survey area we find that we 
should expect $\sim 3-4$ high-amplitude AGNs in the area covered in DR7. We note that the LAS covers 
areas of low extinction, so if we allow for the higher extinction in the Galactic plane (typically
0 to 1 mag in the $K$ band) then this estimate should be considered an upper limit.

\subsection{Concentration of extreme variables in SFRs}\label{sec:conc_ext}

When we add the results of our searches of DR5 and DR7, we find that 29 out of 45 
variables are likely associated with SFRs, thus representing $\sim$66$\%$ of our sample. As discussed above, 4 other stars show characteristics of YSOs but do not seem to be associated with a known star forming region (a discussion on this point is presented below). Thus, the fraction of variables in SFRs could increase.
  
In view of the spatial bias against star formation regions in the mid-plane in this
initial sample (see Section \ref{sec:search}) it is clear that YSOs must dominate the Galactic 
population of high amplitude infrared variables with $\Delta K > 1$~mag, at least in the
magnitude range that this search has explored ($K \approx 11.5$ to 16 mag). The question
then arises as to whether YSOs dominate the total population of high amplitude infrared 
variables. Inspection of the General Catalogue of Variable Stars \citep[GCVS,][]{2010Samus} shows that Mira-type variables are the commonest type at bright magnitudes. These very luminous stars are generally saturated
in the GPS, except for a tiny minority with exceptionally high
extinction and mass loss rate, such as the two OH/IR stars in our DR7
sample. We note that the high variability OH/IR phase is only expected
to last ∼ 1700 yr \citep{2000Lewis}, thus we do not expect to find many of
these type of variables. The GCVS data indicate that 67\% of Miras have
$\Delta K >1$~mag. (Note that the $K$ to $V$ amplitude ratio has an average
value of 0.2, see \citet{2005Lebzelter,2005Soszynski}).
\citet{1996Ortiz} calculated a space density of 265~kpc$^{-3}$ for a complete
sample of AGB stars within 1~kpc of the Sun, with a scale height of 330~pc.
\citet{2011Ishihara} performed the search for AGB stars complete out to
8~kpc and they found a similar space density for sources located at Galactocentric 
radii $3<R_G<8$~kpc, after correcting for colour-based incompleteness (see figure 9 
of that work). They also showed that most O-rich AGB stars, which dominate the 
numbers at $l<90^{\circ}$, are located at Galactocentric radii $R_G<$10~kpc.
A typical inner disc sight line at $l=45^{\circ}$ with an area of 1~deg$^2$ 
encompasses a volume of 1.11~kpc$^{-3}$ at $R_G<$10~kpc. After allowing for 
the effect of scale height, the effective volume comes to 0.21 ~kpc$^{-3}$ at these distances. This volume 
should contain $\approx$50 AGB stars deg$^{-2}$.

Only a small fraction of AGB stars are expected to be Mira variables, which are 
at the end of the thermally pulsating phase of their evolution. \citet{2001Cioni} give the fraction as 8\%. 
Combining this with the 67\% fraction of Miras with $\Delta K >1$~mag then implies a source density of 
$\approx$2.7~deg$^{-2}$ for typical sight lines in quadrants 1 and 4 of the Milky
Way. This is broadly consistent with an estimate of $\approx 0.5$ deg$^{-2}$ in the Galactic 
plane obtained from inspection of the GCVS: the GCVS only includes Mira variables with 
known periods so we expect it to provide a somewhat smaller space density.  

\begin{figure}
\begin{center}
\includegraphics[width=1\columnwidth]{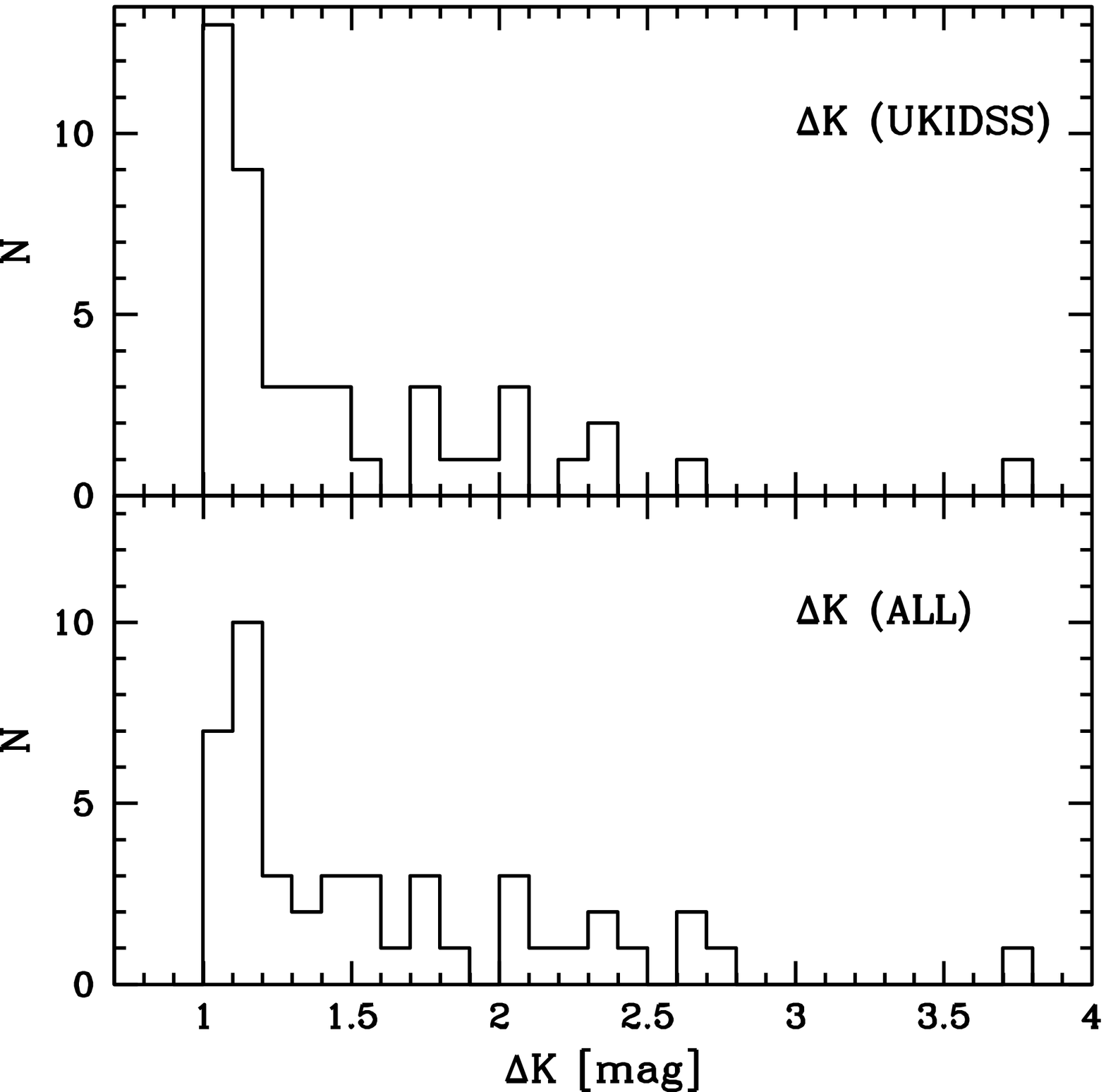}\\
\includegraphics[width=1\columnwidth]{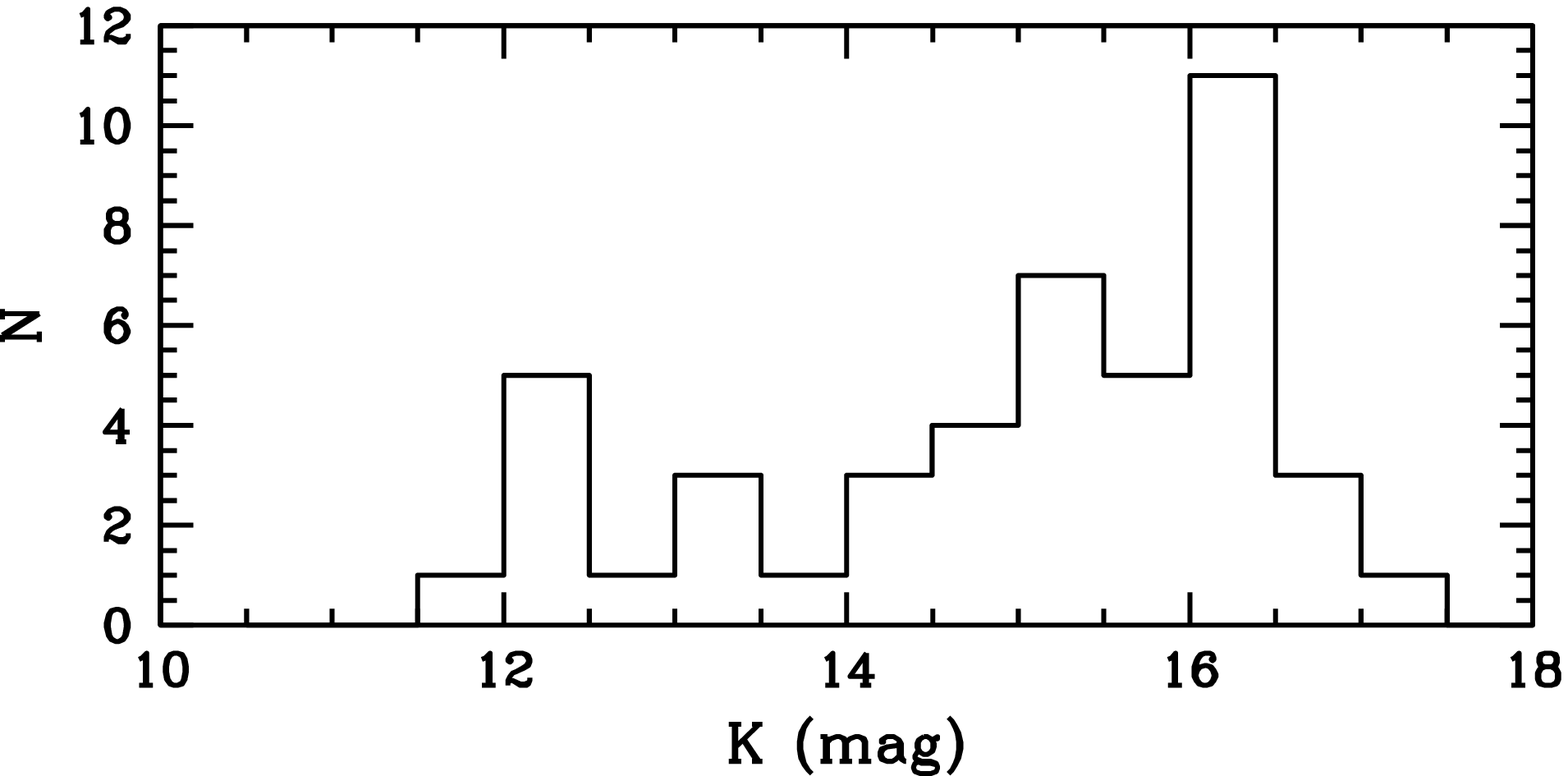}
\caption{Histogram of the absolute $K$ band magnitude difference from GPS ({\it top}) and when taking into account all of the available data ({\it middle}). The middle graph shows an increase of the observed amplitude of the variability when more epochs over a extended baseline are available. {\it bottom} Mean GPS $K$ magnitude distribution for the 45 objects in our sample.}
\label{histgps}
\end{center}
\end{figure}

The 29 YSOs imply a source density of $29/155$ = 0.19~deg$^{-2}$. However, the 
true number is higher because: (1) with only two epochs most high amplitude 
variables will be missed; (2) the source density rises towards our magnitude
cut (see Fig. \ref{histgps}), indicating that most distant, low luminosity PMS variables will
be undetected; and (3) the present dataset excludes the mid-plane and
is therefore strongly biased against SFRs (see Section \ref{sec:search}). The \citet{2002Avedisova} catalogue indicates that item (3) excludes 60\% of Galactic SFRs
(and the true fraction is undoubtedly higher since the catalogue is
incomplete for distant SFRs in the mid-plane). The effect of item (2)
cannot be reliably quantified, but it is likely to raise the source
density by at least an order of magnitude. E.g. typical members of our
sample with $K$ = 14.8 mag, $d$=1.4 to 2~kpc (sources in Cygnus X or
Serpens OB2) would be below our $K$=16 magnitude cut at $d > 3.4$~kpc. The 
overwhelming majority of Galactic SFRs lie at greater distances than 
this. The study of the spiral structure in the Galaxy by \citet{2003Russeil} is said to be complete for star forming complexes with excitation parameters, U, brighter than 60 pc cm$^{-2}$. We find that 85\% of these are found to be at $d>3.4$~kpc. Also, most PMS stars are less luminous than those in our sample but presumably 
can still have high amplitude variability, judging by the rising incidence to
faint magnitudes. Finally, item (1) is believed to cause approximately
$75\%$ incompleteness: our analysis of high-amplitude variables arising from the 2010-2012 VVV data (with at least 14 epochs of K band observations for each variable) reveals that only $\sim$ 25\% of such variables would be detected when comparing the magnitudes of two epochs separated by $\sim$ 2 years. Adopting
factors of 4, 6.7 and 2.5 for items 1, 2 and 3 respectively, our observed
surface density of 0.19~deg$^{-2}$ rises to 12.7~deg$^{-2}$. If we assume that 5 YSOs in our sample are actually 
chance spatial associations of other types of variable (see $\S$2.2), and also remove the 4 stars that did not 
pass our 300$\arcsec$ criterion for SFR association but were still included, the surface density still remains 
high at 8.7~deg$^{-2}$.
This figure is of course highly uncertain, and the figure of 2.7~deg$^{-2}$ for 
Miras in quadrant 1 sight lines is also quite uncertain (by perhaps a factor of two given that
the space density of AGB stars has a $\sim$50\% uncertainty in Ishihara et al.(2011) and the error 
on the 8\% fraction of AGB stars that are Mira variables is not clear in the literature). Nonetheless, we
conclude that PMS high amplitude infrared variables are likely to be
the commonest type of high amplitude infrared variable in most parts of the Milky Way, except in the Bulge 
where AGB stars have higher densities. The true
incidence will no doubt be established more precisely in the near future using
data from VVV \citep[][]{2010Minniti}.

Optical surveys have been used effectively to identify low mass PMS stars even in regions where molecular gas, a common tracer for SFRs, has dissipated \citep{2005Briceno}. Although the near-IR variability of such objects has been less thoroughly studied, previous surveys show that the majority of YSOs are variable at these wavelengths due to different physical processes \citep[see e.g.][]{2012Rice}. However, peak to peak amplitudes have been found to be low,  with a mean of 0.17 mag in the work of e.g. \citet{2001Carpenter}. Only a small fraction show larger amplitudes, e.g. only $\sim 3 \%$ of stars with discs show $\Delta K > 0.5$ mag in the sample of \citet{2012Scholz}. The fraction seems remarkably higher ($\sim50\%$) in the recent time series study of Class I protostars \citet{2012Rice} (see Section \ref{sec:intro}) which suggests that infrared variability surveys can most easily pick out the parts of molecular clouds where star formation has occurred very recently. \citet{2012Scholz} also shows that the fraction of high-amplitude variables increases with extending time baselines. 

It is possible then that high-amplitude variability in the infrared, studied over a long baseline such as the one in GPS, can also effectively trace low mass PMS stars and thus evidence of star formation. This could be supported by the finding of the four objects with YSO colours and not associated with known SFRs. Only marginal and inconclusive evidence for any association to star formation could be found for three of these sources. The projected location of GPSV15 is at the edge of the dark cloud LDN 667 \citep{1962Lynds,2002Dutra}, GPSV36 is within 2 arcmin of reflection Nebula GN 18.08.8 \citep{2003Magakian} and GPSV43 is located in a highly reddened field near the edge of the W40 HII region \citep{2006Quireza}. A search for low-amplitude near-infrared variability in the GPS near these sources does not reveal the presence of a PMS association, however a more statistically thorough study is necessary to firmly discard the presence of such an association.  
 
\section{Follow up Observations}
           
The high variability of the candidates, along with the observed colours and the possible association with SFRs for many of them, makes the stars strong candidates for eruptive variable classification.

This study will focus on the analysis of the eruptive variable candidates arising from DR5 of GPS; the analysis of DR7 objects will be presented on a later paper (Contreras Pe\~{n}a et al., in prep).
        
\begin{table*}
\begin{flushleft}
\caption{Parameters of the high-amplitude variables from UKIDSS GPS. For the description of the columns see Section \ref{sec:search}}\label{table:par}
\begin{tabular}{@{}l@{\hspace{0.15cm}}l@{\hspace{0.1cm}}c@{\hspace{0.15cm}}c@{\hspace{0.15cm}}c@{\hspace{0.2cm}}c@{\hspace{0.2cm}}c@{\hspace{0.2cm}}c@{\hspace{0.2cm}}c@{\hspace{0.2cm}}c@{\hspace{0.15cm}}c@{\hspace{0.15cm}}c@{}}
\hline
n$^{\circ}$ & Object ID & GPS Designation & $\alpha$  & $\delta$  & $K_{1}$ & $J-K_{1}$ & $H-K_{1}$ & $K_2-K_1$ & $\Delta K_{all}$ & $N_{K}$ & DR\\
&  &  & (J2000) & (J2000) & (mag) & (mag) & (mag) & (mag) & (mag) & &  \\
\hline
1 & GPSV36  & UGPS J181140.3$-$152954.2 & 18:11:40.30  & -15:29:54.27 & 12.20(0.02) & 2.56(0.02) & 1.05(0.02)  & 1.43 & 1.76 & 4 & DR7\\
2 & GPSV37  & UGPS J181219.4$-$150304.3 & 18:12:19.43  & -15:03:04.30 & 15.80(0.04) & 1.14(0.04) & 0.30(0.04) & 1.14 & 1.14 & 2 & DR7\\
3 & GPSV1 & UGPS J181452.9$-$115748.7 & 18:14:52.94 & -11:57:48.77 & 15.11(0.02) & 2.72(0.04) & 1.21(0.03) & 3.75 & 3.75 & 3 & DR5 \\
4 & GPSV8 & UGPS J181556.9$-$114113.8 & 18:15:56.91 & -11:41:13.82 & 12.79(0.02) & 1.94(0.02) & 0.75(0.02) & -1.07 & 1.17 & 6 & DR5\\
5 & GPSV10 & UGPS J181612.8$-$113809.1 & 18:16:12.83 & -11:38:09.17 & 13.99(0.02) & 3.36(0.03) & 1.56(0.02) & -1.28 & 1.50 & 6 & DR5\\
6 & GPSV7 & UGPS J181707.1$-$115142.5 & 18:17:07.12 & -11:51:42.55 & 14.15(0.02) & 2.71(0.02) & 1.29(0.02) & -1.00 & 1.00 & 4 & DR5\\
7 & GPSV9 & UGPS J181717.3$-$115854.3 & 18:17:17.38 & -11:58:54.30 & 16.43(0.05) & $>$2.30 & $>$2.56 & -1.16 & 1.16 & 4 & DR5\\
8 & GPSV5 & UGPS J181727.1$-$113531.4 & 18:17:27.11 & -11:35:31.49 & 15.53(0.02) & 2.08(0.04) & 0.77(0.03) & 1.05 & 1.05 & 4 & DR5\\
9 & GPSV4 & UGPS J181753.5$-$120116.6 & 18:17:53.51 & -12:01:16.61 & 15.53(0.02) & 2.01(0.04) & 0.95(0.04) & 1.07 & 1.07 & 4 & DR5\\
10 & GPSV38  & UGPS J181758.5$-$091831.4 & 18:17:58.57  & -09:18:31.40 & 14.38(0.02) & $>$5.43 & $>$4.62 & -2.68 & 2.68 & 4 & DR7\\
11 & GPSV11 & UGPS J181802.1$-$115424.8 & 18:18:02.14 & -11:54:24.85 & 15.84(0.03) & 3.55(0.18) & 1.69(0.07) & -1.34 & 2.62 & 5 & DR5\\
12 & GPSV6 & UGPS J181809.2$-$115519.6 & 18:18:09.24 & -11:55:19.61 & 16.37(0.05) & 1.88(0.08) & 0.67(0.07) & -1.00 & 1.00 & 4 & DR5\\
13 & GPSV3 & UGPS J181844.7$-$113651.9 & 18:18:44.74 & -11:36:51.92 & 12.47(0.02) & $>$7.00 & 3.56(0.02) & 1.28 & 1.72 & 6 & DR5\\
14 & GPSV2 & UGPS J181933.6$-$112800.8 & 18:19:33.64 & -11:28:00.80 & 14.47(0.02) & 1.95(0.02) & 0.72(0.02) & 1.94 & 2.45 & 6 & DR5\\
15 & GPSV39  & UGPS J182126.3$-$121224.7 & 18:21:26.31  & -12:12:24.75 & 15.12(0.02) & 1.03(0.02) & 0.35(0.03)  & 1.11 & 1.11 & 2 & DR7\\
16 & GPSV40  & UGPS J182206.7$-$123254.9 & 18:22:06.71  & -12:32:54.95 & 16.67(0.09) & 3.39(0.32) & 1.78(0.21) & -1.10 & 1.10 & 2 & DR7\\
17 & GPSV41  & UGPS J182642.1$-$062747.3 & 18:26:42.18  & -06:27:47.30 & 17.31(0.11) & 1.28(0.13) & 0.38(0.14) & -1.34 & 1.34 & 2 & DR7\\
18 & GPSV42  & UGPS J183135.3$-$021424.2 & 18:31:35.31  & -02:14:24.21 & 13.99(0.02) & 3.55(0.02) & 1.30(0.02) & 1.45 & 1.45 & 3 & DR7\\
19 & GPSV43$^{a}$  & UGPS J183237.6$-$023102.7 & 18:32:37.67  & -02:31:02.70 & 14.60(0.02) & $>$2.06 & 1.90(0.02) & -1.06 & 1.18 & 4 & DR7\\
20 & GPSV44  & UGPS J183321.4$-$045726.0 & 18:33:21.45  & -04:57:26.06 & 13.46(0.02) & 4.66(0.04) & 2.18(0.02) & 2.30 & 2.30 & 2 & DR7\\
21 & GPSV45  & UGPS J184221.9$-$080459.6 & 18:42:21.94  & -08:04:59.68 & 11.37(0.02) & 7.78(0.16) & 5.11(0.03) & 2.01 & 2.01 & 4 & DR7\\
22 & GPSV18 & UGPS J184834.5$-$053518.3 & 18:48:34.53  & -05:35:18.36 & 15.18(0.02) & 1.83(0.02) & 0.60(0.02) & -1.10 & 1.10 & 2 & DR7\\
23 & GPSV17 & UGPS J184937.6$-$093350.9 & 18:49:37.60 & -09:33:50.94 & 17.77(0.16) & 0.36(0.17) & 0.10(0.18) & -2.35 & 2.35 & 2 & DR5\\
24 & GPSV12 & UGPS J185318.8$-$094329.1 & 18:53:18.82 & -09:43:29.15 & 15.77(0.03) & 0.78(0.03) & 0.16(0.03) & 1.10 & 1.29 & 3 & DR5\\
25 & GPSV19  & UGPS J190003.3+010528.7 & 19:00:03.34  &  01:05:28.73 & 14.19(0.02) & 5.27(0.18) & 2.27(0.03) & 2.08 & 2.08 & 2 & DR7\\
26 & GPSV20  & UGPS J190028.0$-$064918.7 & 19:00:28.02  & -06:49:18.77 & 13.52(0.02) & 0.54(0.02) & 0.16(0.02) & 1.04 & 1.04 & 4 & DR7\\
27 & GPSV13 & UGPS J190122.5+115203.2 & 19:01:22.50 & 11:52:03.24 & 12.92(0.02) & 0.29(0.02) & 0.14(0.02) & -1.02 & 1.23 & 4 & DR5\\
28 & GPSV21 & UGPS J190146.8+011349.3 & 19:01:46.89  &  01:13:49.37 & 17.27(0.09) & $>$2.53 & 1.86(0.18) & -1.81 & 1.81 & 2 & DR7\\
29 & GPSV22  & UGPS J190147.0.021228.0 & 19:01:47.07  &  01:12:28.08 & 13.40(0.02) & 4.89(0.04) & 1.86(0.02) & 1.72 & 1.72 & 2 & DR7\\
30 & GPSV15 & UGPS J190332.1+120557.2 & 19:03:32.11 & 12:05:57.25 & 11.21(0.02) & 6.23(0.02) & 2.93(0.02) & 1.03 & 1.38 & 4 & DR5\\
31 & GPSV14 & UGPS J190815.0+124021.4 & 19:08:15.06 & 12:40:21.44 & 15.98(0.03) & 0.88(0.03) & 0.25(0.03) & 2.08 & 2.08 & 2 & DR5\\
32 & GPSV23  & UGPS J191557.0.022906.7 & 19:15:57.05  &  01:29:06.77 & 15.78(0.02) & 0.61(0.02) & 0.16(0.02) & 1.01 & 1.63 & 3 & DR7\\
33 & GPSV16 & UGPS J201310.3+333128.1 & 20:13:10.31 & 33:31:28.17 & 13.46(0.02) & 3.20(0.02) & 1.47(0.02) & 2.21 & 2.21 & 3 & DR5\\
34 & GPSV24  & UGPS J201613.5+415436.5 & 20:16:13.50  &  41:54:36.51 & 14.83(0.02) & 2.73(0.02) & 0.99(0.02) & 1.07 & 1.07 & 2 & DR7\\
35 & GPSV25 & UGPS J201924.8+410503.0 & 20:19:24.83  &  41:05:03.07 & 14.82(0.02) & 2.33(0.02) & 0.97(0.02) & 1.22 & 1.22 & 2 & DR7\\
36 & GPSV26  & UGPS J202038.9+405606.8 & 20:20:38.91  &  40:56:06.86 & 14.48(0.02) & 3.13(0.02) & 1.19(0.02) & 1.36  & 1.48 & 3 & DR7\\
37 & GPSV27  & UGPS J202117.2+403825.1 & 20:21:17.28  &  40:38:25.18 & 15.73(0.02) & 3.28(0.08) & 1.15(0.03) & 1.14  & 1.14 & 3 & DR7\\
38 & GPSV28  & UGPS J202421.3+421605.6 & 20:24:21.37  &  42:16:05.63 & 11.88(0.02) & $>$7.92 & 2.62(0.02)  & 1.00 & 1.56 & 3 & DR7\\
39 & GPSV29  & UGPS J202504.2+403332.6 & 20:25:04.23  &  40:33:32.63 & 15.76(0.02) & 1.14(0.02) & 0.38(0.02) & 1.42 & 1.42 & 2 & DR7\\
40 & GPSV30  & UGPS J202605.3+420932.9 & 20:26:05.36  &  42:09:32.94 & 15.38(0.02) & $>$4.42 & 2.86(0.11) & 1.51 & 1.51 & 2 & DR7\\
41 & GPSV31  & UGPS J202759.3+415255.7 & 20:27:59.34  &  41:52:55.79 & 16.60(0.05) & $>$3.20 & 1.37(0.09) & -1.16 & 1.16 & 2 & DR7\\
42 & GPSV32  & UGPS J203117.5+413242.5 & 20:31:17.58  &  41:32:42.52 & 15.87(0.03) & 2.86(0.10) & 1.06(0.04) & -1.00 & 1.00 & 2 & DR7\\
43 & GPSV33  & UGPS J203352.9+413844.9 & 20:33:52.90  &  41:38:44.94 & 16.41(0.04) & 2.59(0.14) & 0.81(0.07) & -1.13 & 1.13 & 2 & DR7\\
44 & GPSV34  & UGPS J203427.1+421353.3 & 20:34:27.13  &  42:13:53.31 & 11.48(0.02) & 4.58(0.02) & 1.82(0.02) & 1.74 & 2.11 & 3 & DR7\\
45 & GPSV35  & UGPS J204142.0+393852.0 & 20:41:42.01  &  39:38:52.03 & 16.84(0.05) & $>$2.96 & $>$2.16 & -1.79 & 2.79 & 3 & DR7\\
\hline
\multicolumn{12}{@{}l}{{\footnotesize $^a$ For GPSV43 the J band image is not contemporaneous to H and K$_1$. Given the red H-K$_{1}$ colour we regard the J-K$_{1}$} }\\
\multicolumn{12}{@{}l}{colour as a lower limit.}
\end{tabular}
\end{flushleft}
\end{table*}

\subsection{VLT/ISAAC}\label{isaac:dr}

Spectroscopic and photometric follow-up observations of 15 of the DR5 candidates were 
obtained in visitor mode during two consecutive nights in June 30th/July 1st 2010. 
The data were acquired at the Very Large Telescope (VLT) in Cerro Paranal, Chile with the 
ISAAC imager and spectrograph. 
           
Observations were carried out with the short-wavelength (1-2.5 $\mu$m) arm of ISAAC, equipped with a Hawaii Rockwell 1024$\times$1024 array The imaging mode has a scale of 0.147$\arcsec$/px with a field of view of 152$\arcsec \times$152$\arcsec$.          
           
Low-resolution (0.8$\arcsec$slit, R=$\Delta \lambda / \lambda \sim$ 700) $K_{s}$ spectra were obtained in two different positions along the direction of the slit (positions A and B), in an ABBA sequence with individual integration
times of 120~s.
In order to remove telluric lines, main sequence F-type stars were observed consecutively at similar airmass with the same instrumental setup. Filament lamp images were taken for purposes of flat-fielding of the data, while Xenon-Argon lamp images were acquired for wavelength calibration and to model the slit curvature. In addition a set of 12 STARTRACE images were downloaded from the ESO archive to correct for the tilt of ISAAC spectra.
            
Flat-fielding, wavelength calibration, tilt correction and extraction of the spectrum for each object and the corresponding F-type calibrator were performed with the usual tasks in the NOAO/TWODSPEC package in IRAF. Finally, the calibrator spectrum was divided by a blackbody curve of similar temperature to that of the star and used to correct the target spectrum. 
            
Photometric observations for each target consisted of 12 images in $J_{s}$, $H$ and $K_{s}$  with typical exposure times of 5 s. Each image was observed in a different position within a 15$\arcsec$ wide jitter box. The seeing at $K_{s}$ varied between 0.65-1.1$\arcsec$ for the first night and 1.0-2.04$\arcsec$ for the second night of observations.
           
Dark images and sky-flats were obtained for bias subtraction and to correct for pixel-to-pixel variations in the detector response. The latter consisted of $\sim$ 10-15 images in filters $J_{s}$ and $H$ with exposure times of $5$ s. $K_{s}$ images were not taken for this project so suitable images were acquired from the ESO archive, taking care to select images that were observed with the same instrumental setup, exposure times and a similar observing date. Source magnitudes were obtained through aperture photometry, performed on final sky-subtracted images, using APPHOT in IRAF. The calibration of the final magnitudes was obtained with the process described below. 

\subsubsection{Photometric Calibration}

Firstly, a set of 180 local standards from the IRAF aperture photometry results were selected, corresponding to bright stars, with low photometric errors, no processing errors and with no close companions. Cross-match with the UKIDSS GPS catalog of the same area covered by ISAAC, yielded typically $\sim$ 60-100 matches. 
   
By taking the average (after removal of outlying sources) of  $\Delta m = M_{UKIDSS} - m_{ISAAC}$, we obtain the offset that will be applied to the ISAAC instrumental magnitudes. This process is sufficiently accurate for both $J_{s}$ and $H$, given the similar $\Delta \lambda$ and $\lambda_{c}$ for the UKIDSS and ISAAC filters, however, $K_{s,ISAAC}$ differs from $K_{UKIDSS}$, being similar to 2MASS $K_s$ filter profile.

\begin{figure}
\begin{center}
\resizebox{8cm}{8cm}{\includegraphics{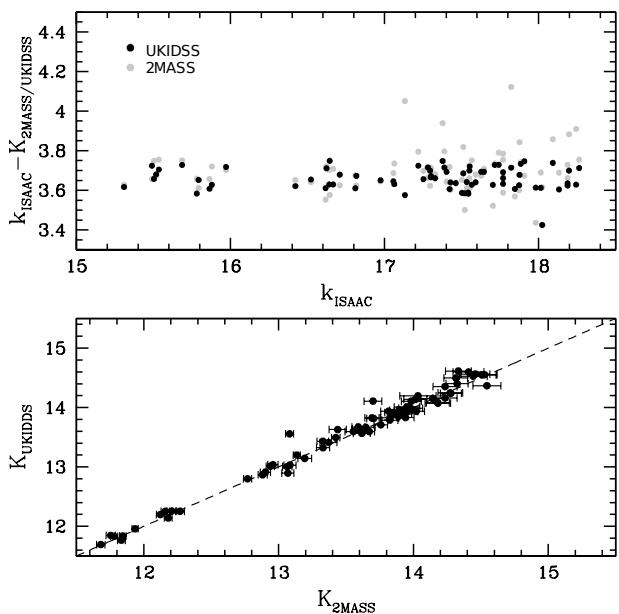}}
\caption{Comparison of 2MASS and UKIDSS $K$ for a set of local standards in the field of view centered on GPSV3, as explained in the text (\textit{bottom}), the dashed line represents the identity line. The \textit{top} graph shows the offset between ISAAC instrumental magnitudes and UKIDSS (\textit{black filled circles}) and 2MASS (\textit{grey filled circles}) photometry.}
\label{uk2m}
\end{center}
\end{figure}

To study the differences between UKIDSS and 2MASS $K$, we cross-matched our local standards with the 2MASS catalog for the same area, finding around 20-40 stars. The differences between UKIDSS and 2MASS are plotted in Fig. \ref{uk2m}, here it can be seen that magnitudes are similar in both systems, especially for brighter sources. Thus the selection of either system for the calibration would not have a severe effect on the calibrated magnitudes.  This was found to be true for each of our objects. Given the greater number of UKIDSS stars found, and the greater precision of UKIDSS magnitudes for the fainter stars, we decide to use UKIDSS photometry for the $K_{s}$ calibration. A magnitude correction, however, needs to be applied for our sample, especially the case of the reddest stars where most of its flux falls in the $K$-band. Using the reduced spectrum of the object, a correction factor was derived by comparing the ratio of the total flux of the star in UKIDSS and ISAAC $K$ filters, to the same ratio for a blackbody spectrum at T = 9700~K (which is appropriate for these Vega-based magnitudes). The correction factor is then given by
 
\begin{equation}
\Delta K = -2.5\times \log \frac{F_{\star, UKIDSS}/F_{\star,ISAAC}}{F_{bb,UKIDSS}/F_{bb,ISAAC}}   
\end{equation}

This correction was largest for the reddest stars in our sample but never exceeded 0.05 mag.

\subsection{Magellan Baade/FIRE}
 
    Follow up of the two reddest objects in the DR5 sample, GPSV3 and GPSV15, was performed on May 8th, 2012 with the FIRE spectrograph mounted on the Magellan Baade Telescope at Las Campanas Observatory, Chile. The observations were carried out in the high-throughput prism mode which provides a continuous coverage from $0.8-2.5~\mu m$ at low resolution of $R \sim 250-350$ with a 0.6$\arcsec$ slit. 
    
    The objects were observed in the usual ABBA pattern along the slit, with individual exposures of 148 s and at airmasses of 1.1 and 1.4 for GPSV3 and GPSV15 respectively. Observations of spectroscopic calibrators were carried out for purposes of telluric correction and flux calibration; the calibrators were observed at similar airmass as the objects and with the same instrumental setup. Quartz lamp images were obtained for flat fielding of the data, they consisted of high-voltage (2.2V) images that provide data for the z/J bands but saturate the red end of the spectrum, and low-voltage(1.1V) that generate counts in H/K but are too faint for the blue end. Additionally, NeNeAr arc lamp images were acquired for wavelength 
calibration.
    
    The images were reduced using the longslit package of the FIREHOSE software. Using the NeNeAr lamp image, FIREHOSE generates an arc solution, which had typical uncertainties of 0.4 pixels or $\sim$ 2.7 $\AA$. The second step consisted of creating a flat image; this is done by combining the 2.2V (blue) and 1.1V (red) with a smooth weighting function spread over $\sim$ 150 pixels centered on a transition pixel defined by the user. Finally the software traces and extracts the spectrum of the object. Telluric correction and flux calibration were performed in the standard mode using NOAO/TWODSPEC package in IRAF, in a similar process to that described in Section \ref{isaac:dr}.
    
    Unfortunately, the selected exposure times saturated part of the K-band data of both objects, more noticeably in GPSV15. However, the data still provide useful information 
for our analysis, particularly at the shorter wavelengths not covered by the ISAAC or 
NIFS spectra.
 
\subsection{Gemini North/NIFS}

   Additional K-band spectroscopic data of four variables, namely GPSV3, GPSV8, GPSV15 and GPSV16, were acquired in August 2012 using the Near-infrared Integral Field Spectrometer (NIFS) of Gemini North telescope, Mauna Kea, Hawaii. The instrument has a field of view of $3\arcsec \times 3\arcsec$ with a pixel scale of $0.1\arcsec$ across image slices and $0.04\arcsec$ within image slices and a resolving power $R\sim 5300$ at 2.2 $\mu$m. 
   
   The observations are part of the queue mode programme GN-2012B-Q-102. Observations of GPSV16 consisted of 4 integrations of 8 minutes, 2 of which were on source. The other, brighter objects were observed with two 8 minute integrations: one on source and one sky offset. Main sequence F-type stars were also observed for telluric correction and relative flux calibration. For each object a standard set of calibration data were acquired, consisting of flat images, XeAr lamp and its corresponding dark images and Ronchi calibration mask to correct for the spatial distortion of integral field unit.
   
   Data reduction was performed in the standard way using the tasks of the GEMINI/NIFS package in IRAF.     

\subsection{UKIRT/WFCAM}

With the aim of investigating the presence of molecular outflows around eruptive variables 
candidates in Serpens OB2, near-infrared $K$ band and 2.12~$\mu$m (1-0)S(1) H$_{2}$
narrow band imaging data were obtained on August 7th 2012 using the Wide Field Camera 
\citep[WFCAM,][]{2007Casali} mounted on UKIRT. The instrument has four Rockwell Hawaii II arrays (2048$\times$2048 pixels) separated by 90\% of a detector width, with a 
pixel scale of 0.4$\arcsec/$pixel. The standard 2$\times$2 tiling pattern was used to
cover a continuous $0.9^{\circ} \times 0.9^{\circ}$ region containing 10 of the 11 
eruptive variables candidates in the Serpens OB2 association.
The total integration times were 2 minutes in $K$ and 25 minutes at 2.12~$\mu$m. The 
observations consisted of 24 images in $K$, with exposure times of 5~s ($K$) and 20
images at 2.12~$\mu$m, with 75 s exposures. A 2$\times$2 microstepping pattern was used
to fully sample the spatial resolution in both filters, yielding a 0.2$\arcsec$ pixel scale
in the reduced images.
Data reduction and photometry were done by the Cambridge Astronomical Survey Unit (CASU) using
the standard reduction pipeline.

Difference ($H_{2}-K$) images are built in order to reveal emission line regions possibly associated with H$_{2}$ jets and outflows. The continuum subtracted images are built with the standard procedure \citep[see e.g.][]{2012Ioannidis}.

\subsection{VST/OmegaCam VPHAS$+$}\label{sec:vphas}

VPHAS$+$ (Drew et al, in prep.) is an ESO public survey, collecting $u^{'}$,
$g^{'}$, $r^{'}$, $i^{'}$ and narrowband $H\alpha$ photometry of the southern
Galactic Plane and Bulge.  It is in execution on the VLT Survey
Telescope (VST), using the 1 square-degree imager, OmegaCam.  The
camera is constructed around a 4$\times$8 mosaic of 2k$\times$4k
CCDs. The pixel size projects to 0.21$\arcsec$ on the sky, permitting
good sampling of the 0.8-1.0$\arcsec$ point-spread function achieved.
The exposure times for the data used here were $u^{'}$: 150~s,
$g^{'}$: 30~s, $r^{'}$,$i^{'}$: 25~s, and $H\alpha$ 120~s.

To ensure more uniform seeing conditions, finally, across the full filter
set, $u^{'}$, $g^{'}$, $r^{'}$ exposures are obtained as a
contemporaneous group, separately from $r$, $i$ and $H\alpha$
exposures.  At each field pointing, each filter is exposed twice
(or three times, in $H\alpha$) at different offsets to compensate
for gaps between the CCDs making up the mosaic.  The data are reduced
and calibrated by the Cambridge Astronomical Survey Unit (CASU), on the basis of nightly standards.  In principle this should yield an
accuracy of $\pm$0.03 magnitudes.  The scale set should be regarded as
provisional, as it is too early to establish a global calibration.
All magnitudes and colours are referred to Vega as the zero-magnitude,
zero-colour reference object.

The Serpens OB2 region, relevant to this study, was observed as part of
VPHAS$+$ on the nights of June 10th 2012 ($r^{'}$, $i^{'}$ and $H\alpha$) and
November 11th 2012 ($u^{'}$, $g^{'}$, $r^{'}$).

\subsection{Public Surveys}

In addition to UKIDSS and ISAAC near-infrared photometry, we search for detections within 1$\arcsec$ of our objects in other near- and mid- IR surveys in the IPAC science archive,  namely {\it Spitzer-}GLIMPSE3D \citep{2003Benjamin}, {\it Spitzer-}GLIMPSE360 \citep{2011Whitney}, WISE \citep{2010Wright}, MSX6C \citep{2003Egan}, {\it Akari} \citep{2007Murakami}, 2MASS \citep{2006Skrutskie} and DENIS \citep{1994Epchtein}. We also searched for counterparts in the initial data release of IPHAS \citep{2005Drew,2008Gonzalez-Solares}. We note
that all of our sample lies outside the area of the {\it Spitzer-}GLIMPSE survey of the mid-plane but the 11 sources in the Serpens OB2 region are included within GLIMPSE3D, whilst GPSV16 is included within GLIMPSE360, a post-cryogenic {\it Spitzer} mission which only uses filters I1 ($[3.6]$) and I2 ($[4.5]$) \citep{2011Whitney}.

\begin{landscape}
\clearpage
\pagestyle{empty}
\setlength\textwidth{702pt}
\setlength\textheight{42pc}
\begin{table*}
\begin{flushleft}
\caption{Photometric measurements for DR5 objects.}\label{table:photom}
\begin{tabular}{@{}l@{\hspace{0.1cm}}c@{\hspace{0.1cm}}c@{\hspace{0.1cm}}c@{\hspace{0.1cm}}c@{\hspace{0.1cm}}c@{\hspace{0.1cm}}c@{\hspace{0.1cm}}c@{\hspace{0.1cm}}c@{\hspace{0.1cm}}c@{\hspace{0.1cm}}c@{\hspace{0.1cm}}c@{\hspace{0.1cm}}c@{\hspace{0.1cm}}c@{\hspace{0.1cm}}c@{\hspace{0.1cm}}c@{\hspace{0.1cm}}c@{\hspace{0.1cm}}c@{\hspace{0.1cm}}c@{\hspace{0.1cm}}c@{\hspace{0.1cm}}c@{\hspace{0.1cm}}c@{\hspace{0.1cm}}c@{\hspace{0.1cm}}c@{\hspace{0.1cm}}c@{\hspace{0.1cm}}c@{\hspace{0.1cm}}c@{\hspace{0.1cm}}c@{\hspace{0.1cm}}c@{\hspace{0.1cm}}c@{\hspace{0.1cm}}c@{\hspace{0.1cm}}c@{}}
\hline
{\footnotesize Dataset}&\multicolumn{3}{c}{{\footnotesize IPHAS/VPHAS$+$}}&\multicolumn{3}{c}{{\footnotesize DENIS}}&\multicolumn{3}{c}{{\footnotesize 2MASS}}&\multicolumn{4}{c}{{\footnotesize GPS}}&\multicolumn{3}{c}{{\footnotesize VLT/ISAAC}}&\multicolumn{2}{c}{{\footnotesize UKIRT}}&\multicolumn{4}{c}{{\footnotesize Spitzer/IRAC}} & \multicolumn{4}{c}{{\footnotesize WISE}} & \multicolumn{3}{c}{{\footnotesize MSX6C}} & \multicolumn{2}{c}{{\footnotesize Akari}}\\
{\footnotesize filter} & {\footnotesize r$^{'}$} & {\footnotesize i$^{'}$} & {\footnotesize H$_{\alpha}$} & {\footnotesize I} & {\footnotesize J} & {\footnotesize K$_{s}$} & {\footnotesize J} & {\footnotesize H} & {\footnotesize K$_{s}$} & {\footnotesize J$^{a}$} & {\footnotesize H$^{a}$} & {\footnotesize K$_{1}^{a}$} & {\footnotesize K$_{2}^{b}$} & {\footnotesize J} & {\footnotesize H} & {\footnotesize K$_{s}$} &{\footnotesize K} & {\footnotesize $[2.12]$} & {\footnotesize $[3.6]$} & {\footnotesize $[4.5]$} & {\footnotesize $[5.8]$} & {\footnotesize $[8.0]$} & {\footnotesize $[3.4]$} & {\footnotesize $[4.6]$} & {\footnotesize $[12]$} & {\footnotesize $[22]$} & {\footnotesize $[8.28]$} & {\footnotesize $[12.13]$} & {\footnotesize $[14.65]$} & {\footnotesize $[9.0]$} & {\footnotesize $[18.0]$}\\
{\scriptsize  } & {\scriptsize (mag)} & {\scriptsize (mag)} & {\scriptsize (mag)} & {\scriptsize (mag)} & {\scriptsize (mag)} & {\scriptsize (mag)} & {\scriptsize (mag)} & {\scriptsize (mag)} & {\scriptsize (mag)} & {\scriptsize (mag)} & {\scriptsize (mag)} & {\scriptsize (mag)} & {\scriptsize (mag)} & {\scriptsize (mag)} & {\scriptsize (mag)} & {\scriptsize (mag)} &{\scriptsize (mag)} & {\scriptsize (mag)} & {\scriptsize (mag)} & {\scriptsize (mag)} &{\scriptsize (mag)} & {\scriptsize (mag)} & {\scriptsize (mag)} & {\scriptsize (mag)} & {\scriptsize (mag)} & {\scriptsize (mag)} & {\scriptsize (mag)} & {\scriptsize (mag)} & {\scriptsize (mag)} &{\scriptsize (mag)} & {\scriptsize (mag)}\\
{\footnotesize 5$\sigma$ Limit} & {\footnotesize \it{21.8}} & {\footnotesize \it{21}} & {\footnotesize \it{20.8}} & {\footnotesize \it{18.5}} & {\footnotesize \it{16.5}} & {\footnotesize \it{14.5}} & {\footnotesize \it{17}} & {\footnotesize \it{16.3}} & {\footnotesize \it{15.5}} & {\footnotesize \it{19.8}} & {\footnotesize \it{19}} & {\footnotesize \it{18.1}} & {\footnotesize \it{18.1}} & {\footnotesize \it{21}} & {\footnotesize \it{19.9}} & {\footnotesize \it{19.5}} &{\footnotesize \it{17.3}} & {\footnotesize \it{17.4}} & {\footnotesize \it{15.4}} & {\footnotesize \it{14.9}} & {\footnotesize \it{13.6}} & {\footnotesize \it{13}} & {\footnotesize \it{16.6}} & {\footnotesize \it{15.6}} & {\footnotesize \it{11.3}} & {\footnotesize \it{8.0}} & {\footnotesize \it{6.9}} & {\footnotesize \it{3.5}} & {\footnotesize \it{3.3}} & {\footnotesize \it{7.4}} & {\footnotesize \it{4.2}}\\
\hline
{\footnotesize GPSV1}& {\footnotesize {\it nd}} & {\footnotesize {\it nd}} & {\footnotesize {\it nd}} &{\footnotesize \it{nd}}&{\footnotesize \it{nd}}&{\footnotesize \it{nd}}& {\footnotesize \it{nd}} & {\footnotesize \it{nd}} & {\footnotesize \it{nd}} & {\footnotesize 17.83} & {\footnotesize 16.32} & {\footnotesize 15.11} & {\footnotesize 18.86:} & {\footnotesize 17.98} & {\footnotesize 16.49} & {\footnotesize 15.38} & {\footnotesize {\it nc}} & {\footnotesize {\it nc}}& {\footnotesize 15.26:} & {\footnotesize 14.47:} & {\footnotesize 13.33:} & {\footnotesize \it{nd}} & {\footnotesize \it{nd}} & {\footnotesize \it{nd}} & {\footnotesize \it{nd}} & {\footnotesize \it{nd}} & {\footnotesize \it{nd}}& {\footnotesize \it{nd}} & {\footnotesize \it{nd}} & {\footnotesize \it{nd}} & {\footnotesize \it{nd}}\\
{\scriptsize } & {\scriptsize --} & {\scriptsize --} & {\scriptsize --} &{\scriptsize -- }&{\scriptsize --}&{\scriptsize --}& {\scriptsize --} & {\scriptsize --} & {\scriptsize --} & {\scriptsize (0.04)} & {\scriptsize (0.02)} & {\scriptsize (0.02)} & {\scriptsize (0.55)} & {\scriptsize (0.05)} & {\scriptsize (0.03)} & {\scriptsize (0.04)} & {\scriptsize --} & {\scriptsize --}& {\scriptsize (0.16)} & {\scriptsize (0.35)} & {\scriptsize (0.34)} & {\scriptsize --} & {\scriptsize --} & {\scriptsize --} & {\scriptsize --} & {\scriptsize --} & {\scriptsize --}& {\scriptsize --} & {\scriptsize --} & {\scriptsize --} & {\scriptsize --}\\
{\footnotesize GPSV2} &{\footnotesize 20.92:} & {\footnotesize 19.06} & {\footnotesize 20.04:}&{\footnotesize 18.14}&{\footnotesize 16.06}&{\footnotesize 13.98}&{\footnotesize 15.72}&{\footnotesize 14.59}&{\footnotesize 13.96}&{\footnotesize 16.43}&{\footnotesize 15.19}&{\footnotesize 14.47}&{\footnotesize 16.41}&{\footnotesize 17.79}&{\footnotesize 16.06}&{\footnotesize 15.08}&{\footnotesize 14.30}&{\footnotesize 14.28}& {\footnotesize 13.11} & {\footnotesize 12.28} & {\footnotesize 11.40} & {\footnotesize 10.42} & {\footnotesize 13.61} & {\footnotesize 11.53} & {\footnotesize 7.92} & {\footnotesize 6.00:} & {\footnotesize \it{nd}}& {\footnotesize \it{nd}} & {\footnotesize \it{nd}} & {\footnotesize \it{nd}} & {\footnotesize \it{nd}}\\
{\scriptsize }  & {\scriptsize (0.10)} & {\scriptsize (0.04)} & {\scriptsize (0.12)} &{\scriptsize (0.18)} &{\scriptsize (0.24)} &{\scriptsize (0.22)} &{\scriptsize (0.08)} &{\scriptsize (0.07)} &{\scriptsize (0.06)} &{\scriptsize (0.02)} &{\scriptsize (0.02)} &{\scriptsize (0.02)} &{\scriptsize (0.06)} & {\scriptsize (0.05)} & {\scriptsize (0.03)} & {\scriptsize (0.04)}&{\scriptsize (0.02)} &{\scriptsize (0.02)} &{\scriptsize (0.08)} &{\scriptsize (0.06)} &{\scriptsize (0.07)} &{\scriptsize (0.05)} &{\scriptsize (0.07)} &{\scriptsize (0.04)} &{\scriptsize (0.05)} &{\scriptsize (0.18)}&{\scriptsize --}& {\scriptsize --} & {\scriptsize --} & {\scriptsize --} & {\scriptsize --}\\
{\footnotesize GPSV3} &{\footnotesize {\it nd}} & {\footnotesize {\it nd}} & {\footnotesize {\it nd}}&{\footnotesize \it{nd}}&{\footnotesize \it{nd}}&{\footnotesize 12.28}&{\footnotesize {\tiny $>$}16.62}&{\footnotesize {\tiny $>$}16.39}&{\footnotesize 13.28}&{\footnotesize \it{nd}}&{\footnotesize 16.03}&{\footnotesize 12.47}&{\footnotesize 13.75}&{\footnotesize \it{nd}}&{\footnotesize 17.24}&{\footnotesize 13.81}&{\footnotesize 12.09}& {\footnotesize 12.51}& {\footnotesize 9.25} & {\footnotesize 7.60} & {\footnotesize 6.30} & {\footnotesize 5.30} &{\footnotesize  9.27} & {\footnotesize 7.02} & {\footnotesize 4.44} & {\footnotesize 3.29:} & {\footnotesize 4.32} & {\footnotesize 3.43:} &{\footnotesize  3.39:}&{\footnotesize  4.77 }&{\footnotesize  3.39}\\
{\scriptsize } &{\scriptsize --} &{\scriptsize --} &{\scriptsize --} &{\scriptsize --} &{\scriptsize --} &{\scriptsize (0.13)} &{\scriptsize --} &{\scriptsize --} &{\scriptsize (0.04)} &{\scriptsize --} &{\scriptsize (0.02)} &{\scriptsize (0.02)} &{\scriptsize (0.02)} &{\scriptsize --} &{\scriptsize (0.04)} &{\scriptsize (0.05)} &{\scriptsize (0.02)} &{\scriptsize (0.02)} &{\scriptsize (0.05)} &{\scriptsize (0.05)} &{\scriptsize (0.03)} &{\scriptsize (0.03)} &{\scriptsize (0.03)} &{\scriptsize (0.02)} &{\scriptsize (0.04)} &{\scriptsize (0.16)} &{\scriptsize (0.11)} &{\scriptsize (0.18)} &{\scriptsize (0.20)} &{\scriptsize (0.47)} &{\scriptsize (0.11)}\\
{\footnotesize GPSV4} &{\footnotesize {\it nd}} & {\footnotesize {\it nd}} & {\footnotesize {\it nd}}&{\footnotesize \it{nd}}&{\footnotesize \it{nd}}&{\footnotesize \it{nd}}&{\footnotesize \it{nd}}&{\footnotesize \it{nd}}&{\footnotesize \it{nd}}&{\footnotesize 17.54}&{\footnotesize 16.48}&{\footnotesize 15.53}& {\footnotesize 16.60} & {\footnotesize 17.80} & {\footnotesize 16.87} & {\footnotesize 16.26} & {\footnotesize 16.42} & {\footnotesize 16.42}& {\footnotesize \it{nd}} & {\footnotesize \it{nd}} & {\footnotesize \it{nd}} & {\footnotesize \it{nd}} & {\footnotesize \it{nd}} & {\footnotesize \it{nd}} & {\footnotesize \it{nd}} & {\footnotesize \it{nd}} & {\footnotesize \it{nd}}& {\footnotesize \it{nd}} & {\footnotesize \it{nd}} & {\footnotesize \it{nd}} & {\footnotesize \it{nd}}\\
{\scriptsize } & {\scriptsize --} &{\scriptsize --} &{\scriptsize --} &{\scriptsize --} &{\scriptsize --} &{\scriptsize --} &{\scriptsize --} &{\scriptsize --} &{\scriptsize --} &{\scriptsize (0.04)} &{\scriptsize (0.03)} &{\scriptsize (0.02)} &{\scriptsize (0.07)} &{\scriptsize (0.06)} &{\scriptsize (0.04)} &{\scriptsize (0.06)} &{\scriptsize (0.02)} &{\scriptsize (0.04)} &{\scriptsize --} &{\scriptsize --} &{\scriptsize --} &{\scriptsize --} &{\scriptsize --} &{\scriptsize --} &{\scriptsize --} &{\scriptsize --} &{\scriptsize --} &{\scriptsize --} &{\scriptsize --} &{\scriptsize --} &{\scriptsize --} \\
{\footnotesize GPSV5} &{\footnotesize {\it nd}} & {\footnotesize {\it nd}} & {\footnotesize {\it nd}}&{\footnotesize \it{nd}}&{\footnotesize \it{nd}}&{\footnotesize \it{nd}}&{\footnotesize \it{nd}}&{\footnotesize \it{nd}}&{\footnotesize \it{nd}}&{\footnotesize 17.61}&{\footnotesize 16.30}&{\footnotesize 15.53}&{\footnotesize 16.58}& {\footnotesize 18.19}&{\footnotesize 16.85}&{\footnotesize 15.97}&{\footnotesize 15.55}&{\footnotesize 15.62}& {\footnotesize \it{nd}} & {\footnotesize \it{nd}} & {\footnotesize \it{nd}} & {\footnotesize \it{nd}} & {\footnotesize \it{nd}} & {\footnotesize \it{nd}} & {\footnotesize \it{nd}} & {\footnotesize \it{nd}} & {\footnotesize \it{nd}}& {\footnotesize \it{nd}} & {\footnotesize \it{nd}} & {\footnotesize \it{nd}} & {\footnotesize \it{nd}}\\
{\scriptsize } & {\scriptsize --} &{\scriptsize --} &{\scriptsize --} &{\scriptsize --} &{\scriptsize --} &{\scriptsize --} &{\scriptsize --} &{\scriptsize --} &{\scriptsize --} &{\scriptsize (0.03)} &{\scriptsize (0.02)} &{\scriptsize (0.03)} &{\scriptsize (0.07)} &{\scriptsize (0.05)} &{\scriptsize (0.04)} &{\scriptsize (0.05)} &{\scriptsize (0.02)} &{\scriptsize (0.02)} &{\scriptsize --} &{\scriptsize --} &{\scriptsize --} &{\scriptsize --} &{\scriptsize --} &{\scriptsize --} &{\scriptsize --} &{\scriptsize --} &{\scriptsize --} &{\scriptsize --} &{\scriptsize --} &{\scriptsize --} &{\scriptsize --} \\
{\footnotesize GPSV6} &{\footnotesize {\it nd}} & {\footnotesize {\it nd}} & {\footnotesize {\it nd}}&{\footnotesize \it{nd}}&{\footnotesize \it{nd}}&{\footnotesize \it{nd}}&{\footnotesize \it{nd}}&{\footnotesize \it{nd}}& {\footnotesize \it{nd}}& {\footnotesize 18.25} & {\footnotesize 17.04} & {\footnotesize 16.37} & {\footnotesize 15.36} & {\footnotesize 17.98} & {\footnotesize 16.89} & {\footnotesize 16.18} & {\footnotesize 15.77} & {\footnotesize 15.80}& {\footnotesize \it{nd}} & {\footnotesize \it{nd}} & {\footnotesize \it{nd}} & {\footnotesize \it{nd}} & {\footnotesize \it{nd}} & {\footnotesize \it{nd}} & {\footnotesize \it{nd}} & {\footnotesize \it{nd}} & {\footnotesize \it{nd}}& {\footnotesize \it{nd}} & {\footnotesize \it{nd}} & {\footnotesize \it{nd}} & {\footnotesize \it{nd}}\\
{\scriptsize } & {\scriptsize --} & {\scriptsize --} & {\scriptsize --} & {\scriptsize --} &{\scriptsize --} & {\scriptsize --} & {\scriptsize --} & {\scriptsize --} &{\scriptsize --} & {\scriptsize (0.07)} & {\scriptsize (0.04)} &{\scriptsize (0.05)} & {\scriptsize (0.03)} & {\scriptsize (0.04)} &{\scriptsize (0.03)} & {\scriptsize (0.04)} & {\scriptsize (0.02)} &{\scriptsize (0.02)} & {\scriptsize --} & {\scriptsize --} & {\scriptsize --} &{\scriptsize --} & {\scriptsize --} & {\scriptsize --} & {\scriptsize --} &{\scriptsize --} & {\scriptsize --} & {\scriptsize --} & {\scriptsize --} &{\scriptsize --} & {\scriptsize --} \\
{\footnotesize GPSV7} &{\footnotesize {\it nd}} & {\footnotesize {\it nd}} & {\footnotesize {\it nd}}&{\footnotesize \it{nd}}&{\footnotesize \it{nd}}&{\footnotesize \it{nd}}& {\footnotesize \it{nd}} & {\footnotesize \it{nd}} & {\footnotesize \it{nd}} & {\footnotesize 16.86} & {\footnotesize 15.44} & {\footnotesize 14.15} & {\footnotesize 13.15} & {\footnotesize 16.53} & {\footnotesize 15.00} & {\footnotesize 13.82} & {\footnotesize 13.30} & {\footnotesize 13.48}&{\footnotesize  11.82 }&{\footnotesize  11.33}&{\footnotesize  10.96 }&{\footnotesize  9.80 }&{\footnotesize  11.20 }& {\footnotesize 10.71 }&{\footnotesize  8.42 }&{\footnotesize {\tiny $>$}5.20}&{\footnotesize \it{nd}}& {\footnotesize \it{nd}} & {\footnotesize \it{nd}} & {\footnotesize \it{nd}} & {\footnotesize \it{nd}}\\
{\scriptsize } &{\scriptsize --} & {\scriptsize --} & {\scriptsize --} & {\scriptsize --} & {\scriptsize --} & {\scriptsize --} & {\scriptsize --} & {\scriptsize --} & {\scriptsize --} & {\scriptsize (0.02)} & {\scriptsize (0.02)} & {\scriptsize (0.02)} & {\scriptsize (0.02)} & {\scriptsize (0.04)} & {\scriptsize (0.03)} & {\scriptsize (0.04)} & {\scriptsize (0.02)} & {\scriptsize (0.02)} & {\scriptsize (0.07)} & {\scriptsize (0.10)} & {\scriptsize (0.09)} & {\scriptsize (0.04)} & {\scriptsize (0.03)} & {\scriptsize (0.03)} & {\scriptsize (0.08)} & {\scriptsize } & {\scriptsize --} & {\scriptsize --} & {\scriptsize --} & {\scriptsize --} & {\scriptsize --} \\
{\footnotesize GPSV8} &{\footnotesize 19.85} & {\footnotesize 18.16} & {\footnotesize 18.85}&{\footnotesize 16.70}&{\footnotesize 14.29}&{\footnotesize 12.03}& {\footnotesize 14.59} & {\footnotesize 13.12} & {\footnotesize 12.11} & {\footnotesize 14.72} & {\footnotesize 13.54} & {\footnotesize 12.79} & {\footnotesize 11.72} & {\footnotesize 14.71} & {\footnotesize 13.53} & {\footnotesize 12.89} & {\footnotesize 12.88} & {\footnotesize 12.96}&{\footnotesize  11.28 }&{\footnotesize  10.49 }&{\footnotesize  9.80 }&{\footnotesize 8.72}&{\footnotesize  12.06}&{\footnotesize  11.08 }&{\footnotesize  7.87 }&{\footnotesize 5.62:}&{\footnotesize \it{nd}}& {\footnotesize \it{nd}} & {\footnotesize \it{nd}} & {\footnotesize \it{nd}} & {\footnotesize \it{nd}}\\
{\scriptsize } & {\scriptsize (0.04)} & {\scriptsize (0.02)} & {\scriptsize (0.04)} & {\scriptsize (0.11)} & {\scriptsize (0.11)} & {\scriptsize (0.10)} & {\scriptsize (0.04)} & {\scriptsize (0.03)} & {\scriptsize (0.03)} & {\scriptsize (0.02)} & {\scriptsize (0.02)} & {\scriptsize (0.02)} & {\scriptsize (0.02)} & {\scriptsize (0.04)} & {\scriptsize (0.03)} & {\scriptsize (0.04)} & {\scriptsize (0.02)} & {\scriptsize (0.02)} & {\scriptsize (0.05)} & {\scriptsize (0.05)} & {\scriptsize (0.05)} & {\scriptsize (0.03)} & {\scriptsize (0.04)} & {\scriptsize (0.03)} & {\scriptsize (0.06)} & {\scriptsize (0.47)} & {\scriptsize --} & {\scriptsize --} & {\scriptsize --} & {\scriptsize --} & {\scriptsize --} \\
{\footnotesize GPSV9} &{\footnotesize {\it nd}} & {\footnotesize {\it nd}} & {\footnotesize {\it nd}}&{\footnotesize \it{nd}}&{\footnotesize \it{nd}}&{\footnotesize \it{nd}}&{\footnotesize \it{nd}}&{\footnotesize \it{nd}}&{\footnotesize \it{nd}}&{\footnotesize \it{nd}}&{\footnotesize \it{nd}}&{\footnotesize 16.43}&{\footnotesize 15.28}&{\footnotesize \it{nd}}&{\footnotesize \it{nd}}&{\footnotesize 16.19}&{\footnotesize 15.77}&{\footnotesize 16.08}&{\footnotesize  13.52:}&{\footnotesize  12.73:}&{\footnotesize  \it{nd} }&{\footnotesize  \it{nd} }&{\footnotesize \it{nd}} & {\footnotesize \it{nd}} & {\footnotesize \it{nd}} & {\footnotesize \it{nd}} & {\footnotesize \it{nd}}& {\footnotesize \it{nd}} & {\footnotesize \it{nd}} & {\footnotesize \it{nd}} & {\footnotesize \it{nd}}\\
{\scriptsize } &{\scriptsize --} & {\scriptsize --} & {\scriptsize --} & {\scriptsize --} & {\scriptsize --} & {\scriptsize --} & {\scriptsize --} & {\scriptsize --} & {\scriptsize --} & {\scriptsize --} & {\scriptsize --} & {\scriptsize (0.05)} & {\scriptsize (0.02)} & {\scriptsize --} & {\scriptsize --} & {\scriptsize (0.05)} & {\scriptsize (0.02)} & {\scriptsize (0.03)} & {\scriptsize (0.19)} & {\scriptsize (0.16)} & {\scriptsize --} & {\scriptsize --} & {\scriptsize --} & {\scriptsize --} & {\scriptsize --} & {\scriptsize --} & {\scriptsize --} & {\scriptsize --} & {\scriptsize --} & {\scriptsize --} & {\scriptsize --} \\
{\footnotesize GPSV10} &{\footnotesize {\it nd}} & {\footnotesize {\it nd}} & {\footnotesize {\it nd}}&{\footnotesize \it{nd}}&{\footnotesize \it{nd}}&{\footnotesize 13.04}& {\footnotesize {\tiny $>$}16.56}&{\footnotesize 14.60}&{\footnotesize 13.10}&{\footnotesize 17.36}&{\footnotesize 15.55}&{\footnotesize 13.99}&{\footnotesize 12.71}&{\footnotesize 17.15}&{\footnotesize 15.39}&{\footnotesize 14.01}&{\footnotesize 14.21}&{\footnotesize 14.36}&{\footnotesize  11.05 }&{\footnotesize  10.40 }&{\footnotesize 9.71}&{\footnotesize  8.75 }&{\footnotesize  12.16 }& {\footnotesize 10.78 }&{\footnotesize  8.52 }&{\footnotesize  4.16 }&{\footnotesize \it{nd}}& {\footnotesize \it{nd}} & {\footnotesize \it{nd}} & {\footnotesize \it{nd}} & {\footnotesize \it{nd}}\\
{\scriptsize } & {\scriptsize --} & {\scriptsize --} & {\scriptsize --} & {\scriptsize --} & {\scriptsize --} & {\scriptsize (0.16)} & {\scriptsize --} & {\scriptsize (0.08)} & {\scriptsize (0.05)} & {\scriptsize (0.03)} & {\scriptsize (0.02)} & {\scriptsize (0.02)} & {\scriptsize (0.02)} & {\scriptsize (0.04)} & {\scriptsize (0.03)} & {\scriptsize (0.04)} & {\scriptsize (0.02)} & {\scriptsize (0.02)} & {\scriptsize (0.03)} & {\scriptsize (0.06)} & {\scriptsize (0.03)} & {\scriptsize (0.03)} & {\scriptsize (0.04)} & {\scriptsize (0.03)} & {\scriptsize (0.10)} & {\scriptsize (0.08)} & {\scriptsize --} & {\scriptsize --} & {\scriptsize --} & {\scriptsize --} & {\scriptsize --} \\
{\footnotesize GPSV11} &{\footnotesize {\it nd}} & {\footnotesize {\it nd}} & {\footnotesize {\it nd}}&{\footnotesize \it{nd}}&{\footnotesize 16.06}&{\footnotesize \it{nd}}&{\footnotesize 16.05}&{\footnotesize 14.80}&{\footnotesize 14.39}&{\footnotesize 19.38:}&{\footnotesize 17.53}&{\footnotesize 15.84}&{\footnotesize 14.49}&{\footnotesize 20.28:}&{\footnotesize 18.31:}&{\footnotesize 16.99}&{\footnotesize 14.37}&{\footnotesize 14.40}&{\footnotesize  13.02}&{\footnotesize  12.45}&{\footnotesize 11.51:}&{\footnotesize  10.30:}&{\footnotesize \it{nd}} & {\footnotesize \it{nd}} & {\footnotesize \it{nd}} & {\footnotesize \it{nd}} & {\footnotesize \it{nd}}& {\footnotesize \it{nd}} & {\footnotesize \it{nd}} & {\footnotesize \it{nd}} & {\footnotesize \it{nd}}\\
{\scriptsize } &{\scriptsize --} & {\scriptsize --} & {\scriptsize --} & {\scriptsize --} & {\scriptsize (0.24)} & {\scriptsize --} & {\scriptsize (0.13)} & {\scriptsize (0.13)} & {\scriptsize (0.13)} & {\scriptsize (0.18)} & {\scriptsize (0.06)} & {\scriptsize (0.03)} & {\scriptsize (0.02)} & {\scriptsize (0.2)} & {\scriptsize (0.12)} & {\scriptsize (0.08)} & {\scriptsize (0.02)} & {\scriptsize (0.02)} & {\scriptsize (0.09)} & {\scriptsize (0.11)} & {\scriptsize (0.16)} & {\scriptsize (0.19)} & {\scriptsize (0.08)} & {\scriptsize (0.06)} & {\scriptsize (0.15)} & {\scriptsize --} & {\scriptsize --} & {\scriptsize --} & {\scriptsize --} & {\scriptsize --} & {\scriptsize --} \\
{\footnotesize GPSV12} &{\footnotesize {\it nc}} & {\footnotesize {\it nc}} & {\footnotesize {\it nc}}&{\footnotesize \it{nd}}&{\footnotesize \it{nd}}&{\footnotesize \it{nd}}&{\footnotesize 16.26:}&{\footnotesize {\tiny $>$}16.75}&{\footnotesize {\tiny $>$}16.36}&{\footnotesize 16.55}&{\footnotesize 15.93}&{\footnotesize 15.77}&{\footnotesize 16.87}&{\footnotesize 16.32}&{\footnotesize 15.77}&{\footnotesize 15.58}&{\footnotesize {\it nc}}&{\footnotesize {\it nc}}& {\footnotesize {\it nc}} & {\footnotesize {\it nc}} & {\footnotesize {\it nc}} & {\footnotesize {\it nc}} & {\footnotesize \it{nd}} & {\footnotesize \it{nd}} & {\footnotesize \it{nd}} & {\footnotesize \it{nd}} & {\footnotesize \it{nd}}& {\footnotesize \it{nd}} & {\footnotesize \it{nd}} & {\footnotesize \it{nd}} & {\footnotesize \it{nd}}\\
{\scriptsize } &{\scriptsize --} & {\scriptsize --} & {\scriptsize --} & {\scriptsize --} & {\scriptsize --} & {\scriptsize --} & {\scriptsize (0.12)} & {\scriptsize --} & {\scriptsize --} & {\scriptsize (0.02)} & {\scriptsize (0.02)} & {\scriptsize (0.03)} & {\scriptsize (0.08)} & {\scriptsize (0.04)} & {\scriptsize (0.03)} & {\scriptsize (0.04)} & {\scriptsize --} & {\scriptsize --} & {\scriptsize --} & {\scriptsize --} & {\scriptsize --} & {\scriptsize --} & {\scriptsize --} & {\scriptsize --} & {\scriptsize --} & {\scriptsize --} & {\scriptsize --} & {\scriptsize --} & {\scriptsize --} & {\scriptsize --} & {\scriptsize --} \\
{\footnotesize GPSV13} &{\footnotesize 13.98} & {\footnotesize 13.38} & {\footnotesize 13.45}&{\footnotesize \it{nc}}&{\footnotesize \it{nc}}&{\footnotesize \it{nc}}&{\footnotesize 12.46}&{\footnotesize 12.13}&{\footnotesize 11.90}&{\footnotesize 13.21}&{\footnotesize 13.06}&{\footnotesize 12.92}&{\footnotesize 11.90}&{\footnotesize 13.33}&{\footnotesize 13.21}&{\footnotesize 13.12}&{\footnotesize {\it nc}}&{\footnotesize {\it nc}}&{\footnotesize {\it nc}} & {\footnotesize {\it nc}} & {\footnotesize {\it nc}} & {\footnotesize {\it nc}}&{\footnotesize  13.23 }&{\footnotesize  13.29 }&{\footnotesize {\tiny $>$}12.30}&{\footnotesize  {\tiny $>$}8.66}&{\footnotesize \it{nd}}& {\footnotesize \it{nd}} & {\footnotesize \it{nd}} & {\footnotesize \it{nd}} & {\footnotesize \it{nd}}\\
{\scriptsize } & {\scriptsize (0.02)} & {\scriptsize (0.02)} & {\scriptsize (0.02)} & {\scriptsize --} & {\scriptsize --} & {\scriptsize --} & {\scriptsize (0.02)} & {\scriptsize (0.02)} & {\scriptsize (0.02)} & {\scriptsize (0.02)} & {\scriptsize (0.02)} & {\scriptsize (0.02)} & {\scriptsize (0.02)} & {\scriptsize (0.03)} & {\scriptsize (0.02)} & {\scriptsize (0.04)} & {\scriptsize --} & {\scriptsize --} & {\scriptsize --} & {\scriptsize --} & {\scriptsize --} & {\scriptsize --} & {\scriptsize (0.03)} & {\scriptsize (0.04)} & {\scriptsize --} & {\scriptsize --} & {\scriptsize --} & {\scriptsize --} & {\scriptsize --} & {\scriptsize --} & {\scriptsize --} \\
{\footnotesize GPSV14} &{\footnotesize {\it nd}} & {\footnotesize {\it nd}} & {\footnotesize {\it nd}}&{\footnotesize {\it nc}}&{\footnotesize {\it nc}}&{\footnotesize {\it nc}}&{\footnotesize \it{nd}}&{\footnotesize \it{nd}}&{\footnotesize \it{nd}}&{\footnotesize 16.88}&{\footnotesize 16.23}&{\footnotesize 15.99}& {\footnotesize 18.07:} & {\footnotesize {\it nc} } & {\footnotesize {\it nc}} & {\footnotesize {\it nc}} & {\footnotesize {\it nc}} & {\footnotesize {\it nc}}& {\footnotesize {\it nc}} & {\footnotesize {\it nc}} & {\footnotesize {\it nc}} & {\footnotesize {\it nc}} & {\footnotesize \it{nd}} & {\footnotesize \it{nd}} & {\footnotesize \it{nd}} & {\footnotesize \it{nd}} & {\footnotesize \it{nd}}& {\footnotesize \it{nd}} & {\footnotesize \it{nd}} & {\footnotesize \it{nd}} & {\footnotesize \it{nd}}\\
{\scriptsize } &{\scriptsize --} & {\scriptsize --} & {\scriptsize --} & {\scriptsize --} & {\scriptsize --} & {\scriptsize --} & {\scriptsize --} & {\scriptsize --} & {\scriptsize --} & {\scriptsize (0.02)} & {\scriptsize (0.02)} & {\scriptsize (0.03)} & {\scriptsize (0.23)} & {\scriptsize --} & {\scriptsize --} & {\scriptsize --} & {\scriptsize --} & {\scriptsize --} & {\scriptsize --} & {\scriptsize --} & {\scriptsize --} & {\scriptsize --} & {\scriptsize --} & {\scriptsize --} & {\scriptsize --} & {\scriptsize --} & {\scriptsize --} & {\scriptsize --} & {\scriptsize --} & {\scriptsize --} & {\scriptsize --} \\
{\footnotesize GPSV15} &{\footnotesize {\it nd}} & {\footnotesize {\it nd}} & {\footnotesize {\it nd}}&{\footnotesize {\it nc}}&{\footnotesize {\it nc}}&{\footnotesize {\it nc}}&{\footnotesize {\tiny $>$}17.21}&{\footnotesize 14.50}&{\footnotesize 11.61}&{\footnotesize 17.45}&{\footnotesize 14.14}&{\footnotesize 11.21}&{\footnotesize 12.24}&{\footnotesize 18.29}&{\footnotesize 15.39}&{\footnotesize 12.59}&{\footnotesize {\it nc}}&{\footnotesize {\it nc}} &{\footnotesize {\it nc}} & {\footnotesize {\it nc}} & {\footnotesize {\it nc}} & {\footnotesize {\it nc}}&{\footnotesize  8.41} &{\footnotesize  6.43 }&{\footnotesize  3.72 }& {\footnotesize 2.51 }&{\footnotesize  4.07 }&{\footnotesize  2.87:}&{\footnotesize  2.95:}&{\footnotesize  4.07 }&{\footnotesize  2.67}\\
{\scriptsize } &{\scriptsize --} & {\scriptsize --} & {\scriptsize --} & {\scriptsize --} & {\scriptsize --} & {\scriptsize --} & {\scriptsize --} & {\scriptsize (0.04)} & {\scriptsize (0.02)} & {\scriptsize (0.02)} & {\scriptsize (0.02)} & {\scriptsize (0.02)} & {\scriptsize (0.02)} & {\scriptsize (0.04)} & {\scriptsize (0.02)} & {\scriptsize (0.03)} & {\scriptsize --} & {\scriptsize --} & {\scriptsize --} & {\scriptsize --} & {\scriptsize --} & {\scriptsize --} & {\scriptsize (0.02)} & {\scriptsize (0.02)} & {\scriptsize (0.02)} & {\scriptsize (0.02)} & {\scriptsize (0.11)} &{\scriptsize (0.17)} &{\scriptsize (0.20)} & {\scriptsize (0.25)} & {\scriptsize (0.21)} \\
{\footnotesize GPSV16} &{\footnotesize {\it nd}} & {\footnotesize {\it nd}} & {\footnotesize {\it nd}}&{\footnotesize {\it nc}}&{\footnotesize {\it nc}}&{\footnotesize {\it nc}}&{\footnotesize \it{nd}}&{\footnotesize \it{nd}}&{\footnotesize \it{nd}}&{\footnotesize 16.66}&{\footnotesize 14.93}&{\footnotesize 13.46}&{\footnotesize 15.67}&{\footnotesize 16.90}&{\footnotesize 15.12}&{\footnotesize 13.83}&{\footnotesize {\it nc}}&{\footnotesize {\it nc}}&{\footnotesize 11.73}&{\footnotesize 10.75}&{\footnotesize {\it nc}}&{\footnotesize {\it nc}}&{\footnotesize  12.20 }&{\footnotesize  10.80 }&{\footnotesize  7.97 }&{\footnotesize  5.25 }&{\footnotesize \it{nd}}& {\footnotesize \it{nd}} & {\footnotesize \it{nd}} & {\footnotesize \it{nd}} & {\footnotesize \it{nd}}\\
{\scriptsize } &{\scriptsize --} & {\scriptsize --} & {\scriptsize --} & {\scriptsize --} & {\scriptsize --} & {\scriptsize --} & {\scriptsize --} & {\scriptsize --} & {\scriptsize --} & {\scriptsize (0.02)} & {\scriptsize (0.02)} & {\scriptsize (0.02)} & {\scriptsize (0.03)} & {\scriptsize (0.07)} & {\scriptsize (0.04)} & {\scriptsize (0.04)} & {\scriptsize --} & {\scriptsize --} & {\scriptsize (0.03)} & {\scriptsize (0.03)} & {\scriptsize --} & {\scriptsize --} & {\scriptsize (0.03)} & {\scriptsize (0.02)} & {\scriptsize (0.03)} & {\scriptsize (0.04)} & {\scriptsize --} & {\scriptsize --} & {\scriptsize --} & {\scriptsize --} & {\scriptsize --} \\
{\footnotesize GPSV17} &{\footnotesize {\it nc}} & {\footnotesize {\it nc}} & {\footnotesize {\it nc}}&{\footnotesize \it{nd}}&{\footnotesize \it{nd}}&{\footnotesize \it{nd}}&{\footnotesize \it{nd}}&{\footnotesize \it{nd}}&{\footnotesize \it{nd}}&{\footnotesize 18.14}&{\footnotesize 17.88}&{\footnotesize 17.78:}& {\footnotesize 15.43} & {\footnotesize {\it nc} } & {\footnotesize {\it nc}} & {\footnotesize {\it nc}} & {\footnotesize {\it nc}} & {\footnotesize {\it nc}}& {\footnotesize {\it nc}} & {\footnotesize {\it nc}} & {\footnotesize {\it nc}} & {\footnotesize {\it nc}} & {\footnotesize \it{nd}} & {\footnotesize \it{nd}} & {\footnotesize \it{nd}} & {\footnotesize \it{nd}} & {\footnotesize \it{nd}}& {\footnotesize \it{nd}} & {\footnotesize \it{nd}} & {\footnotesize \it{nd}} & {\footnotesize \it{nd}}\\
{\scriptsize } & {\scriptsize --} & {\scriptsize --} & {\scriptsize --} & {\scriptsize --} & {\scriptsize --} & {\scriptsize --} & {\scriptsize --} & {\scriptsize --} & {\scriptsize --} & {\scriptsize (0.05)} & {\scriptsize (0.08)} & {\scriptsize (0.16)} & {\scriptsize (0.02)} & {\scriptsize --} & {\scriptsize --} & {\scriptsize --} & {\scriptsize --} & {\scriptsize --} & {\scriptsize --} & {\scriptsize --} & {\scriptsize --} & {\scriptsize --} & {\scriptsize --} & {\scriptsize --} & {\scriptsize --} & {\scriptsize --} & {\scriptsize --} & {\scriptsize --} & {\scriptsize --} & {\scriptsize --} & {\scriptsize --} \\
\hline
\multicolumn{32}{l}{{\footnotesize $a$ 2008 observations.}}\\ 
\multicolumn{32}{l}{{\footnotesize $b$ 2005 observations.}}\\ 
\multicolumn{32}{l}{{\footnotesize {\it nc} Not covered in the observations.}} \\
\multicolumn{32}{l}{{\footnotesize {\it nd} Not detected in the observations.}}\\
\multicolumn{32}{l}{{\footnotesize $>$ The values represent upper limits.}}\\
\multicolumn{32}{l}{{\footnotesize $:$ Represent uncertain values in the photometry as explained in the text.}}\\
\hline
\end{tabular}
\end{flushleft}
\end{table*}
\end{landscape}

\section{Results}\label{sec:res}

\subsection{Near- and mid-infrared photometry}

Table \ref{table:photom} shows the results of the 2010 ISAAC and 2012 UKIRT photometry, along with the magnitudes found for targets in other near- and mid-IR surveys. The errors are typically small ($\leq 0.08$~mag), although given the faintness of some of our objects and significant source confusion in some fields, some of the detections in 2MASS, and WISE have moderate to poor signal to noise ratios (SNR). Detections with SNR $<10$ or errors larger than the typical values are marked with ``:'' in Table \ref{table:photom}. We also note that even though errors in DENIS photometry are in the range $0.07-0.24$~mag, the photometric quality flag had a value of 100 for all of the detections in Table \ref{table:photom} and are considered to be reliable measurements. Only one measurement is left out of the table and it corresponds to a VPHAS$+$ $g^{'}$ detection of GPSV8 with $g^{'}=22.84\pm0.15$~ mag.

\begin{figure*}
\begin{center}
\includegraphics[width=1.4\columnwidth]{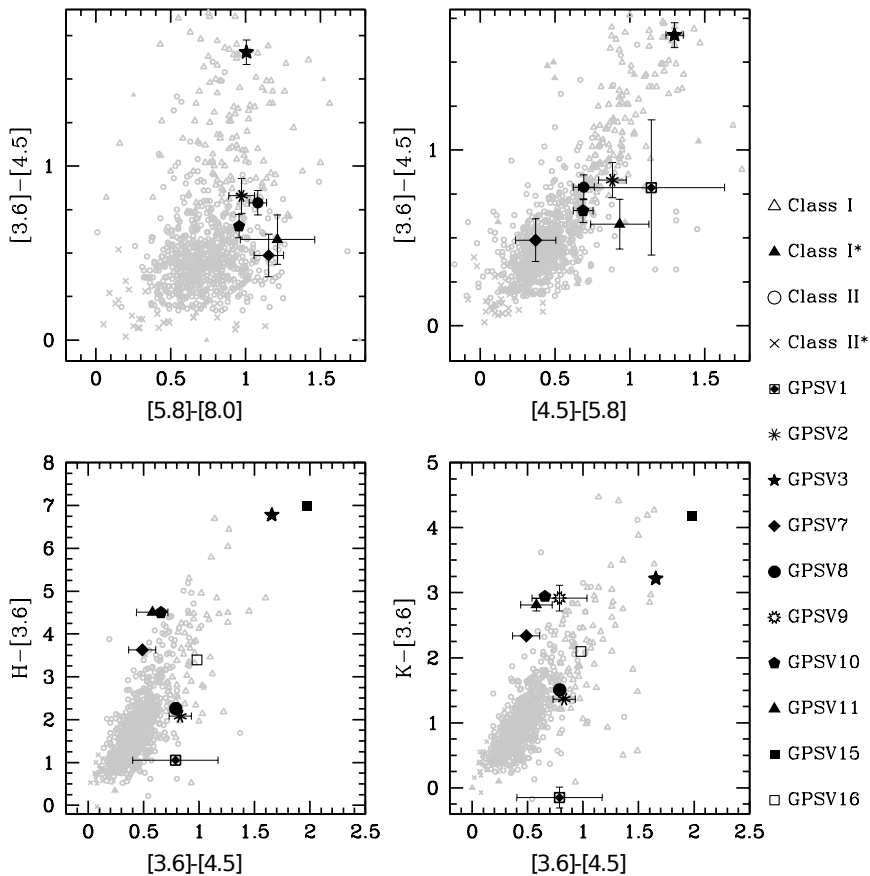}
\caption[]{Colour-colour diagrams of GPS candidates based on {\it Spitzer} and UKIDSS photometry. Young stellar objects from \citet{2009Gutermuth} are shown for comparison. The latter are divided in: protostars with infalling envelopes (Class I, {\it open triangles}), pre-main-sequence stars with optically thick discs (Class II, {\it open circles}), deeply embedded sources (Class I*, {\it filled triangles}) and transition disc candidates (Class II*, {\it crosses}).  Objects GPSV3 and GPSV15 are located in the Class I region of the two bottom graphs, however SEDs of the sources show a flatter distribution than expected for Class I objects (see Figure \ref{lc_seds:erup}). Errors are plotted only for objects that present significant uncertainties on their measurements. } 
\label{irac:ysos}
\end{center}
\end{figure*}

{\it Spitzer} GLIMPSE detections yield more evidence that most of our objects are PMS stars. Figure \ref{irac:ysos} shows the location of our sample on colour-colour diagrams often used for YSO classification when only near-infrared and IRAC 3.6-8.0~$\mu$m fluxes are available \citep[see e.g.][]{2005Hartmann}. For comparison we show the sample of YSOs in young stellar clusters of \citet{2009Gutermuth}, the classification scheme shown in the diagrams is the same as the one used by these authors. Due to the high variability of our sample, the location of the objects will be affected in the two lower panels, which use $H$ and $K$ photometry. The latter are not contemporary with GLIMPSE3D ($\sim$ 2006), GLIMPSE360 ($\sim$ 2010-2012) nor WISE ($\sim$ 2010). Therefore we use the closest epoch with simultaneous observations in these two bands, i.e. GPS $H$ and $K_{1}$ ($\sim$ 2008) for the Serpens objects, and VLT/ISAAC $H$ and $K$ (2010) for both GPSV15 and GPSV16. 

Figure \ref{irac:ysos} shows that most of our objects fall in what appears to be an area of transition between class I to class II objects, especially on the diagrams based purely on {\it Spitzer} photometry. Two sources stand out as they are located in the region where Class I objects are found: these are GPSV3 and GPSV15. However, the latter is not covered by {\it Spitzer} surveys and filters $W1$ and $W2$ of WISE are used instead. The use of these filters should not have much effect on location in the diagrams as differences between WISE and IRAC filters are only observed for $W1\geq14$ magnitudes and $W2>13$ magnitudes \citep{2012Cutri}, which is fainter than the measured magnitudes for GPSV15.  

\subsection{\citeauthor{2007Robitaille} SED fits}

Spectral energy distributions are constructed with the help of data from public surveys.

Fits to stars in Table \ref{table:sedfits} were performed with the SED fitting tool of \citet{2007Robitaille}, which explores the parameter space of Class 0 to Class III YSOs at all system inclinations. In order to perform the fits we did not include every available data point, and in general we only used the information arising from UKIDSS $JHK_{1}$, {\it Spitzer} $I1-I4$ and {\it WISE} $22 \mu m$. We note that these are not contemporaneous observations, so the fits could be unreliable due to the variability of our objects. In addition we are not certain whether \citeauthor{2007Robitaille} models  can accurately describe highly variable YSOs, since their discs may have an unusual structure.
 
The distances are set to vary between 1-5 kpc and $A_v$ within 0-30 mag. We note that varying the distances in the given range instead of using the known distances for the Serpens objects or GPSV16, had a negligible effect on the measured parameters, i.e. the change in the values were never larger than their estimated errors. The weighted mean values and standard deviation of the parameters (in logarithmic scale) shown in Table \ref{table:sedfits} were determined using the models for which $\chi^{2}-\chi_{best}^{2} < 3N$ \citep[as suggested in][]{2007Robitaille}, where $\chi_{best}^{2}$ is the $\chi^{2}$ of the best-fitting model and $N$ represents the number of data points used in the fitting process. The last column of Table \ref{table:sedfits} shows the number of models that fulfilled this condition for each of the objects in our sample. The discussion on the results of the fits can be found in the analysis of the individual objects presented below.

\subsection{Eruptive Variables}\label{subsec:ev}

Given the faintness of many of our objects, the low resolution VLT/ISAAC spectra were of insufficient quality to clearly detect characteristics of eruptive variables. The subsequent Gemini/NIFS and Magellan/FIRE spectroscopy included only relatively bright sources whose ISAAC spectra suggested a possible eruptive variable classification. The following analysis is then divided between stars where spectral characteristics of eruptive variables could be found and those for which there is only non-spectroscopic evidence for a YSO or eruptive PMS 
variable classification.

Four stars are defined as likely to be eruptive variables. These correspond to GPSV3, GPSV15, 
GPSV8 and GPSV16.

\subsubsection{GPSV3 and GPSV15}\label{sec:erup_3_15}

\begin{table*}
\begin{center}
\begin{minipage}{120mm}
\caption{Parameters derived from the \citeauthor{2007Robitaille} models SED fitting as explained in the text. The standard deviations are shown in brackets and they are often large.}\label{table:sedfits}
\begin{tabular}{@{}l@{\hspace{0.5cm}}c@{\hspace{0.2cm}}c@{\hspace{0.2cm}}c@{\hspace{0.2cm}}c@{\hspace{0.2cm}}c@{\hspace{0.2cm}}c@{}}
\hline
Object & $\log M_{\ast}$ & $\log\dot{M}_{disc}$  & $\log \dot{M}_{env}$  & $\log L_{tot}$  & $\chi^{2}_{best}/N_{data}$ & $N_{fits}$\\
& \footnotesize{($M_{\odot}$)} & \footnotesize{$(M_{\odot}$~yr$^{-1})$} & \footnotesize{$(M_{\odot}$~yr$^{-1})$} & \footnotesize{$(L_{\odot})$} & & \\
\hline
GPSV1 &  0.06[0.43] &    -7.62[1.32] & -3.80[1.05] &  1.17[0.75] &  0.07 &  2187 \\  
GPSV2 & -0.23[0.44] &    -7.91[1.50] & -4.95[0.61] &  0.74[0.53] &  0.20 &   333 \\  
GPSV3 &  0.79[0.17] &    -8.54[1.83] & -0.13[0.82] &  3.00[0.56] &  1.64 &   473 \\  
GPSV7 &  0.21[0.43] &    -8.99[1.93] & -1.12[1.95] &  1.42[0.62] &  0.12 &  1829 \\  
GPSV8 &  0.35[0.26] &    -8.54[1.60] & -2.40[2.41] &  1.46[0.46] &  0.14 &   518 \\  
GPSV10 & -0.01[0.52] &    -7.02[1.25] & -4.44[0.93] &  1.26[0.64] &  0.67 &   237 \\  
GPSV11 & -0.01[0.45] &    -7.81[1.54] & -3.70[1.33] &  1.15[0.60] &  0.48 &  1073 \\  
GPSV15 &  0.92[0.14] &    -9.56[1.53] &  0    &  3.49[0.47] &  0.86 &    94 \\  
GPSV16 &  0.07[0.53] &    -7.82[1.83] & -4.38[1.35] &  1.26[0.69] &  0.57 &   349 \\
\hline
\end{tabular}
\end{minipage}
\end{center}
\end{table*}

\begin{figure*}
\begin{center}
\includegraphics[width=1\columnwidth]{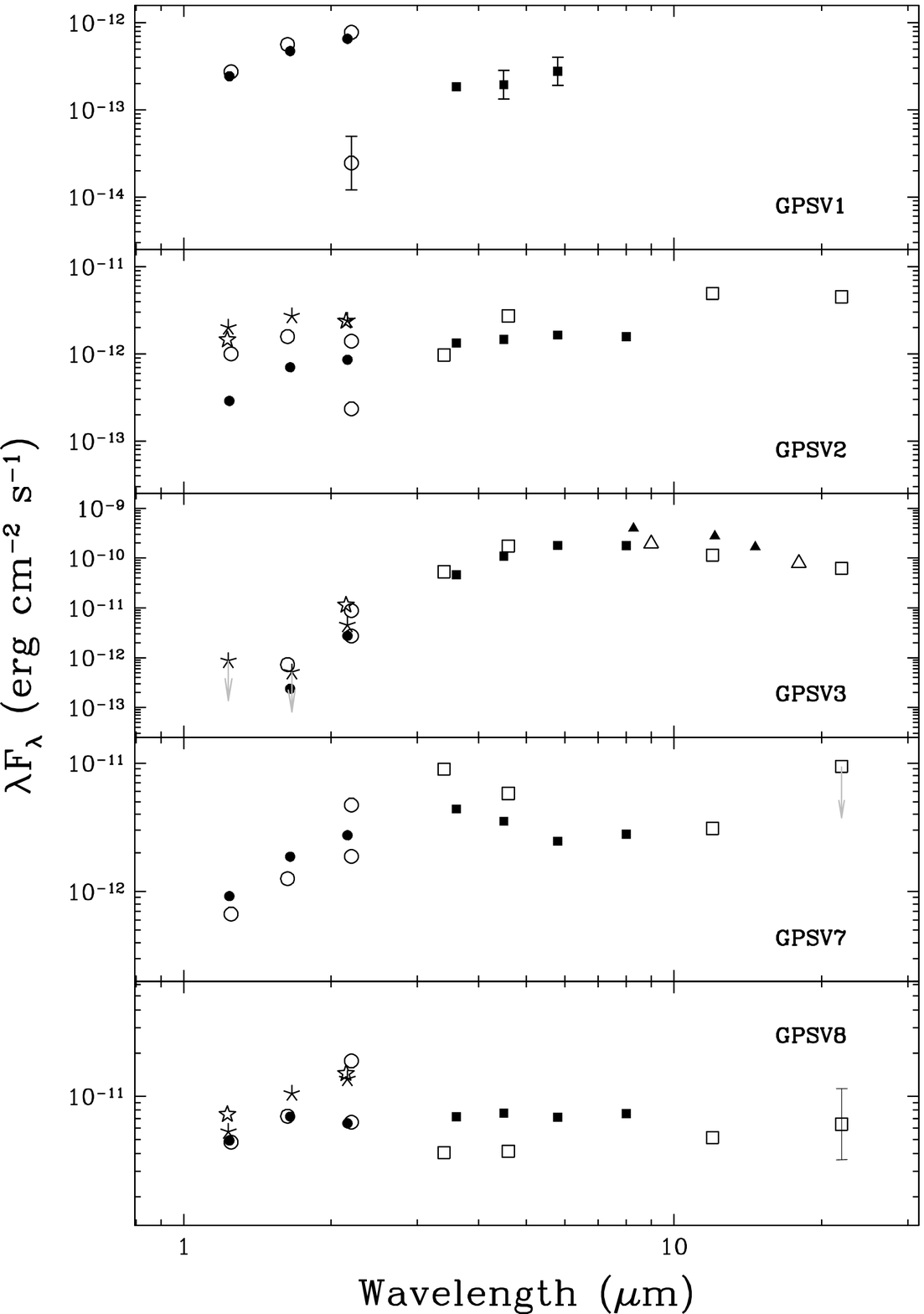}
\includegraphics[width=1\columnwidth]{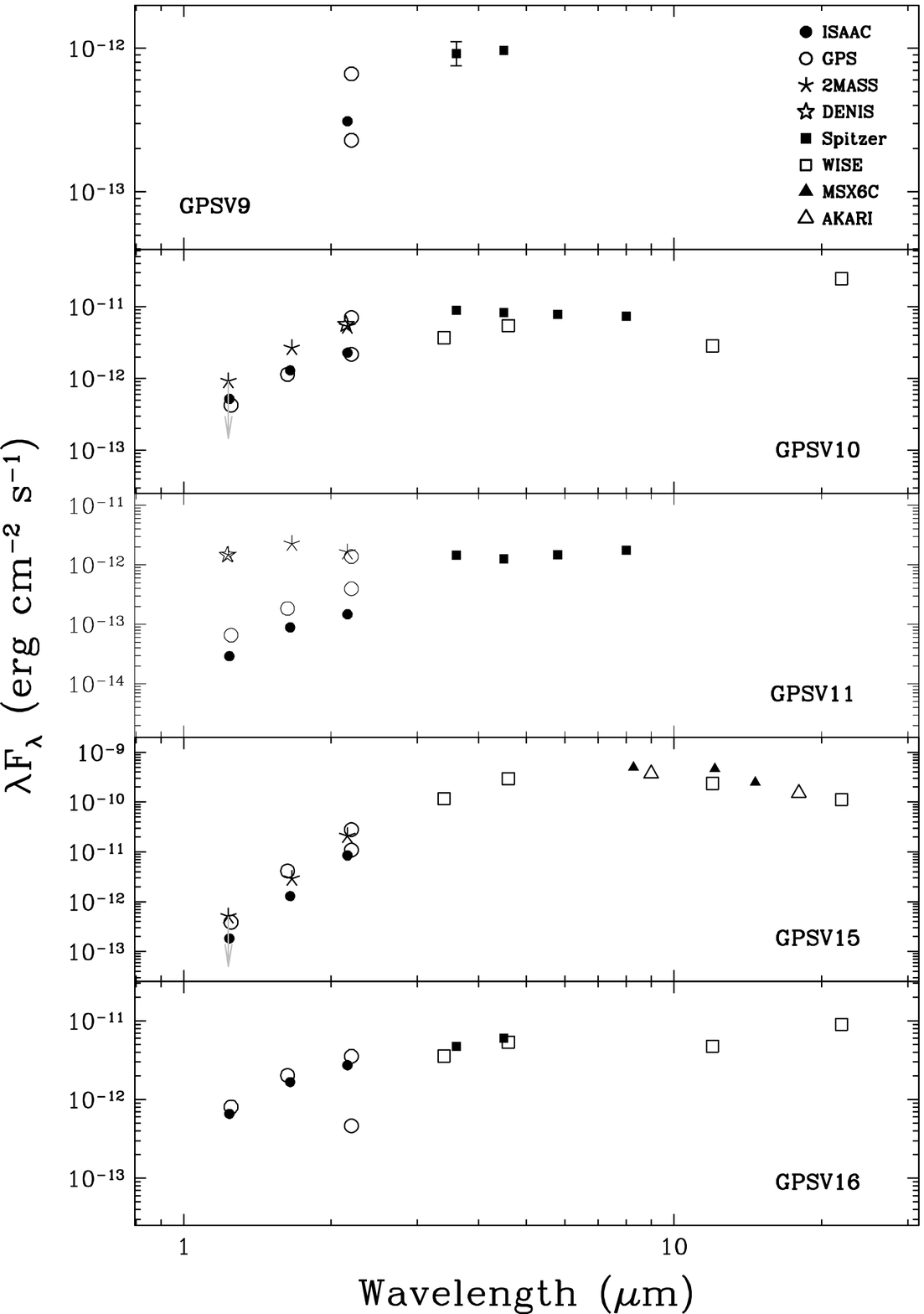}
\caption{Spectral energy distributions of 10 out of 13 DR5 objects within star forming regions. Errors are plotted only for objects that present significant uncertainties on their measurements.}
\label{lc_seds:erup_test}
\end{center}
\end{figure*}

\begin{figure*}
\begin{center}
\includegraphics[width=1\columnwidth]{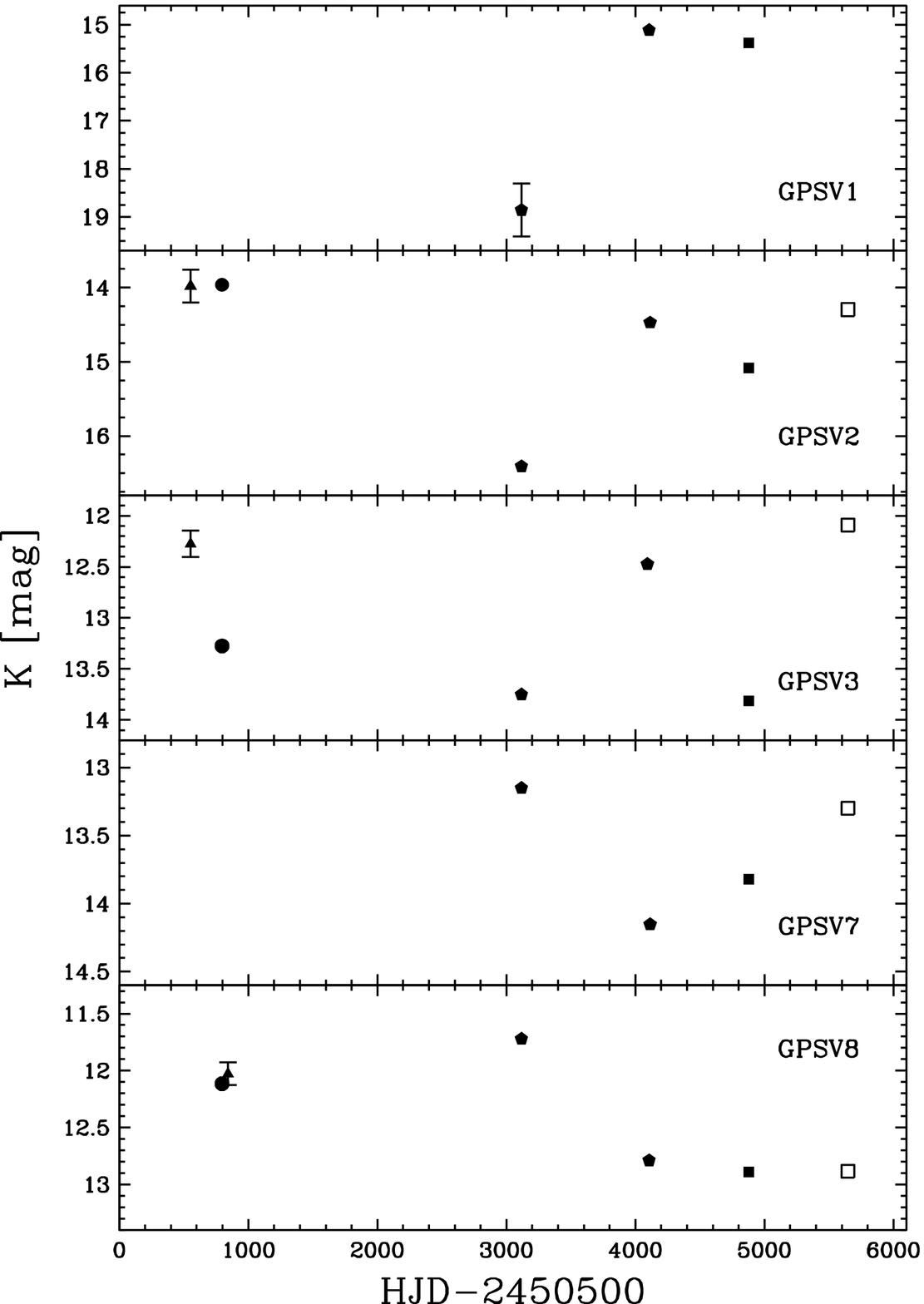}
\includegraphics[width=1\columnwidth]{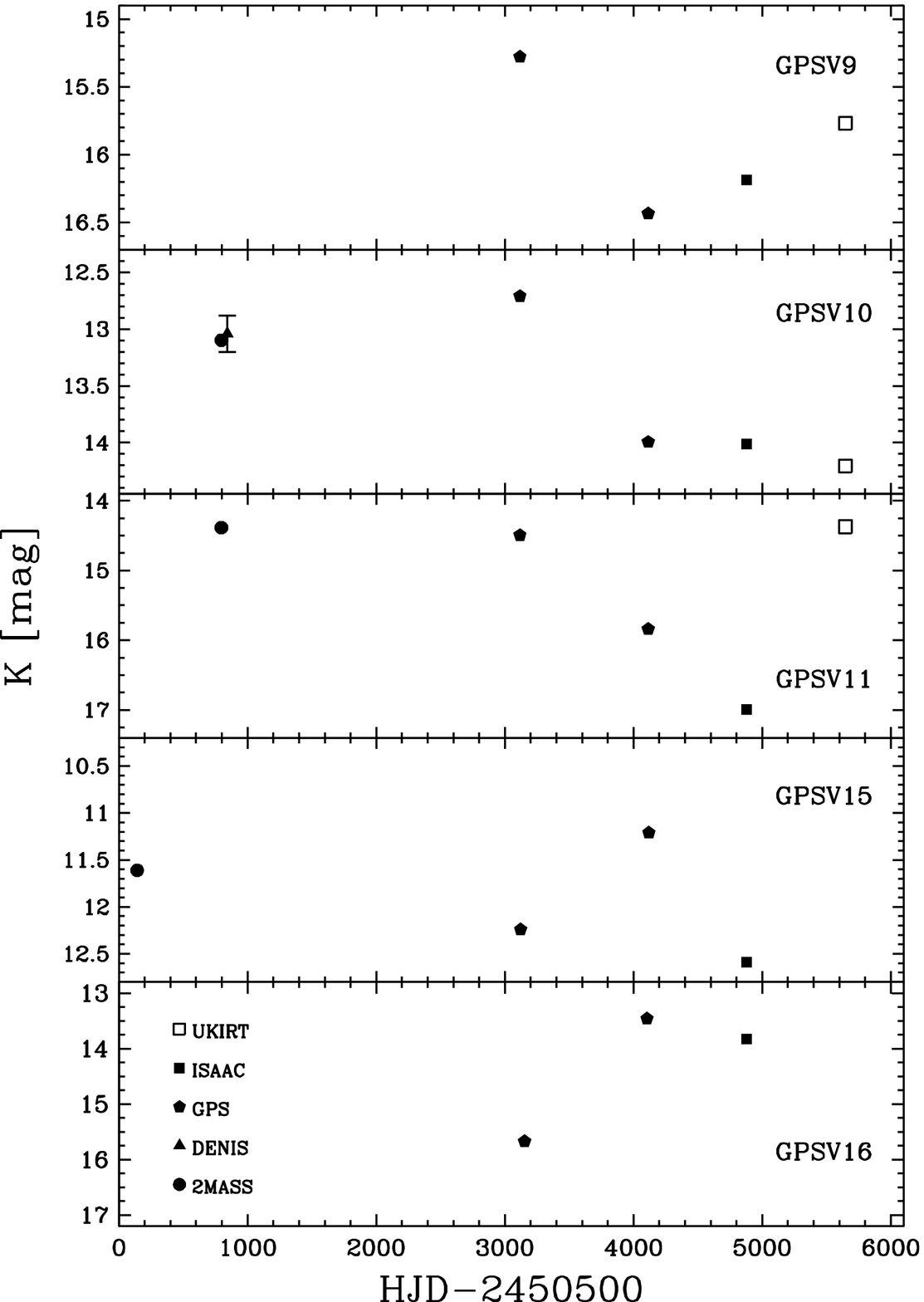}
\caption{K-band light curves for the same DR5 objects in figure \ref{lc_seds:erup_test}, where errors are plotted only for objects that present significant uncertainties on their measurements.}
\label{lc_seds:erup}
\end{center}
\end{figure*}

These two objects are the reddest stars in our DR5 sample (see Fig. \ref{dr57cmd}), with 
GPSV3 in fact not being detected in $J$ band in any of the near-infrared surveys nor 
in our ISAAC imaging. Both stars also show a large ($H$-$K$) excess in 
colour-colour diagrams (Fig. \ref{nircdr57}), beyond what is observed for Classical T Tauri 
stars and in agreement with what is expected for deeply embedded Class I YSOs. The 
colours are also similar to those of the deeply embedded FUor-like objects PP13S and 
AR6B \citep[][]{2001Aspin,2003Aspin}, and the deeply embedded outburst sources OO Ser and GM Cha \citep[][]{1996Hodapp,2007Persi}. The $K$ band light curves of both objects show similar behaviour, with repetitive increases in magnitude (Fig. \ref{lc_seds:erup}).

\begin{figure*}
\centering
\includegraphics[width=1\columnwidth]{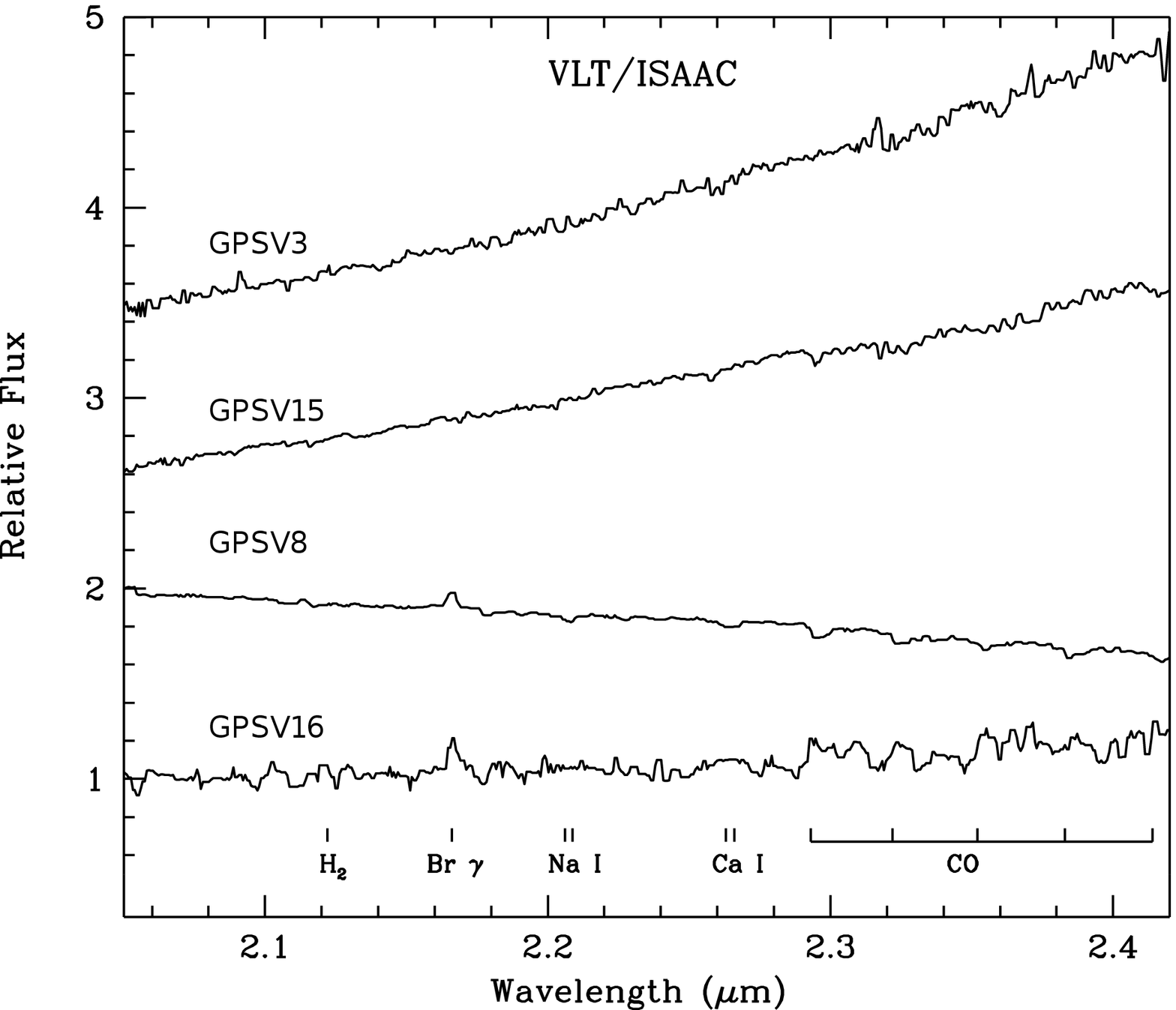}
\includegraphics[width=1\columnwidth]{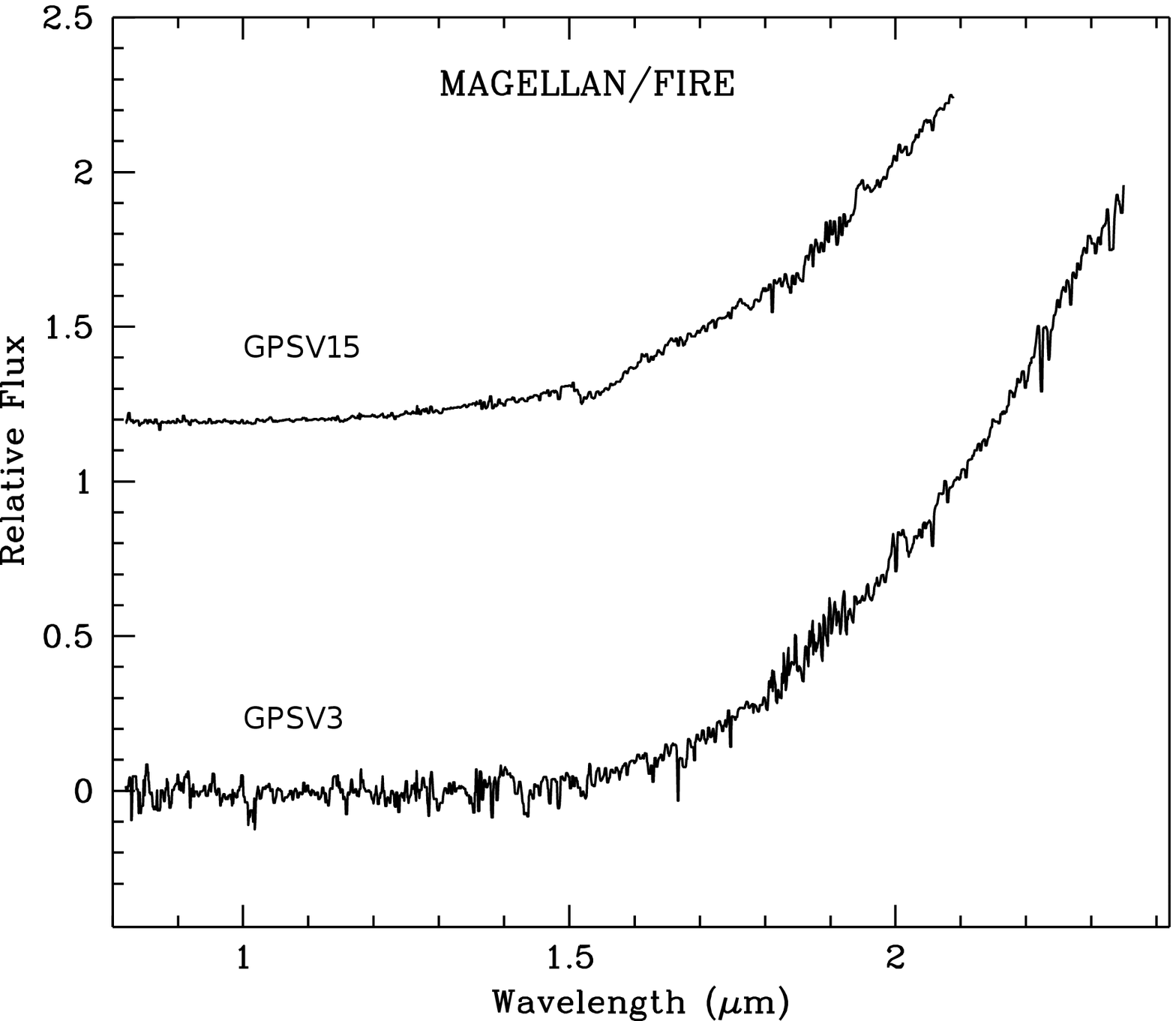}\\
\includegraphics[width=1.5\columnwidth]{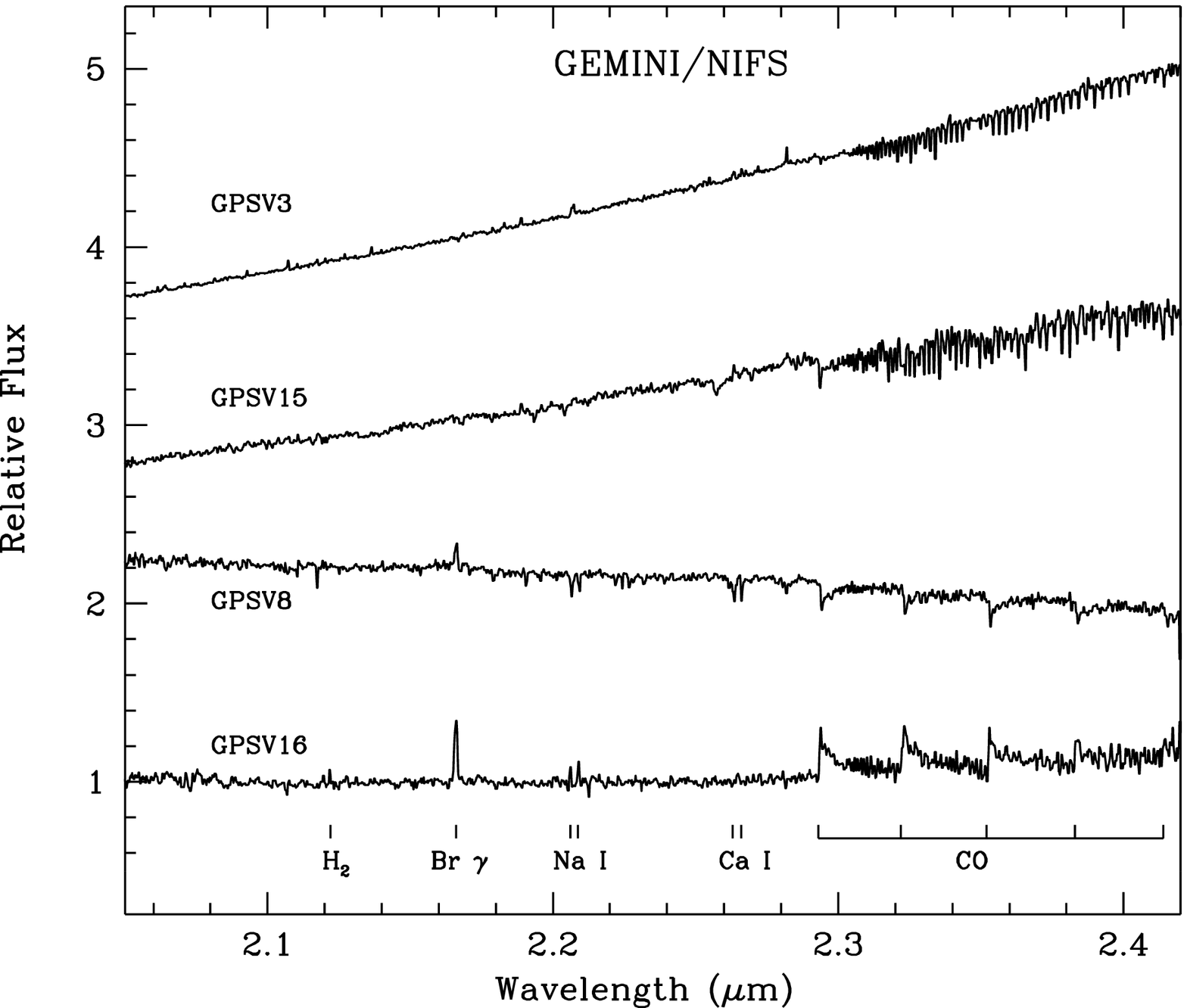}
\caption{2.2 $\mu$m ISAAC ({\it top left}) and NIFS ({\it bottom}) spectra for the four likely eruptive PMS variables. Spectral features usually found for these type of variable stars are shown in the bottom of each of the plots. It is readily apparent that the R=5000 NIFS spectra reveal far more information than the R=700 ISAAC spectra.({\it top right}) FIRE 0.8-2.5 $\mu$m spectra of red sources GPSV3 and GPSV15.}
\label{sp:erup}
\end{figure*}

Their ISAAC spectra (Fig. \ref{sp:erup}) also show a remarkable resemblance, where we observe a featureless continuum that rises steeply toward longer wavelengths. This is similar to the aforementioned embedded object OO Ser, and not too dissimilar to PP13S and AR6B, which differ only in that they show a strong $v$=2-0 CO absorption trough in the 2.3~$\mu$m region \citep[see Fig. 5 in][]{1996Hodapp}. The same steep rise is observed in the FIRE spectra of the sources, where GPSV15 is detected in $H$ band but with no sign of the H$_{2}$O absorption that is observed in a number of FUor and FUor-like stars. 

NIFS higher resolution spectra reveal a more complicated structure in the CO region. The spectra of the objects only show clearly the $v=2-0$ bandhead (stronger in GPSV15) with strong absorption that can be identified with the low $J$-lines of the $P$ and $R$ branches of the $v=2-0$ transition. The weakness of the bandheads, relative to the lower J transitions, indicates that we are observing gas that is somewhat cooler than is typical of M-type stellar atmospheres, so that only the lower energy levels are populated. Moreover, the spectra in figure \ref{sp:erup} strongly indicate that the photosphere is obscured from view and that the K band spectra are dominated by circumstellar matter. This is seen especially in the Magellan/FIRE spectra (upper right panel of Fig. \ref{sp:erup}), with a lack of flux at wavelengths shorter than $\sim 1.5 \mu m$ (where the stellar component would be expected to contribute to the observed flux of the system) and steeply rising flux at longer wavelengths, in agreement with where a disc following a T $\leq 1200$ K blackbody distribution is expected to begin contributing to the system.

\begin{figure*}
\begin{center}
\includegraphics[width=1\columnwidth]{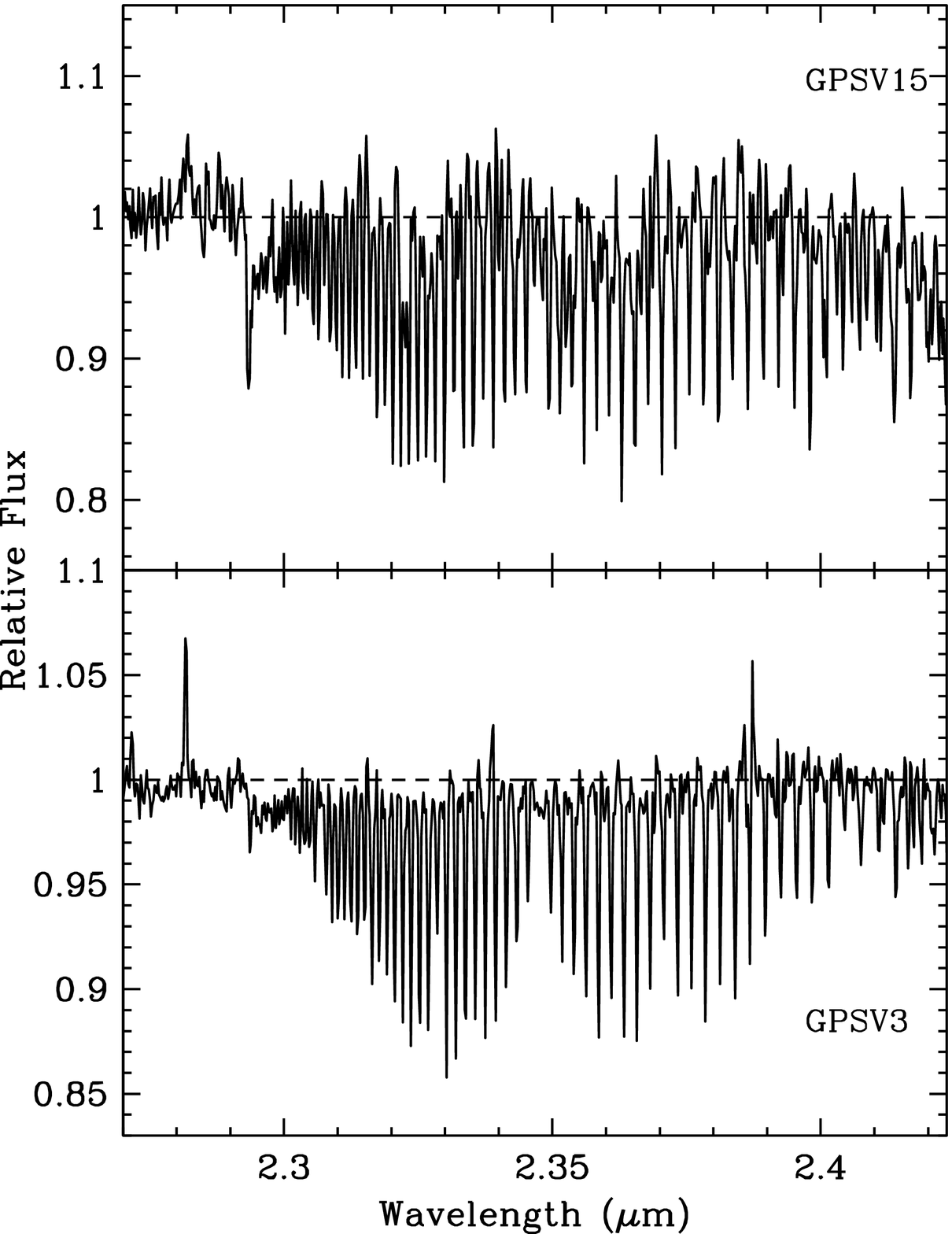}
\includegraphics[width=1\columnwidth]{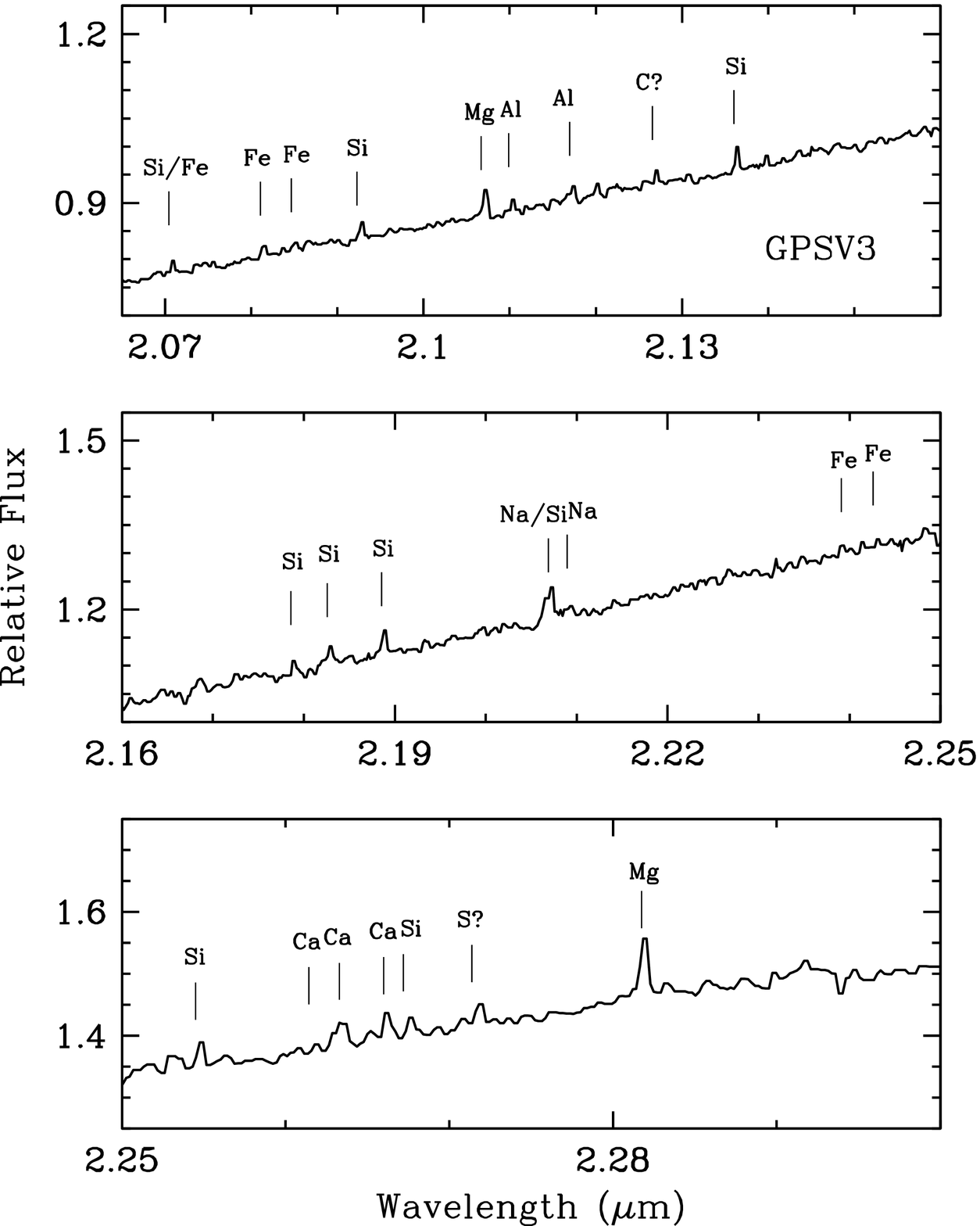}
\caption{Continuum normalized spectra of GPSV3 ({\it left, bottom}) and GPSV15 ({\it left, top}) showing the first overtone CO emission/absorption region. The continuum level is marked by a short-dashed line in both spectra. There is a clear emission feature at 2.28 $\mu$m that corresponds to a Mg I transition and is discussed in the text. {\it Right} Atomic lines identified in the NIFS spectrum of GPSV3.}
\label{nifs_co}
\end{center}
\end{figure*}

We removed the slope of the higher resolution spectra by fitting a straight line to the 2.18-2.28 $\mu$m  region and extrapolating towards the CO region. The results are shown in Fig. \ref{nifs_co}. In GPSV15 we see numerous features rising above the normalised continuum level in between the CO absorption lines. This could be due to emission from the rapidly rotating hot portion of the inner disc, which would show as the wings of rotationally broadened emission lines in between the stronger but narrower absorption lines. Rotation velocities would be of  the order of 200 km/s within a few stellar radii of a low mass YSO, which is slightly higher than the NIFS velocity resolution of 60 km/s. Since features in excess of the continuum are not observed in between every pair of adjacent CO absorption lines we cannot be certain of this interpretation. However, the wavelengths where we see no excess above the continuum are mostly located close to one of the higher order bandheads (3-1, 4-2 and 5-3) where the close proximity of numerous CO absorption lines disfavours observation of weak rotationally broadened emission features, at this spectral resolution. Observations at higher spectral resolution are therefore desirable in order to resolve the profile of rotationally broadened lines from the disc and verify their existence.

The presence of any such features in GPSV3 is marginal and strongly dependent on the choice of wavelengths used for the removal of the slope of the continuum. However, several emission lines are observed in the high resolution spectrum of GPSV3 (see fig. \ref{nifs_co}, right panel). These lines can be identified with atomic features observed in the photospheric absorption spectra of K-M dwarf stars \citep[see e.g.,][]{1997Wallace}. Specifically we detect lines of Mg~I, Na~I, Ca~I, Fe~I, Si~I, Al~I, and perhaps S~I and C~I, most of which have rarely been reported in emission. A temperature inversion is required to observe lines in emission, which could occur in a hot disc or wind near the star, making these features of circumstellar origin. One of the strongest lines is observed at $\sim 2.28~\mu m$ (also observed in GPSV15, see fig. \ref{nifs_co}), this is identified as the $4d^{3}D_{3,2,1}-6f^{3}F^{\circ}_{2,3,4}$ transition of Mg I \citep{1986Kleinmann}. Emission at  this wavelength has also been observed in a sample of embedded protostars by \citet{2011Davis} but the authors mark this feature as unknown in their analysis. Evidence for the presence of a hot inner disc in GPSV3 is given by the observed emission of the Na I doublet revealed in the NIFS spectra. This characteristic is observed in EXors when in outburst phase and is explained as arising from a hot inner disc \citep[see e.g.,][]{2012Loren}. However, it is unclear whether we are observing emission arising from a hot inner disc, wind or a combination of both.

In the case of GPSV15, blueshifted absorption of $\sim$ 40 km s$^{-1}$ is observed in the CO bandhead and individual lines. Blueshifted CO absorption has been observed previously in the FU Ori objects V1057 Cyg \citep*{2004Hartmann} and AR6A \citep{2003Aspin}. In the latter, it was suggested that the blueshift might be caused by rotational broadening, as described by \citet{1996Najita}, but this was unclear. In V1057 Cyg, the observed (50 km s$^{-1}$) blueshifted absorption was interpreted by \citet{2004Hartmann} as arising from an ejected dense and low temperature shell (T$\sim$~620 K). We note that the 40 km s$^{-1}$ blueshift seen in GPSV15 is too small to significantly affect our discussion concerning the presence of rotationally broadened emission features in between the absorption features.

In summary, the observed CO structure of GPSV15 could be explained by an emission component from an inner disc, combined with absorption by a cooler gas arising in an ejected shell. It is also conceivable that cooler parts of the disc at larger radii are contributing to the absorption (which could also be the case in GPSV3). However, the presence of both emission and absorption components in an individual object is more readily understood in terms of two separate structures. 

The SEDs of GPSV3 and GPSV15 are exceptionally red in the 1 to 5$~\mu$m region, with $J$-W2$>$11~mag. This is consistent with a Class I YSO classification but at longer wavelengths their SEDs become roughly flat (from $\sim$5-12~$\mu$m) and then decline towards 23~$\mu$m, as measured by {\it Akari} and WISE. The SED flattening and lack of cold dust emission is consistent with an accretion disc.
 
We find that for GPSV15, only a relatively small number of models could be used to estimate the values of the parameters in Table \ref{table:sedfits}. In addition the $\chi^2$ per data point for the best-fitting model of GPSV3 was found to be larger than the derived values for the remaining objects. Thus, the parameters derived by the fits for GPSV3 and GPSV15, shown in Table \ref{table:sedfits} can be considered as highly unreliable. However the \citeauthor{2007Robitaille} fits support our impression from inspection of the SEDs that GPSV3 and GPSV15 are not systems with optically thick envelopes given the lack of emission of cold dust at longer wavelengths.

Even though both sources show clear YSO characteristics, the fact that GPSV15 is not associated with any star forming region raises the doubt that these stars could actually be highly reddened post-AGB stars, or OH/IR stars similar to those mentioned in Section \ref{sec:search_res}. 
Using K band photometry and {\it Akari 9~$\mu$m} observations \citet{2011Ishihara} were able to determine mass loss rates and distances for carbon- and oxygen-rich AGB stars in the Galaxy. The [K]-[9] colours for both GPSV3 and GPSV15 would correspond to AGB stars with very large mass loss rates, as high as 10$^{-4.1}$ M$_{\odot}$yr$^{-1}$  according to Fig. A.2 in \citet{2011Ishihara}. However, using these values and their observed fluxes at 9~$\mu$m in Eq. A.2 of \citet{2011Ishihara}, yields distances larger than 25~kpc from the Sun, which at their Galactic latitude ($b \sim 2-3^{\circ}$) would put them 800 pc above the Galactic plane and well beyond the Galactic disc edge 
\citep[$R \approx 14$~kpc, see e.g.][]{2011Minniti}, thus ruling out a post-AGB classification for both GPSV3 and GPSV15.

%

\subsubsection{GPSV8 and GPSV16}

K band photometry of GPSV8 shows that the star was brighter during 2MASS and GPS $K_{2}$ observations in 1999 and 2005 but faded by the epoch of GPS $K_{1}$ in 2008 and remained at that state during ISAAC and UKIRT observations in 2010 and 2012 (see Fig. \ref{lc_seds:erup}). The fading is also apparent when comparing {\it Spitzer} $I1,I2$ ($\sim 2006$) with {\it WISE} $W1,W2$ ($\sim$ 2010) photometry, where the difference of both $I1-W1$,$I2-W2$ are larger than the typical difference of $\sim$ 0.2 mag found for field stars with a similar magnitude to GPSV8. The ISAAC spectrum of GPSV8 (Fig. \ref{sp:erup}) shows strong Br$\gamma$ emission along with CO absorption, with Na I and Ca I also in absorption. The same spectroscopic characteristics can be observed in the recent 2012 NIFS spectroscopic data. They are typical of EXor variables in 
the quiescent state \citep{2012Loren} and consistent with the K band light curve.


GPSV8 is also detected in VPHAS$+$. Direct comparison of its $r^{'}-i^{'},r^{'}-H\alpha$ colours places it in the $H\alpha$ emitters region of \citet{2008Witham}. We do note that their actual selection of $H\alpha$ emitters depended on the magnitude of the source and field where it was located. The aforementioned colours also place it near the unreddened mid M star colours in Figure 6 of \citet{2005Drew}. However this classification is not possible given its observed $g^{'}-r^{'},r^{'}-i^{'}$ colours and the SED of the star does not agree with a mid M star classification. The r$^{'}$-H$\alpha$ colour implies emission with maximum equivalent width of 40-50 \AA, for the case of a heavily veiled YSO or object with a continuum that is free of TiO bands.  With near-infrared Br$\gamma$ emission as well and a low mass Class II YSO fit with the \citeauthor{2007Robitaille} models (see Table \ref{table:sedfits}) an accreting YSO scenario for GPSV8 is well supported.

The light curve of GPSV16 (see Fig. \ref{lc_seds:erup}) supports an outburst scenario with the star still close to maximum brightness at the moment of our ISAAC observations (2010). Further evidence to support this scenario is given by the fact that the object is not apparent in the 2MASS (1999) images. The ISAAC spectrum shows Br$\gamma$ emission as well as CO in emission. These emission features are much more clearly seen in the 2012 NIFS spectrum of the source (Fig. \ref{sp:erup}). These are characteristics of an EXor variable in outburst. The fact that the spectrum of GPSV16 has not changed noticeably between the 2010 and 2012 observations suggests that the star was either at a bright state or been fortuitously observed during two separate outbursts.

The SED of GPSV16 shows a strong rise towards 2.2 $\mu$m and a flat distribution towards longer wavelengths, similar to what is observed in classical T Tauri stars (CTTS). Supporting a YSO scenario is the fact that the star is located at a projected distance of $\sim$ 170$\arcsec$ from the star forming region G71.52-0.39 of the \citet{2002Avedisova} catalog (see Figure \ref{fig:serp}), and although it is not conclusive this could be evidence for a possible association. SED fitting of GPSV16 with the tool of \citet{2007Robitaille} suggests a classification as low mass Class II YSOs (see Table \ref{table:sedfits}).

\subsection{Young stars in Serpens OB2}\label{rc:sec}

In addition to GPSV3 and GPSV8, the rest of GPSV1 to GPSV11 are also located in the Serpens OB2 region.

The SEDs of GPSV1, GPSV2, GPSV7, GPSV10 and GPSV11 are consistent with them being YSOs, rising until typically 3.6~$\mu$m and then showing flatter SEDs towards longer wavelengths. 
We note the cases of GPSV7 and GPSV10, which show a dip in their SEDs in the mid-infrared 
before rising towards $24 \mu$m. YSOs with this type of SED have been interpreted as having optically thin or evacuated inner holes and an optically thick edge at a radius of $\sim$10 AU \citep{2005Hartmann}. Alternatively, extinction of the hot inner parts of the disc by an edge-on 
system orientation can produce a similar effect. We note that the GLIMPSE3D catalogue photometry flags indicate that contamination from a nearby source may have affected the photometry of GPSV7. The same source is probably responsible for the non-detection in 2MASS and DENIS and may well have affected the WISE data.

\begin{figure}
\begin{center}
\includegraphics[width=1\columnwidth]{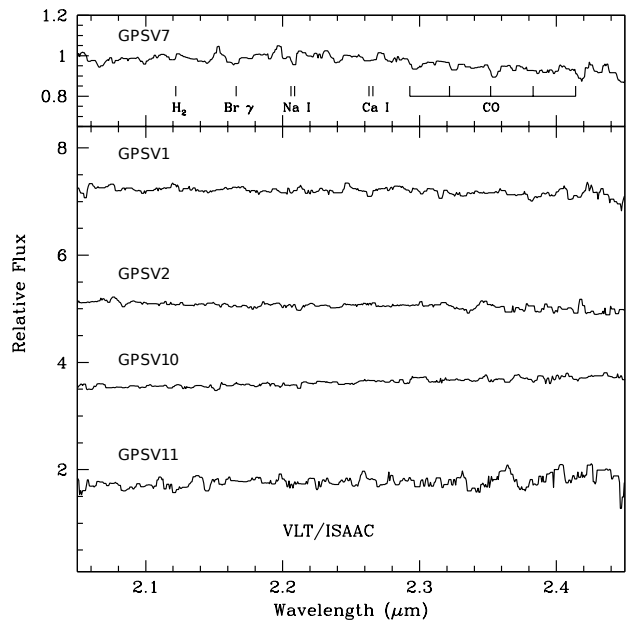}
\caption{Spectra for stars discussed in Section \ref{rc:sec}. Spectral features usually found for eruptive variable stars are shown only for GPSV7, the only object where CO absorption bands can be identified. Spectra for the other four objects are dominated by noise.}
\label{lc:ysos}
\end{center}
\end{figure}

From the aforementioned sources, only GPSV7 shows interesting spectral features in our low resolution ISAAC spectra, with CO absorption bandheads in the 2.3~$\mu$m region (Fig. \ref{lc:ysos}) and no sign of Br$\gamma$ emission. However, its peak-to-trough amplitude in $K$ barely exceeds 1 magnitude, making its eruptive nature questionable. 

GPSV1 and GPSV11 correspond to the most interesting cases. The magnitude of GPSV1 rose by 3.75 mag between the two epochs of GPS, then declined by 0.26 mag in our ISAAC photometry. The fact that the star is not visible in the 2MASS image provides some support for a possible recent outburst scenario. Unfortunately even at its brightest state the star is still faint for our ISAAC spectroscopic observations ($K\sim$ 15.4 mag) and no characteristic features for FUor/EXors could be identified from the spectrum of this object. However, the flat slope of the spectrum rules out a normal main sequence classification and, as noted above, the SED rises steeply towards 5.8 $\mu$m, consistent with a YSO interpretation. If the variability is due to a YSO outburst, the rate of decline between GPS $K1$ and ISAAC is 
$\sim$0.01 mag/month and the star would return to its pre-outburst magnitude (assuming the first epoch of UKIDSS GPS photometry as the pre-outburst state) in $\sim$ 30 yrs. This falls in the expected timescale for an FUor outburst \citep[see e.g.][]{2011Kospal}. 

GPSV11 shows similar magnitudes in both 2MASS (1999) and the first epoch of GPS (2005), followed by a strong decline towards the GPS K$_{1}$ (2008) and ISAAC (2010) epochs of $\sim$ 2.5 magnitudes. The recent UKIRT observations (2012) show that the object has returned to a magnitude level similar to that of 2MASS. The magnitude change in this source is probably not related to  variable extinction along the line of sight (see below). It is unfortunate however that the spectrum, also dominated by noise, makes it impossible to observe any spectral characteristics that could clarify the nature of the object, however the observed slope in the spectrum of GPSV11 goes against a main sequence star classification which supports the YSO scenario for this star.

GPSV2 and GPSV10 are also highly variable objects with variations of 2.4 and 1.5 mag respectively. However, the variability of the former is more likely related to variable extinction in the line of sight (see below). Similar to GPSV8, GPSV2 is also detected in VPHAS$+$ and is located on the lower envelope of the $H\alpha$ emitters distribution from \citet{2008Witham}. The case for line emission is also marginal when comparing to Figure 6 in \citet{2005Drew}. The $r^{'}-i^{'}$ and $r^{'}-H\alpha$ could be compatible with an unreddened mid-M star. The slope of the observed K band spectrum is incompatible with this scenario. The observed SED also goes against the mid M classification. Then the $H\alpha$ equivalent width could be 30-40 \AA~if the underlying SED is free of TiO bands. This supports a YSO classification. The slope of the ISAAC spectrum and the SED of GPSV10 are inconsistent with a normal main sequence star and support the YSO classification.

GPSV9 is a very red (and faint) object. Its relatively low observed amplitude of 1.16 magnitudes makes an eruptive YSO classification questionable, similar to objects GPSV4, GPSV5 and GPSV6. However, the apparent spatial association of these objects with Serpens OB2, makes them likely YSOs. Spectra of the four candidates did not help in the classification of these objects.

\section{Discussion}

\subsection{Physical Mechanisms}\label{sec:phys_mec}

We note that for two of the objects discussed as probable eruptive variables above, GPSV8 and GPSV15, peak-to-trough K-band amplitudes do not exceed $\sim$1.5~mag, bringing some conflict with one of the main characteristics used to define eruptive variables, which is an increase of brightness larger than at least $\sim$2~mag  \footnote{This is the limit usually found in the literature. We note that the original classification for these type of objects was done in the optical and amplitudes are usually smaller in the infrared. The $K$ band amplitudes of the (new detections) EXor sample from \citet{2012Loren} (see their Table 1) show a limit of $\Delta K > 1.2$~mag, with the majority of the sample having $\Delta K>1.95$~mag.}. Although this is probably due to the incomplete coverage of the light curves, we want to discard other possible physical mechanisms that could be causing large variability.  

\begin{figure}
\centering
\includegraphics[width=1\columnwidth]{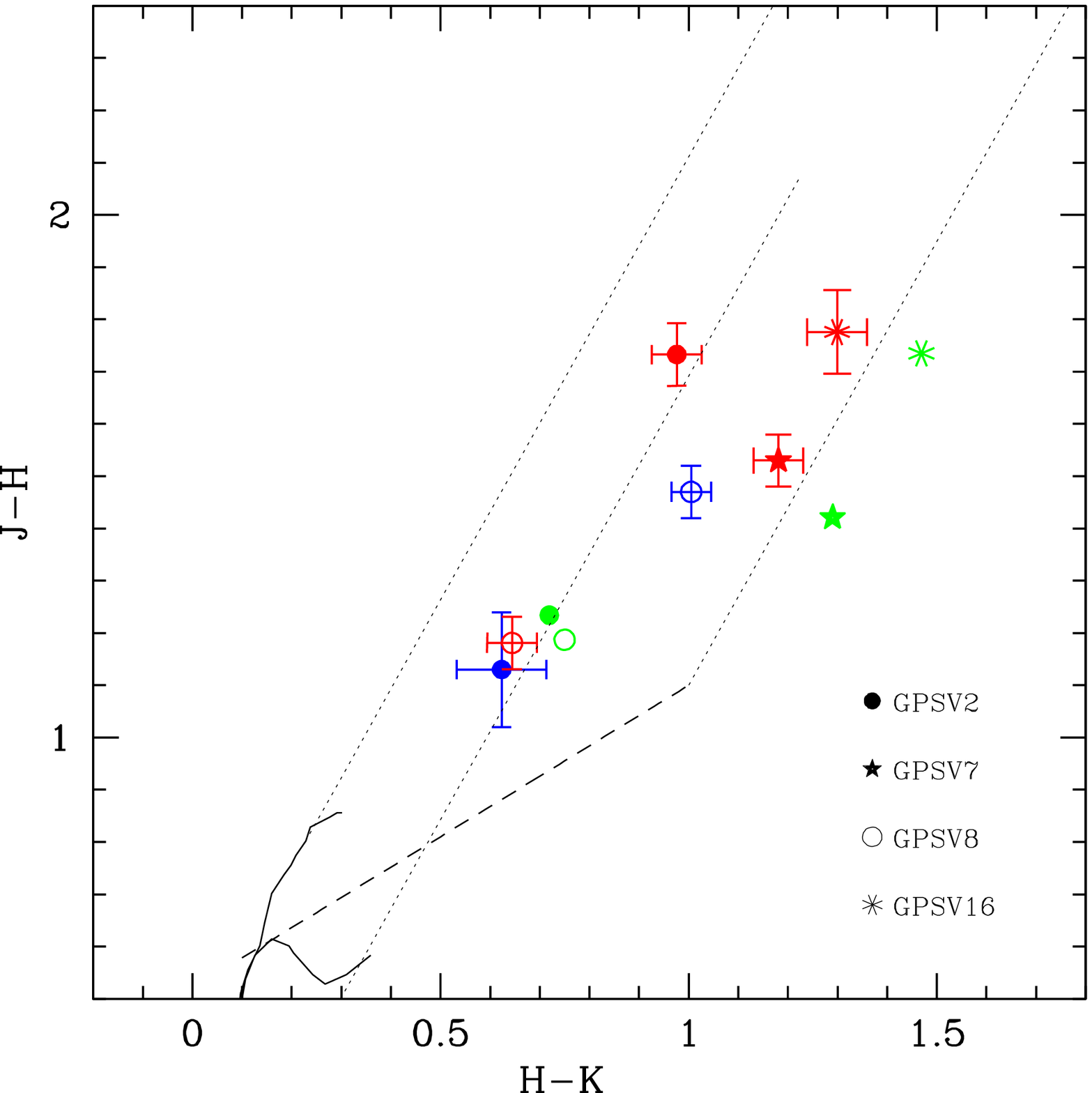}\\
\includegraphics[width=1\columnwidth]{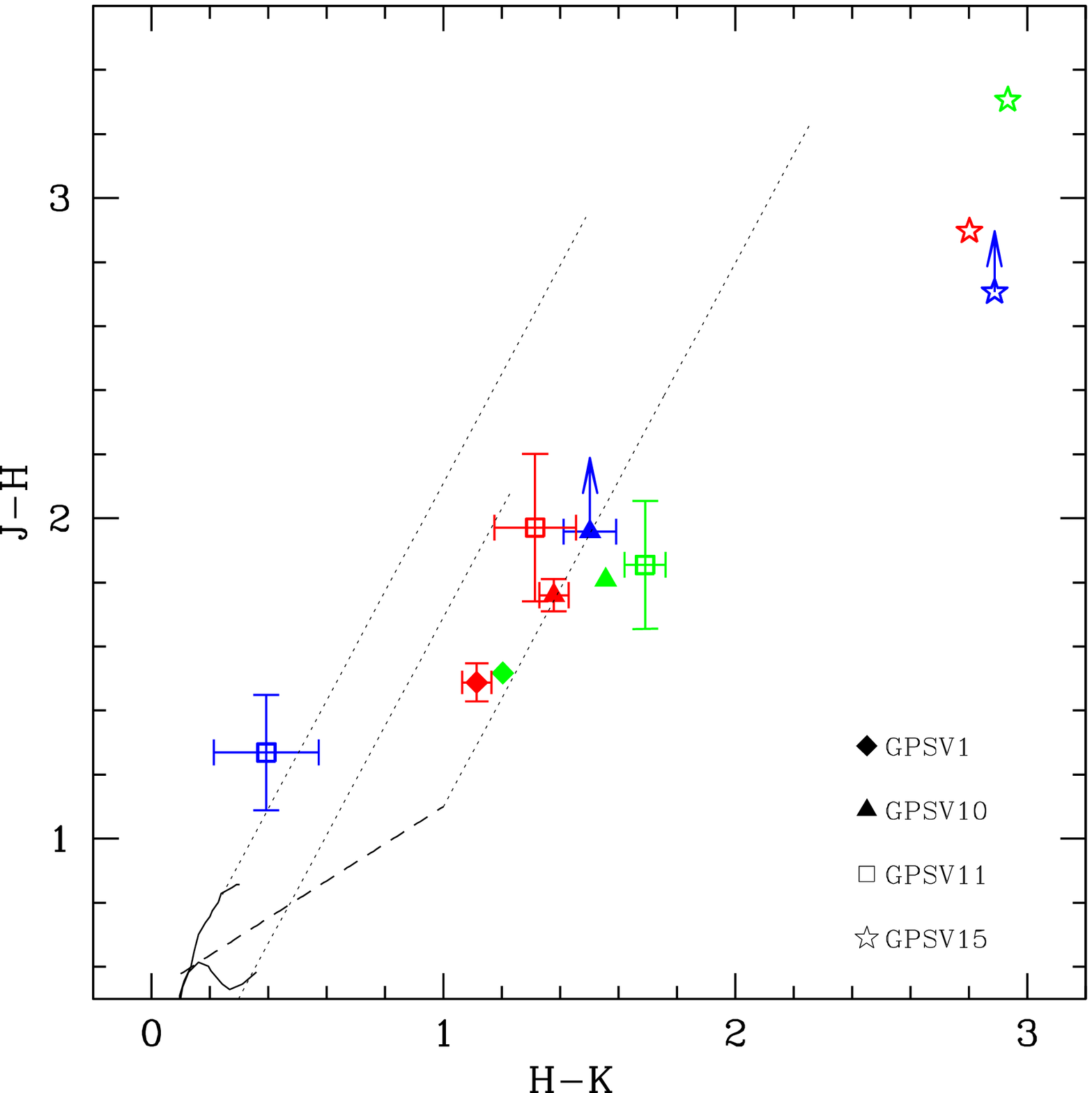}
\caption{$J-H,H-K$ colour-colour plot comparing the 2MASS (\textit{blue}), GPS (\textit{green}) and ISAAC (\textit{red}) colours for some of the objects in our sample. The errors are plotted only for objects that present significant uncertainties on their measurements.}
\label{chcol:erup}
\end{figure}

Several physical mechanisms can explain the near-infrared variability observed in YSOs. Some of these mechanisms are cool spots induced by magnetic activity, hotspots formed by the accretion flow \citep[see e.g.][]{2009Scholz}, magnetic field interaction between the disc and the star \citep{2013Romanova} and accretion driven wind and outflow \citep{2012Bans}. However, these mechanisms often produce short-term variability with amplitudes that are not expected to exceed 1 magnitude in K band. Changes in the extinction along the line of sight can produce larger changes in magnitude and on longer timescales \citep[see e.g. the long-lasting deepening of AA Tau,][]{2013Bouvier}. In this case the colour of the star would be expected to move along the reddening vector in colour-colour plots. 

The fact that we use $\Delta K=1$~mag to select our candidates already discards some of the mechanisms stated before. The changes in colour of the sources discussed above (Fig. \ref{chcol:erup} ) can provide further information on the nature of the objects; EXor variables for example are expected to have bluer colours as their magnitude increases. However, this is not observed in our sample. We note that due to the limited coverage we are not able to directly compare colours from quiescent and bright states. We can only discard variable extinction as a possible explanation because the stars do not move along the reddening vectors, except in the case of GPSV2.

\citet{2001Carpenter} find 1235 variable stars in their study of the Orion molecular cloud. Visual inspection of their light curves yields 20 variables with $K$ peak-to-trough amplitudes $>$ 1 mag. The variability of most of these stars is of the order of days and is likely to be explained by the physical processes discussed above. However, a handful of stars present large variability (up to 1.7 mag) on longer timescales, driven by a few points in the light curves and resembling what one would expect to observe in eruptive variables. We note that \citet{2001Carpenter} do not claim to have observed any FU Ori type outbursts.

\begin{figure}
\centering
\includegraphics[width=1\columnwidth]{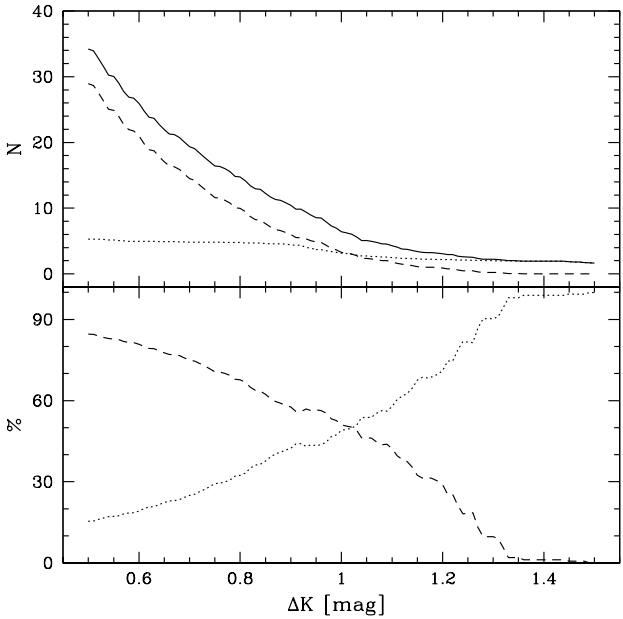}
\caption{\textit{top} Expected number of stars to be observed in Serpens OB2 as a function of amplitude (\textit{solid line}). The results are also divided in stars with short-term variability (\textit{dashed line}) and long-term variability (\textit{dotted line}). \textit{bottom} Percentage of detections as a function of $\Delta K$ comparing short-term variables  (\textit{dashed line}) with long-term and more likely to be accretion-related variables (\textit{dotted line}). }
\label{lc:carp}
\end{figure}

We want to determine the likelihood of detecting any of the 20 high-amplitude \citet{2001Carpenter} variables, by just observing two epochs and as a function of a defined amplitude cut. In order to do so, we selected two random points of the light curve of each star in the high amplitude \citet{2001Carpenter} sample, determine the magnitude difference between the two points and marked a detection if the difference was larger than the defined amplitude cut. The process was iterated a few thousand times. The likelihood of the detection of each star is then defined as the number of detections divided by the number of iterations. The sum of the expected detections over the 20 variables from the \citeauthor{2001Carpenter} sample was multiplied by the ratio between the number of OB stars in Serpens OB2 and those within the region of Orion studied by \citet{2001Carpenter}. 

The described process yields the number of variable stars we would expect to have selected in Serpens OB2 as a function of the amplitude cut (see Fig. \ref{lc:carp} ). Even though there are a number of caveats to this analysis, we find that for an amplitude cut of $\Delta K=1$~mag, we should detect 7 variables, which is remarkably close to our actual number (11). At this amplitude cut, a large part of our sample would be formed by stars which variability is explained by other processes than abrupt changes in the accretion rate. However, we observe that for $\Delta K>1.1$~mag, the sample is mostly formed by stars which vary over longer timescales driven by a few points in their light curves (dotted line in Fig. \ref{lc:carp}), and thus more likely associated with accretion-related variability. Even though our analysis uses sources in Serpens OB2, we can assume that the observed photometric changes in YSOs in our sample which present $\Delta K>1.1$~mag are more likely related to this type of variability.


\subsection{Molecular Hydrogen outflows}

The visual inspection of the UKIRT $H_{2}-K$ difference images reveals the presence of only one pair of
bright H$_{2}$ molecular outflows in the area of Serpens covered by our study. We identify them as the 
previously discovered emission features MHO2201 and MHO2202, which are associated with the deeply embedded 
luminous YSO IRAS 18151-1208 \citep[][]{2010Varricatt}. There is also widespread diffuse
H$_{2}$ emission in parts of the region, but no other jet-like features that can be associated with a single 
star. This suggests that the observations were not deep enough to detect jets associated with low mass YSOs. No $H_2$ outflows could be observed near the eruptive variable candidates GPSV2 to GPSV11. Similarly, 
there is no sign of an outflow in the NIFS integral field data for the four objects spectroscopically verified
as eruptive variables (GPSV3, GPSV8, GPSV15 and GPSV16).

Most FU Orionis objects are known to be associated with Herbig-Haro objects \citep{1997Reip} and drive molecular outflows \citep{1994Evans}, whilst EXors generally do not show evidence of shocks as a major mechanism of excitation \citep{2009Loren}. FU Orionis itself, although having massive winds, does not drive a molecular outflow \citep{1994Evans} which could be explained by the lack of organization of magnetic fields required for jet launching and collimation in the binary scenario of \citep{2004Reip} or by previous outflows having cleared up the remnant envelope evidencing the lack of swept up material which constitutes the molecular outflows \citep{1994Evans}.

From the objects associated with the deeply embedded subclass of eruptive YSOs, OO Ser is associated with faint 
H$_{2}$ shock fronts \citep{2012Hodapp}, corresponding to MHO 3245 in the catalog of \citet{2010Davis}. GM Cha is the driving source of a CO outflow and shows a $K$ band spectrum dominated by H$_{2}$ emission with several shocked H$_{2}$ emission knots observed along the CO flow \citep{2007Persi}. Finally, V2775 Ori is not associated with any outflows nor H$_{2}$ shock emission. 

Given the limitation imposed by the sensitivity of the dataset and the fact that outflows are not 
ubiquitous in eruptive variable YSOs, the lack of observable H$_{2}$ emission near our somewhat more distant 
candidates is not a strong argument against their eruptive variable classification. 

\subsection{Eruptive variable classification}

Although four stars in our sample show strong characteristics of eruptive variability, it is difficult to classify them as either of the two known classes of this kind of variables.

GPSV3 and GPSV15 show similar spectra to known deeply-embedded FUor-like objects and that of OO Ser. However, the amplitude and time-scale of the variation, with what appear to be repetitive outbursts, are not in line with the observed behaviour of OO Ser itself and the outbursting protostar [CFT93]216-2 \citep{2011Caratti}. In fact, the photometric behaviour of both candidates is similar to that observed for deeply embedded objects V371 Ser \citep{2012Hodapp} and source 90 in the study from \citet*{2005Varricatt} of SFR 173.58+2.45. The latter has large infrared colours and it is the driving source of an H$_{2}$ molecular outflow. Its variability was first thought to be of the FU Orionis kind but this classification was discarded because of the shorter time scale of the variation. \citet{2005Varricatt} still conclude that its variability probably arises from variable accretion. V371 Ser is classified as a Class I object, also driving a H$_{2}$ molecular outflow. Its variability is periodic ($P=543~d$) and is 
thought to be explained by periodic variation of the accretion rate, modulated by the presence of a close companion \citep{2012Hodapp}. A binary companion in an elliptic orbit is a possible explanation for the FUor and 
EXor phenomena, given that some repetitive EXors have been seen (though not FUors). \citet{2012Hodapp} proposed 
that V371 Ser represents an extension of eruptive variability to shorter periods.

The spectrum and photometry of GPSV8 from 2010-2012 is consistent with an EXor during quiescent states. However, the star appears to have been at a bright state during 1998-2005 (although this conclusion is based on only two epochs of observations), if so the time-scale is inconsistent with EXor variables. The spectrum of GPSV16 shows strong emission lines (Br$\gamma$, CO) during 2010 and 2012 observations, which is usually observed in EXor variables during bright states. However, the photometric data of the object would point towards an FUor type outburst given the longer timescale of the variation.

A number of stars in which the observed eruptive variability is thought to be accretion related 
have shown characteristics that can be associated with both sub-classes of eruptive variables. e.g. V1647 Ori, 
V2775 Ori and OO Ser. These objects have been proposed to be intermediate objects between EXors and FUors or 
being the extension of the phenomenon towards lower luminosities. 

This raises the scenario where eruptive variability cannot be simply divided in two distinctive sub-classes but it is instead formed by a continuum of outburst events of different luminosities, amplitudes and time-scales, but triggered by a similar physical mechanism \citep[see e.g.][]{2007Fedele,2011Caratti,2007Kospal,2006Gibb}.   

Thus, our results seem consistent with this scenario, where GPSV3 and GPSV15 would be new additions to the deeply-embedded extension of the FUor phenomenon.

Future photometric monitoring of the sources will allow us to better  constrain the timescales of the observed variability. Also, spectroscopic analysis of sources in later data releases from GPS (DR7,DR8), with addition of future searches in the multiple $K$ band epochs data from VVV will help us in our effort to characterise and understand the subclasses of eruptive variables.

\section{Variability outside Star Forming Regions}

Five stars from our DR5 sample are marked as variables not associated with SFRs. GPSV15 has already been discussed at length and has been classified as a probable eruptive variable in our analysis. From the remaining four stars,  GPSV13 and GPSV17 correspond to blue objects, with GPSV17 already being identified as Nova Sct 2003. The location of GPSV12 in its local CMD is consistent with a late type dwarf or a distant giant in the Galactic halo, whilst GPSV14 seems to be projected against the dwarf sequence on its CMD in Figure \ref{dr57cmd}. 

High amplitude variability can be produced by a great number of physical mechanisms. Inspection of Figure 2 of \citep{2008Corradi} shows that symbiotic stars, classical Be stars and cataclysmic variables, among others, can be found at the location of the objects in near-infrared colour-colour plots. We also note that Algol-type eclipsing variables show large variability at optical wavelengths and while amplitudes at near-infrared wavelengths are lower, it is conceivable that we could detect an extreme case of this type of variable stars. 

Not much information is available for objects GPSV12 and GPSV14. The latter was not included in the VLT/ISAAC follow up and only has magnitude information from the UKIDSS GPS data. GPSV12 was faint at the time of our ISAAC observations and the spectrum did not help to draw any conclusion regarding its classification. GPSV12 could 
plausibly be an AGN (see $\S$2.2) but the higher amplitude of variability in GPSV14 (2.1 mag) makes this 
interpretation unlikely for that object.

\begin{figure*}
\begin{center}
\includegraphics[width=1.7\columnwidth]{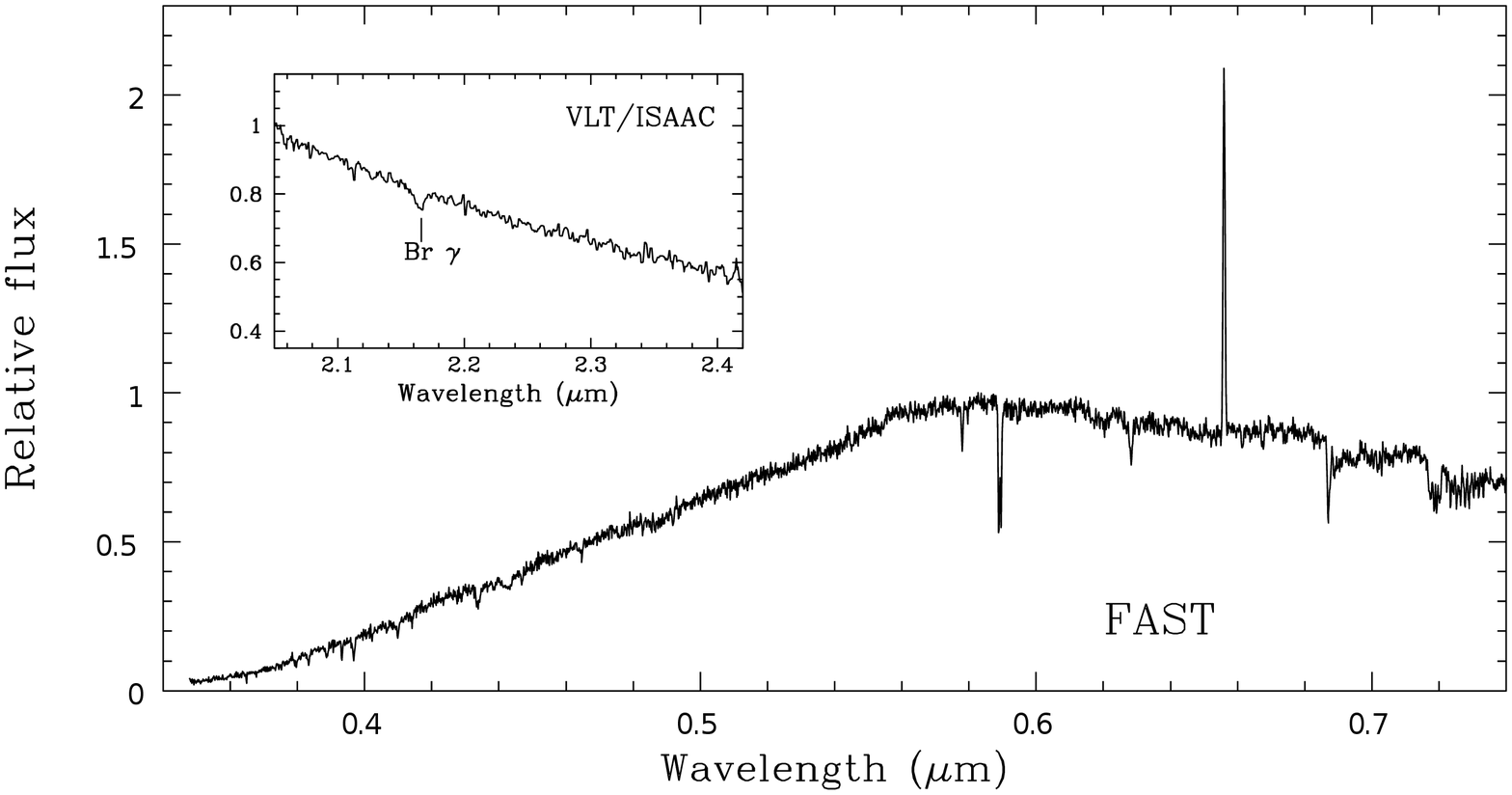}\\
\includegraphics[width=1\columnwidth]{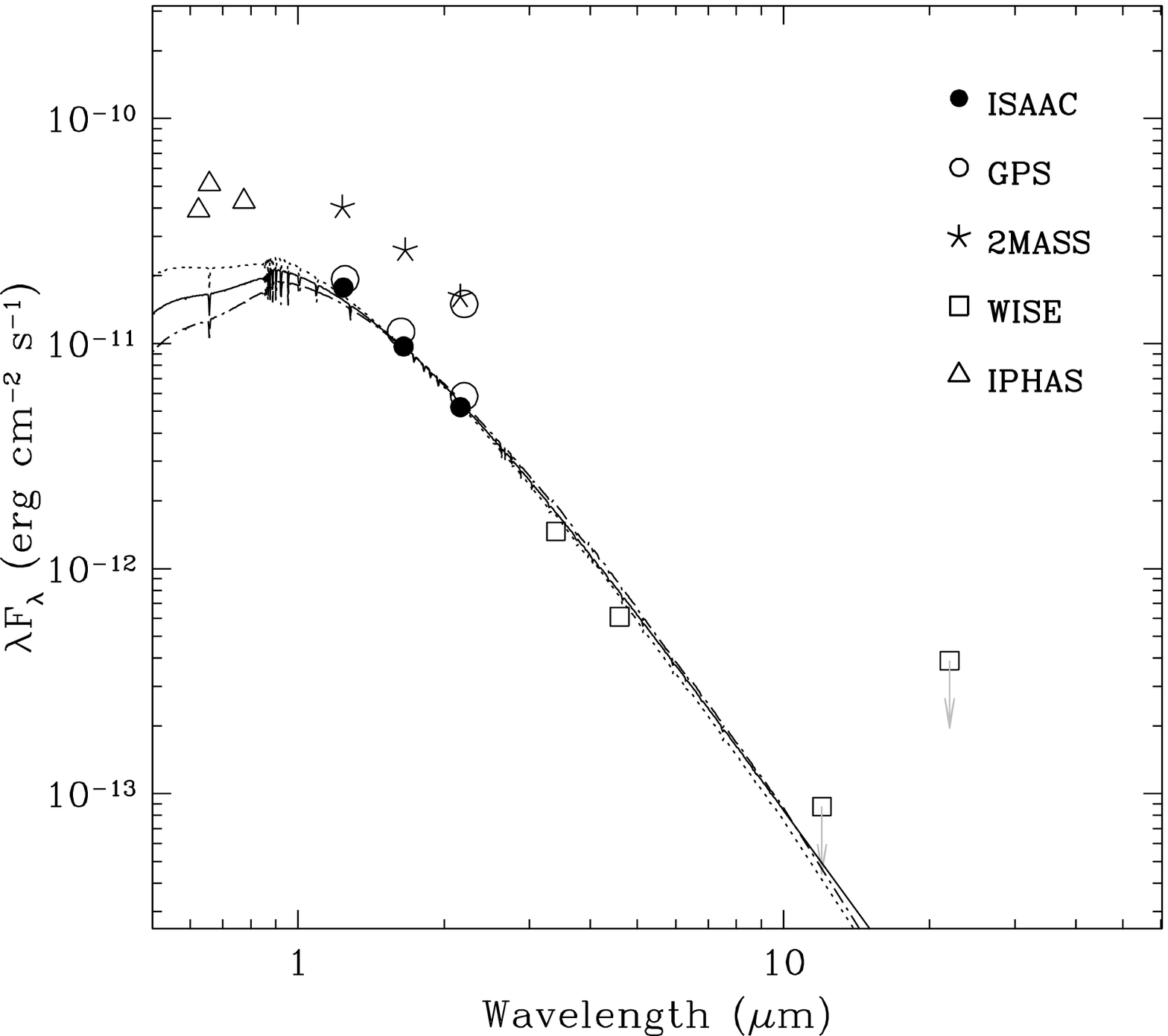}
\includegraphics[width=1\columnwidth]{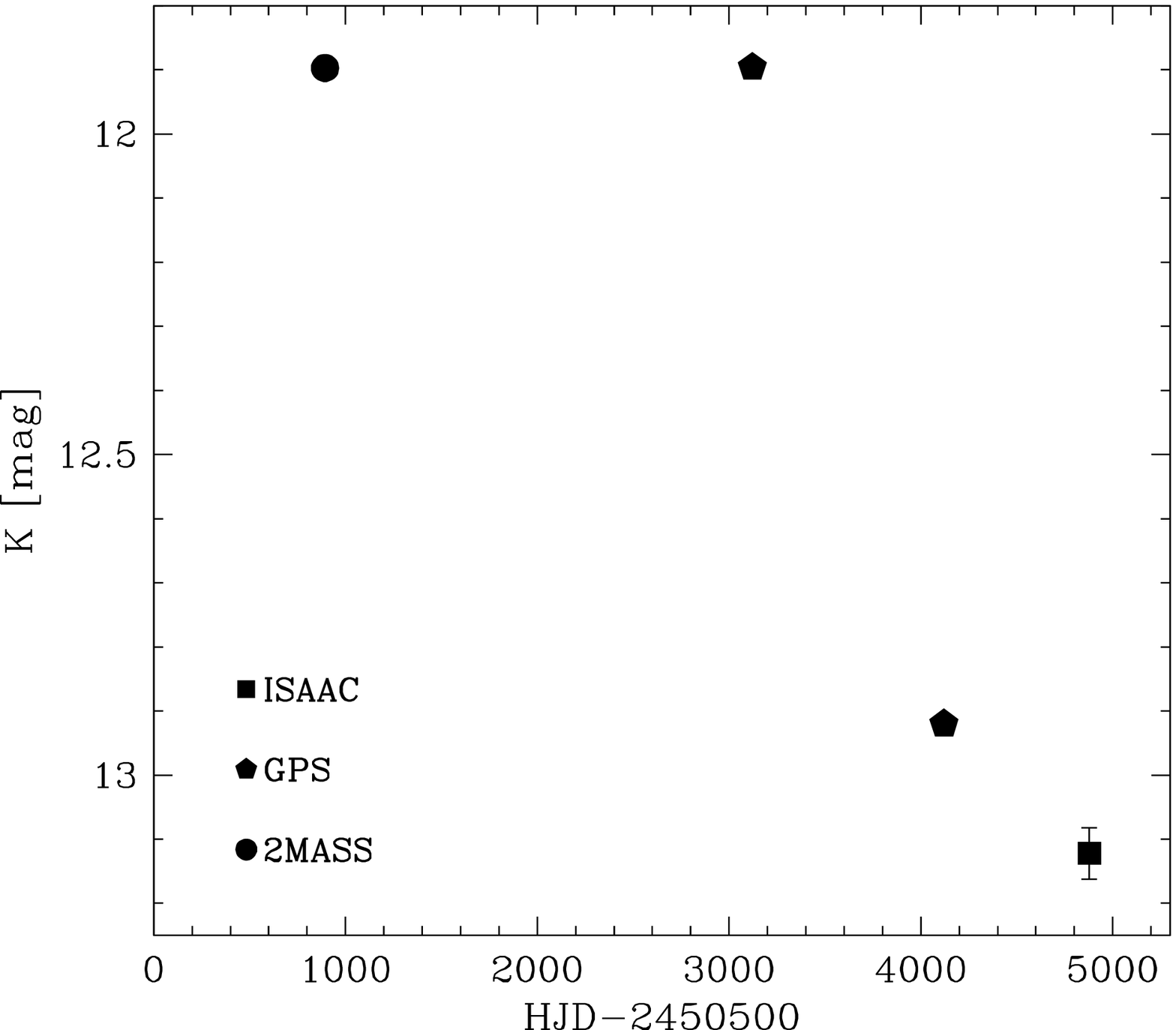}
\caption{The optical FAST spectrum of June 2009, along with the VLT/ISAAC observations of  July 2010 ({\it top}). K band light curve ({\it bottom right}) and spectral energy distribution ({\it bottom left}) of the object GPSV13. The latter is compared to a B5V (15400 K) \citeauthor{2004Castelli} model atmosphere ({\it solid line}) and the same model reddened to $A_{V}=2.55$~mag ({\it dotted line}) and $A_{V}=3.45$~mag ({\it dashed-dotted line}). Fluxes of the model where arbitrarily set to match the 2010 $H$ observations of VLT/ISAAC. The models are reddened using the \citet*{1989Cardelli} extinction law for wavelengths in the range $0.3 \mu$m$ < \lambda < 3.3 \mu$m. Extinction towards longer wavelengths is derived following the \citet{2009Chapman} extinction law for {\it Spitzer} bands. }
\label{figs:gps13}
\end{center}
\end{figure*}

\subsection{GPSV13, A highly variable classical Be star?}

GPSV13 is one of the two blue DR5 sources and it is also one of the brightest variable stars in our sample. Its 1999 2MASS K$_{s}$ magnitude is very similar to the first K band observation of GPS in 2005. The star was fainter at the second GPS epoch (2008) and continues fading towards our VLT/ISAAC observations (2010). The spectrum of the star at this last epoch shows Br$\gamma$ in absorption (see figure \ref{figs:gps13}).
   
The object was found to be classified as an H$\alpha$ emitter in the \citet{2008Witham} catalogue. The IPHAS observation for this object took place in 2004/2005 during the bright state of the object. Figure \ref{figs:gps13} shows the optical spectra of GPSV13 collected in June 2009 at the 1.5 m Fred Laurence Whipple Observatory (FLWO) Tillinghast Telescope using the FAst Spectrograph for the Tillinghast Telescope \citep[FAST, ][]{1998Fabricant}, with a spectral resolution of $\Delta \lambda \sim 6$ \AA, covering the wavelength range 3500$-$7500 \AA. The equivalent width (EW) of the H$\alpha$ emission is 9.7 \AA, and the spectral typing of the object yields a B5 spectral type based on the EW of the HeI 4471, and just resolved but weak MgII 4481 absorption lines. The reddening is estimated to be A$_{V}=3.00\pm0.15$. Assuming a B5V classification yields a distance of $\sim$ 2.8 kpc.
   
The 2004/2005 $r^{'}-i^{'},r^{'}-H\alpha$ colours of GPSV13 allow us to estimate the EW of the H$\alpha$ emission as $\sim$20-22 \AA, which is higher than the measured value of 2009 and consistent with the photometric behaviour of the object under the classical Be interpretation described below.
    
Classical Be stars show variability over long timescales that can be attributed to the dissipation of the disc \citep{2003Porter}. During this process, the star will go from having an emission line spectrum to a normal B type star spectrum \citep*[see e.g.][]{2008McSwain,2009McSwain}. This process could explain the observed behaviour of the spectrum of GPSV13. In addition, dissipation of the disc could explain the large (1.2 mag) decline in brightness in the K band from 2005 to 2010. Inspection of Figure 9 of \citet{2006Carciofi} shows that the dissipation of a pole-on disc in the models of these authors just reproduce the observed amplitude of the change. This scenario also explains the apparent lack of IR excess in the 2010 WISE W1, W2 magnitudes.
    
An alternative scenario would be that GPSV13 is a fairly evolved Herbig Ae/Be star with a far IR excess not detectable in WISE data. However the star is not close to any known SFR, there is no observable nebulosity in  the WISE colour image of the region and the estimated distance to the object implies that GPSV13 would be $\sim$ 160 pc above the Galactic plane, at least three molecular cloud scale heights for an inner Galaxy object. Figure 2 in  \citet{2005Hernandez} compares the location of classical Be and Herbig Ae/Be stars in near-infrared colour-colour plots, the authors find that these types of variables seem to occupy different places in these diagrams. The observed colours of GPSV13 would place the object in the classical Be region. Also, its SED resembles those of classical Be stars from the sample of \citet{2005Hernandez}, see \ref{figs:gps13}. Thus the classical Be scenario seems the most likely explanation for GPSV13. 

If GPSV13 is in fact a classical Be star, it would correspond to one of the most (if not the most) extreme variable star of this class. To our knowledge classical Be stars have not been observed to vary by more than 1 magnitude \citep[see e.g.,][]{1984Ashok,1994Dougherty}.

\section{Summary}
 
We have presented the results of the first panoramic search for high-amplitude near-infrared variables in the Galactic plane. We summarize our principal results as follows.

\begin{itemize}

\item The search for high-amplitude infrared variables in data releases DR5 and DR7 of GPS yields 45 stars which display $\Delta K > 1$ mag. The sample included one known nova and two OH/IR stars. Most notably, two thirds of our candidates are located within known areas of star formation and are likely YSOs. Four other stars also show characteristics of YSOs but do not appear to be close to areas of star formation. Thus we discuss the possibility of high-amplitude IR variables as a tracer of star formation.\\

\item The high YSO fraction found in our study implies an average source density of 0.19 deg$^{-2}$. We 
conclude that this number would be likely to rise to $\sim$13 deg$^{-2}$ in the mid-plane, if we had 
sufficient sensitivity to detect them across the Galaxy, after allowing for the effects of 2-epoch sampling 
and scale height upon our sample.
This would make PMS stars the commonest type of high amplitude variable in the near-infrared.\\

\item  Considering the DR5 sample, we assign an eruptive variable classification to four stars based on their observed spectral characteristics and their resemblance to known members of the FUor/EXor class. Two of these are likely new members of the deeply embedded subclass of eruptive variables. Nine stars likely associated with the Serpens OB2 association are confirmed as having YSO characteristics, but an eruptive variable classification cannot be confirmed. In this context, physical mechanisms that could  explain IR variability in YSOs are discussed. Variable extinction along the line of sight is found to be the likely cause of the variation for only one of our variable objects.\\ 

\item The comparison with the variable YSO sample of \citet{2001Carpenter} allows us to estimate that the 
variability in stars with $\Delta K>1.1$ mag in two epochs of observation separated by a few years is most likely related to longer timescale variability like that observed in eruptive variables.\\

\item Stars confirmed as eruptive variables show characteristics that could be associated with either of the two known classes of these type of variable stars (EXors or FUors). This is found to be consistent with recent studies that suggest that the episodic accretion phenomena is comprised of a continuum of events with different properties but triggered by a similar physical mechanism.\\

\item Variable stars in the Serpens OB2 region do not show any signs of molecular outflows.\\

\item Four stars in the DR5 sample appear not to be associated with star forming regions. We argue in favour of a classical Be classification for one of such objects. If so, the star would be one of the most variable classical Be stars.

\end{itemize}

\section*{Acknowledgments}

This paper is based in part on observations obtained in programme GN-2012B-Q-102 at the Gemini Observatory, which is operated by the Association of Universities for Research in Astronomy Inc., under a cooperative agreement with the NSF on behalf of the Gemini partnership: The National Science Foundation (USA), the Science and Technology Facilities Council (UK), the National Research Council (Canada), CONICYT (Chile), the Australian Research Council (Australia). CNPq (Brazil) and CONICET (Argentina).

This paper is also based on observations taken as part of programme 085.C-0352(A) with the Very Large 
Telescope in Cerro Paranal, which is operated by the European Southern Observatory.

We thank our colleagues in the Spitzer Legacy Survey of the Cygnus-X Complex for their work in providing the public data for many of the
DR7 objects. We also thank our colleagues in the GLIMPSE360 Spitzer
legacy survey team for providing the IRAC data for GPSV16.

This research has made use of the SIMBAD database, operated at CDS, Strasbourg, France.

This research has made use of the NASA/ IPAC Infrared Science Archive, which is operated by the Jet Propulsion Laboratory, California Institute of Technology, under contract with the National Aeronautics and Space Administration.

We thank the referee, Jacco van Loon, for a thorough and helpful report.

We would also like to thank J. Najita for the very helpful discussions on some of our CO spectra and 
D. Steeghs for his assistance in providing the FAST data.

C. Contreras Pe\~{n}a is supported by a University of Hertfordshire PhD studentship.

\bibliographystyle{mn2e}

\begin{thebibliography}{99}
\bibitem[\protect\citeauthoryear{Alves de Oliveira 
\& Casali}{2008}]{2008Oliveira} Alves de Oliveira C., Casali M., 2008, A\&A, 485, 155
\bibitem[\protect\citeauthoryear{Ashok et al.}{1984}]{1984Ashok} 
Ashok N.~M., Bhatt H.~C., Kulkarni P.~V., Joshi S.~C., 1984, MNRAS, 211, 
471 
\bibitem[\protect\citeauthoryear{Aspin 
\& Sandell}{2001}]{2001Aspin} Aspin C., Sandell G., 2001, MNRAS, 328, 751
\bibitem[\protect\citeauthoryear{Aspin 
\& Reipurth}{2003}]{2003Aspin} Aspin C., Reipurth B., 2003, AJ, 126, 2936 
\bibitem[\protect\citeauthoryear{Avedisova}{2002}]{2002Avedisova} 
Avedisova V.~S., 2002, ARep, 46, 193 
\bibitem[\protect\citeauthoryear{Bans K{\"o}nigl}{2012}]{2012Bans} Bans A., K{\"o}nigl A., 2012, ApJ, 758, 100 
\bibitem[\protect\citeauthoryear{Baraffe, Chabrier, 
\& Gallardo}{Baraffe et al.}{2009}]{2009Baraffe} Baraffe I., Chabrier G., Gallardo J., 2009, ApJ, 702, L27
\bibitem[\protect\citeauthoryear{Baraffe, Vorobyov, 
\& Chabrier}{Baraffe et al.}{2012}]{2012Baraffe} Baraffe I., Vorobyov E., Chabrier G., 2012, ApJ, 756, 118 
\bibitem[\protect\citeauthoryear{Bedijn}{1987}]{1987Bedijn} Bedijn P.~J., 1987, A\&A, 186, 136
\bibitem[\protect\citeauthoryear{Benjamin et 
al.}{2003}]{2003Benjamin} Benjamin R.~A., et al., 2003, PASP, 115, 
953
\bibitem[\protect\citeauthoryear{Bessell 
\& Brett}{1988}]{1988Bessell} Bessell M.~S., Brett J.~M., 1988, PASP, 100, 1134 
\bibitem[\protect\citeauthoryear{Bouvier et 
al.}{2013}]{2013Bouvier} Bouvier J., Grankin K., Ellerbroek L., 
Bouy H., Barrado D., 2013, arXiv, arXiv:1304.1487
\bibitem[\protect\citeauthoryear{Brice{\~n}o et 
al.}{2005}]{2005Briceno} Brice{\~n}o C., Calvet N., Hern{\'a}ndez 
J., Vivas A.~K., Hartmann L., Downes J.~J., Berlind P., 2005, AJ, 129, 907
\bibitem[\protect\citeauthoryear{Caratti o Garatti et 
al.}{2011}]{2011Caratti} Caratti o Garatti A., et al., 2011, A\&A, 526, L1
\bibitem[\protect\citeauthoryear{Caratti o Garatti et 
al.}{2012}]{2012Caratti} Caratti o Garatti A., et al., 2012, A\&A, 538, A64
\bibitem[\protect\citeauthoryear{Carciofi 
\& Bjorkman}{2006}]{2006Carciofi} Carciofi A.~C., Bjorkman J.~E., 2006, ApJ, 639, 1081
\bibitem[\protect\citeauthoryear{Cardelli, Clayton, 
\& Mathis}{Cardelli et al.}{1989}]{1989Cardelli} Cardelli J.~A., Clayton G.~C., Mathis J.~S., 1989, ApJ, 345, 245
\bibitem[\protect\citeauthoryear{Carpenter, Hillenbrand, 
\& Skrutskie}{Carpenter et al.}{2001}]{2001Carpenter} Carpenter J.~M., Hillenbrand L.~A., Skrutskie M.~F., 2001, AJ, 121, 3160
\bibitem[\protect\citeauthoryear{Casali et 
al.}{2007}]{2007Casali} Casali M., et al., 2007, A\&A, 467, 777
\bibitem[\protect\citeauthoryear{Castelli 
\& Kurucz}{2004}]{2004Castelli} Castelli F., Kurucz R.~L., 2004, astro, arXiv:astro-ph/0405087 
\bibitem[\protect\citeauthoryear{Catelan et 
al.}{2013}]{2013Catelan} Catelan M., et al., 2013, arXiv, 
arXiv:1310.1996
\bibitem[\protect\citeauthoryear{Chapman et 
al.}{2009}]{2009Chapman} Chapman N.~L., Mundy L.~G., Lai S.-P., 
Evans N.~J., II, 2009, ApJ, 690, 496
\bibitem[\protect\citeauthoryear{Cioni et 
al.}{2001}]{2001Cioni} Cioni M.-R.~L., Marquette J.-B., Loup C., Azzopardi M., Habing H.~J., Lasserre T., Lesquoy E., 2001, A\&A, 377, 945
\bibitem[\protect\citeauthoryear{Cioni et 
al.}{2013}]{2013Cioni} Cioni M.-R.~L., et al., 2013, A\&A, 549, A29 
\bibitem[\protect\citeauthoryear{Connelley 
\& Greene}{2010}]{2010Conelley} Connelley M.~S., Greene T.~P., 2010, AJ, 140, 1214 
\bibitem[\protect\citeauthoryear{Corradi et 
al.}{2008}]{2008Corradi} Corradi R.~L.~M., et al., 2008, A\&A, 480, 409 
\bibitem[\protect\citeauthoryear{Corradi et 
al.}{2010}]{2010Corradi} Corradi R.~L.~M., et al., 2010, A\&A, 509, A41 
\bibitem[\protect\citeauthoryear{Crutcher 
\& Chu}{1982}]{1982Crutcher} Crutcher R.~M., Chu Y.~H., 1982, ASSL, 93, 53 
\bibitem[\protect\citeauthoryear{Cutri et al.}{2012}]{2012Cutri} 
Cutri R.~M., et al., 2012, wise.rept, 1
\bibitem[\protect\citeauthoryear{Davis et 
al.}{2010}]{2010Davis} Davis C.~J., Gell R., Khanzadyan T., Smith M.~D., Jenness T., 2010, A\&A, 511, A24
\bibitem[\protect\citeauthoryear{Davis et 
al.}{2011}]{2011Davis} Davis C.~J., et al., 2011, A\&A, 528, A3 
\bibitem[\protect\citeauthoryear{D{\"o}rr et 
al.}{2013}]{2013Dorr} D{\"o}rr M., Chini R., Haas M., Lemke R., N{\"u}rnberger D., 2013, A\&A, 553, A48 
\bibitem[\protect\citeauthoryear{Dougherty 
\& Taylor}{1994}]{1994Dougherty} Dougherty S.~M., Taylor A.~R., 1994, MNRAS, 269, 1123 
\bibitem[\protect\citeauthoryear{Drew et al.}{2005}]{2005Drew} 
Drew J.~E., et al., 2005, MNRAS, 362, 753
\bibitem[\protect\citeauthoryear{Dutra 
\& Bica}{2002}]{2002Dutra} Dutra C.~M., Bica E., 2002, A\&A, 383, 631
\bibitem[\protect\citeauthoryear{Egan et al.}{2003}]{2003Egan} 
Egan M.~P., et al., 2003, yCat, 5114, 0
\bibitem[\protect\citeauthoryear{Dye et al.}{2006}]{2006Dye} 
Dye S., et al., 2006, MNRAS, 372, 1227 
\bibitem[\protect\citeauthoryear{Elias, Cabrera-Ca{\~n}o, 
\& Alfaro}{Elias et al.}{2006}]{2006Elias} Elias F., Cabrera-Ca{\~n}o J., Alfaro E.~J., 2006, AJ, 131, 2700
\bibitem[\protect\citeauthoryear{Enya et al.}{2002}]{2002Enya} 
Enya K., Yoshii Y., Kobayashi Y., Minezaki T., Suganuma M., Tomita H., 
Peterson B.~A., 2002, ApJS, 141, 45
\bibitem[\protect\citeauthoryear{Epchtein et 
al.}{1994}]{1994Epchtein} Epchtein N., et al., 1994, Ap\&SS, 217, 3
\bibitem[\protect\citeauthoryear{Evans et al.}{1994}]{1994Evans} 
Evans N.~J., II, Balkum S., Levreault R.~M., Hartmann L., Kenyon S., 1994, 
ApJ, 424, 793 
\bibitem[\protect\citeauthoryear{Evans et al.}{2009}]{2009Evans} 
Evans N.~J., II, et al., 2009, ApJS, 181, 321
\bibitem[\protect\citeauthoryear{Fabricant et 
al.}{1998}]{1998Fabricant} Fabricant D., Cheimets P., Caldwell N., 
Geary J., 1998, PASP, 110, 79
\bibitem[\protect\citeauthoryear{Fedele et 
al.}{2007}]{2007Fedele} Fedele D., van den Ancker M.~E., Petr-Gotzens M.~G., Rafanelli P., 2007, A\&A, 472, 207
\bibitem[\protect\citeauthoryear{Forbes}{2000}]{2000Forbes} Forbes 
D., 2000, AJ, 120, 2594
\bibitem[\protect\citeauthoryear{Gibb et al.}{2006}]{2006Gibb} 
Gibb E.~L., Rettig T.~W., Brittain S.~D., Wasikowski D., Simon T., Vacca 
W.~D., Cushing M.~C., Kulesa C., 2006, ApJ, 641, 383
\bibitem[\protect\citeauthoryear{Gonz{\'a}lez-Solares et 
al.}{2008}]{2008Gonzalez-Solares} Gonz{\'a}lez-Solares E., Walton N., Irwin 
M., Lewis J., Rixon G., Greimel R., Drew J., 2008, ASPC, 394, 197
\bibitem[\protect\citeauthoryear{Grinin et 
al.}{2001}]{2001Grinin} Grinin V.~P., Kozlova O.~V., Natta A., Ilyin I., Tuominen I., Rostopchina A.~N., Shakhovskoy D.~N., 2001, A\&A, 379, 482 
\bibitem[\protect\citeauthoryear{Gutermuth et 
al.}{2009}]{2009Gutermuth} Gutermuth R.~A., Megeath S.~T., Myers 
P.~C., Allen L.~E., Pipher J.~L., Fazio G.~G., 2009, ApJS, 184, 18
\bibitem[\protect\citeauthoryear{Hambly et al.}{2008}]{2008Hambly} 
Hambly N.~C., et al., 2008, MNRAS, 384, 637 
\bibitem[\protect\citeauthoryear{Hartmann 
\& Kenyon}{1996}]{1996Hartmann} Hartmann L., Kenyon S.~J., 1996, ARA\&A, 34, 207 
\bibitem[\protect\citeauthoryear{Hartmann, Hinkle, 
\& Calvet}{Hartmann et al.}{2004}]{2004Hartmann} Hartmann L., Hinkle K., Calvet N., 2004, ApJ, 609, 906
\bibitem[\protect\citeauthoryear{Hartmann et 
al.}{2005}]{2005Hartmann} Hartmann L., Megeath S.~T., Allen L., 
Luhman K., Calvet N., D'Alessio P., Franco-Hernandez R., Fazio G., 2005, 
ApJ, 629, 881
\bibitem[\protect\citeauthoryear{Hashimoto}{1994}]{1994Hashimoto} Hashimoto O., 1994, A\&AS, 107, 445
\bibitem[\protect\citeauthoryear{Herbig}{1989}]{1989Herbig} Herbig 
G.~H., 1989, ESOC, 33, 233
\bibitem[\protect\citeauthoryear{Hern{\'a}ndez et 
al.}{2005}]{2005Hernandez} Hern{\'a}ndez J., Calvet N., Hartmann L., 
Brice{\~n}o C., Sicilia-Aguilar A., Berlind P., 2005, AJ, 129, 856
\bibitem[\protect\citeauthoryear{Hodapp et al.}{1996}]{1996Hodapp} 
Hodapp K.-W., Hora J.~L., Rayner J.~T., Pickles A.~J., Ladd E.~F., 1996, 
ApJ, 468, 861
\bibitem[\protect\citeauthoryear{Hodapp et al.}{2012}]{2012Hodapp} 
Hodapp K.~W., Chini R., Watermann R., Lemke R., 2012, ApJ, 744, 56 
\bibitem[\protect\citeauthoryear{Hunter, Thronson, 
\& Wilton}{1990}]{1990Hunter} Hunter D.~A., Thronson H.~A., Jr., Wilton C., 1990, AJ, 100, 1915
\bibitem[\protect\citeauthoryear{Ioannidis 
\& Froebrich}{2012}]{2012Ioannidis} Ioannidis G., Froebrich D., 2012, MNRAS, 425, 1380
\bibitem[\protect\citeauthoryear{Ishihara et 
al.}{2011}]{2011Ishihara} Ishihara D., Kaneda H., Onaka T., Ita Y., Matsuura M., Matsunaga N., 2011, A\&A, 534, A79 
\bibitem[\protect\citeauthoryear{Jim{\'e}nez-Esteban et 
al.}{2006}]{2006Jimenez} Jim{\'e}nez-Esteban F.~M., Garc{\'{\i}}a-Lario P., Engels D., Perea Calder{\'o}n J.~V., 2006, A\&A, 446, 773 
\bibitem[\protect\citeauthoryear{Kenyon et al.}{1990}]{1990Kenyon} 
Kenyon S.~J., Hartmann L.~W., Strom K.~M., Strom S.~E., 1990, AJ, 99, 869
\bibitem[\protect\citeauthoryear{Kleinmann 
\& Hall}{1986}]{1986Kleinmann} Kleinmann S.~G., Hall D.~N.~B., 1986, ApJS, 62, 501 
\bibitem[\protect\citeauthoryear{Knapp}{1987}]{1987Knapp} Knapp 
G.~R., 1987, PASP, 99, 1134
\bibitem[\protect\citeauthoryear{K{\'o}sp{\'a}l et 
al.}{2007}]{2007Kospal} K{\'o}sp{\'a}l {\'A}., {\'A}brah{\'a}m P., Prusti T., Acosta-Pulido J., Hony S., Mo{\'o}r A., Siebenmorgen R., 2007, A\&A, 470, 211
\bibitem[\protect\citeauthoryear{K{\'o}sp{\'a}l et 
al.}{2011}]{2011Kospal} K{\'o}sp{\'a}l {\'A}., et al., 2011, A\&A, 527, A133
\bibitem[\protect\citeauthoryear{Kouzuma 
\& Yamaoka}{2012}]{2012Kouzuma} Kouzuma S., Yamaoka H., 2012, ApJ, 747, 14
\bibitem[\protect\citeauthoryear{Lamm et 
al.}{2004}]{2004Lamm} Lamm M.~H., Bailer-Jones C.~A.~L., Mundt R., Herbst W., Scholz A., 2004, A\&A, 417, 557 
\bibitem[\protect\citeauthoryear{Lawrence et 
al.}{2007}]{2007Lawrence} Lawrence A., et al., 2007, MNRAS, 379, 
1599
\bibitem[\protect\citeauthoryear{Lebzelter 
\& Wood}{2005}]{2005Lebzelter} Lebzelter T., Wood P.~R., 2005, A\&A, 441, 1117 
\bibitem[\protect\citeauthoryear{Lewis}{2000}]{2000Lewis} Lewis 
B.~M., 2000, ApJ, 533, 959 
\bibitem[\protect\citeauthoryear{Lorenzetti et 
al.}{2009}]{2009Loren} Lorenzetti D., Larionov V.~M., Giannini 
T., Arkharov A.~A., Antoniucci S., Nisini B., Di Paola A., 2009, ApJ, 693, 
1056
\bibitem[\protect\citeauthoryear{Lorenzetti et 
al.}{2012}]{2012Loren} Lorenzetti D., et al., 2012, ApJ, 749, 188
\bibitem[\protect\citeauthoryear{Lucas et al.}{2008}]{2008Lucas} 
Lucas P.~W., et al., 2008, MNRAS, 391, 136 
\bibitem[\protect\citeauthoryear{Lynds}{1962}]{1962Lynds} Lynds 
B.~T., 1962, ApJS, 7, 1 
\bibitem[\protect\citeauthoryear{Maddox et al.}{2008}]{2008Maddox} 
Maddox N., Hewett P.~C., Warren S.~J., Croom S.~M., 2008, MNRAS, 386, 1605
\bibitem[\protect\citeauthoryear{Magakian}{2003}]{2003Magakian} Magakian T.~Y., 2003, A\&A, 399, 141
\bibitem[\protect\citeauthoryear{Magakian et 
al.}{2013}]{2013Magakian} Magakian T.~Y., et al., 2013, MNRAS, 432, 
2685 
\bibitem[\protect\citeauthoryear{McSwain et 
al.}{2008}]{2008McSwain} McSwain M.~V., Huang W., Gies D.~R., 
Grundstrom E.~D., Townsend R.~H.~D., 2008, ApJ, 672, 590
\bibitem[\protect\citeauthoryear{McSwain, Huang, 
\& Gies}{McSwain et al.}{2009}]{2009McSwain} McSwain M.~V., Huang W., Gies D.~R., 2009, ApJ, 700, 1216
\bibitem[\protect\citeauthoryear{Meyer, Calvet, 
\& Hillenbrand}{Meyer et al.}{1997}]{1997Meyer} Meyer M.~R., Calvet N., Hillenbrand L.~A., 1997, AJ, 114, 288 
\bibitem[\protect\citeauthoryear{Minniti et 
al.}{2010}]{2010Minniti} Minniti D., et al., 2010, NewA, 15, 433
\bibitem[\protect\citeauthoryear{Minniti et 
al.}{2011}]{2011Minniti} Minniti D., Saito R.~K., 
Alonso-Garc{\'{\i}}a J., Lucas P.~W., Hempel M., 2011, ApJ, 733, L43
\bibitem[\protect\citeauthoryear{Murakami et 
al.}{2007}]{2007Murakami} Murakami H., et al., 2007, PASJ, 59, 369
\bibitem[\protect\citeauthoryear{Najita et al.}{1996}]{1996Najita} 
Najita J., Carr J.~S., Glassgold A.~E., Shu F.~H., Tokunaga A.~T., 1996, 
ApJ, 462, 919 
\bibitem[\protect\citeauthoryear{Nakano et al.}{2003}]{2003Nakano} 
Nakano S., Sato H., Nishimura H., Nakamura T., Wakuda S., Yamaoka H., 
Pearce A., 2003, IAUC, 8190, 1
\bibitem[\protect\citeauthoryear{Olivier, Whitelock, 
\& Marang}{Olivier et al.}{2001}]{2001Olivier} Olivier E.~A., Whitelock P., Marang F., 2001, MNRAS, 326, 490
\bibitem[\protect\citeauthoryear{Ortiz 
\& Maciel}{1996}]{1996Ortiz} Ortiz R., Maciel W.~J., 1996, A\&A, 313, 180
\bibitem[\protect\citeauthoryear{Parihar et 
al.}{2009}]{2009Parihar} Parihar P., Messina S., Distefano E., 
Shantikumar N.~S., Medhi B.~J., 2009, MNRAS, 400, 603 
\bibitem[\protect\citeauthoryear{Persi et al.}{2007}]{2007Persi} 
Persi P., Tapia M., G{\`o}mez M., Whitney B.~A., Marenzi A.~R., Roth M., 
2007, AJ, 133, 1690
\bibitem[\protect\citeauthoryear{Porter 
\& Rivinius}{2003}]{2003Porter} Porter J.~M., Rivinius T., 2003, PASP, 115, 1153
\bibitem[\protect\citeauthoryear{Quireza et 
al.}{2006}]{2006Quireza} Quireza C., Rood R.~T., Bania T.~M., 
Balser D.~S., Maciel W.~J., 2006, ApJ, 653, 1226
\bibitem[\protect\citeauthoryear{Reed}{2000}]{2000Reed} Reed 
B.~C., 2000, AJ, 120, 314
\bibitem[\protect\citeauthoryear{Reipurth 
\& Aspin}{1997}]{1997Reip} Reipurth B., Aspin C., 1997, AJ, 114, 2700
\bibitem[\protect\citeauthoryear{Reipurth 
\& Aspin}{2004}]{2004Reip} Reipurth B., Aspin C., 2004, ApJ, 606, L119
\bibitem[\protect\citeauthoryear{Reipurth 
\& Aspin}{2010}]{2010Reip} Reipurth B., Aspin C., 2010, vaoa.conf, 19 
\bibitem[\protect\citeauthoryear{Rice, Wolk, 
\& Aspin}{Rice et al.}{2012}]{2012Rice} Rice T.~S., Wolk S.~J., Aspin C., 2012, ApJ, 755, 65
\bibitem[\protect\citeauthoryear{Robitaille et 
al.}{2007}]{2007Robitaille} Robitaille T.~P., Whitney B.~A., 
Indebetouw R., Wood K., 2007, ApJS, 169, 328
\bibitem[\protect\citeauthoryear{Robitaille et 
al.}{2008}]{2008Robitaille} Robitaille T.~P., et al., 2008, AJ, 136, 
2413
\bibitem[\protect\citeauthoryear{Romanova et 
al.}{2013}]{2013Romanova} Romanova M.~M., Ustyugova G.~V., Koldoba 
A.~V., Lovelace R.~V.~E., 2013, MNRAS, 430, 699
\bibitem[\protect\citeauthoryear{Russeil}{2003}]{2003Russeil} Russeil D., 2003, A\&A, 397, 133
\bibitem[\protect\citeauthoryear{Saito et 
al.}{2012}]{2012Saito} Saito R.~K., et al., 2012, A\&A, 537, A107 
\bibitem[\protect\citeauthoryear{Saito et 
al.}{2013}]{2013Saito} Saito R.~K., et al., 2013, A\&A, 554, A123
\bibitem[\protect\citeauthoryear{Samus et al.}{2010}]{2010Samus} 
Samus N.~N., Kazarovets E.~V., Kireeva N.~N., Pastukhova E.~N., Durlevich 
O.~V., 2010, OAP, 23, 102
\bibitem[\protect\citeauthoryear{Scholz et al.}{2009}]{2009Scholz} 
Scholz A., Xu X., Jayawardhana R., Wood K., Eisl{\"o}ffel J., Quinn C., 
2009, MNRAS, 398, 873
\bibitem[\protect\citeauthoryear{Scholz}{2012}]{2012Scholz} Scholz 
A., 2012, MNRAS, 420, 1495
\bibitem[\protect\citeauthoryear{Scholz, Froebrich, 
\& Wood}{Scholz et al.}{2013}]{2013Scholz} Scholz A., Froebrich D., Wood K., 2013, MNRAS, 430, 2910
\bibitem[\protect\citeauthoryear{Scoville et 
al.}{1987}]{1987Scoville} Scoville N.~Z., Yun M.~S., Sanders D.~B., 
Clemens D.~P., Waller W.~H., 1987, ApJS, 63, 821
\bibitem[\protect\citeauthoryear{Skrutskie et 
al.}{2006}]{2006Skrutskie} Skrutskie M.~F., et al., 2006, AJ, 131, 
1163
\bibitem[\protect\citeauthoryear{Soszynski et 
al.}{2005}]{2005Soszynski} Soszy\'nski I., et al., 2005, AcA, 55, 331
\bibitem[\protect\citeauthoryear{Stamatellos, Whitworth, 
\& Hubber}{Stamatellos et al.}{2012}]{2012Stamatellos} Stamatellos D., Whitworth A.~P., Hubber D.~A., 2012, MNRAS, 427, 1182
\bibitem[\protect\citeauthoryear{Tisserand et 
al.}{2004}]{2004Tisserand} Tisserand P., et al., 2004, A\&A, 424, 245
\bibitem[\protect\citeauthoryear{van Loon et 
al.}{1997}]{1997Vanloon} van Loon J.~T., Zijlstra A.~A., Whitelock P.~A., Waters L.~B.~F.~M., Loup C., Trams N.~R., 1997, A\&A, 325, 585
\bibitem[\protect\citeauthoryear{Varricatt, Davis, 
\& Adamson}{Varricatt et al.}{2005}]{2005Varricatt} Varricatt W.~P., Davis C.~J., Adamson A.~J., 2005, MNRAS, 359, 2
\bibitem[\protect\citeauthoryear{Varricatt et 
al.}{2010}]{2010Varricatt} Varricatt W.~P., Davis C.~J., Ramsay S., 
Todd S.~P., 2010, MNRAS, 404, 661
\bibitem[\protect\citeauthoryear{Vijh et al.}{2009}]{2009Vijh} 
Vijh U.~P., et al., 2009, AJ, 137, 3139
\bibitem[\protect\citeauthoryear{Volk, Kwok, 
\& Langill}{Volk et al.}{1992}]{1992Volk} Volk K., Kwok S., Langill P.~P., 1992, ApJ, 391, 285
\bibitem[\protect\citeauthoryear{Wallace 
\& Hinkle}{1997}]{1997Wallace} Wallace L., Hinkle K., 1997, ApJS, 111, 445 
\bibitem[\protect\citeauthoryear{Whitelock, Feast, 
\& Catchpole}{Whitelock et al.}{1991}]{1991Whitelock} Whitelock P., Feast M., Catchpole R., 1991, MNRAS, 248, 276
\bibitem[\protect\citeauthoryear{Whitelock 
\& Feast}{2000}]{2000Whitelock} Whitelock P., Feast M., 2000, MNRAS, 319, 759
\bibitem[\protect\citeauthoryear{Whitelock, Feast, 
\& van Leeuwen}{Whitelock et al.}{2008}]{2008Whitelock} Whitelock P.~A., Feast M.~W., van Leeuwen F., 2008, MNRAS, 386, 313
\bibitem[\protect\citeauthoryear{Whitney et al.}{2011}]{2011Whitney} Whitney B., et al., 2011, AAS, 43,\#241.16
\bibitem[\protect\citeauthoryear{Witham et al.}{2008}]{2008Witham} 
Witham A.~R., Knigge C., Drew J.~E., Greimel R., Steeghs D., G{\"a}nsicke 
B.~T., Groot P.~J., Mampaso A., 2008, MNRAS, 384, 1277
\bibitem[\protect\citeauthoryear{Wright et al.}{2010}]{2010Wright} 
Wright E.~L., et al., 2010, AJ, 140, 1868
\bibitem[\protect\citeauthoryear{Zhang et al.}{2009}]{2009Zhang} 
Zhang B., Zheng X.~W., Reid M.~J., Menten K.~M., Xu Y., Moscadelli L., 
Brunthaler A., 2009, ApJ, 693, 419
\end{thebibliography}

\newpage

\appendix

\section{SEDs of DR7 objects in SFRs}\label{app:dr7}

\begin{figure*}
\begin{center}
\includegraphics[width=1\columnwidth]{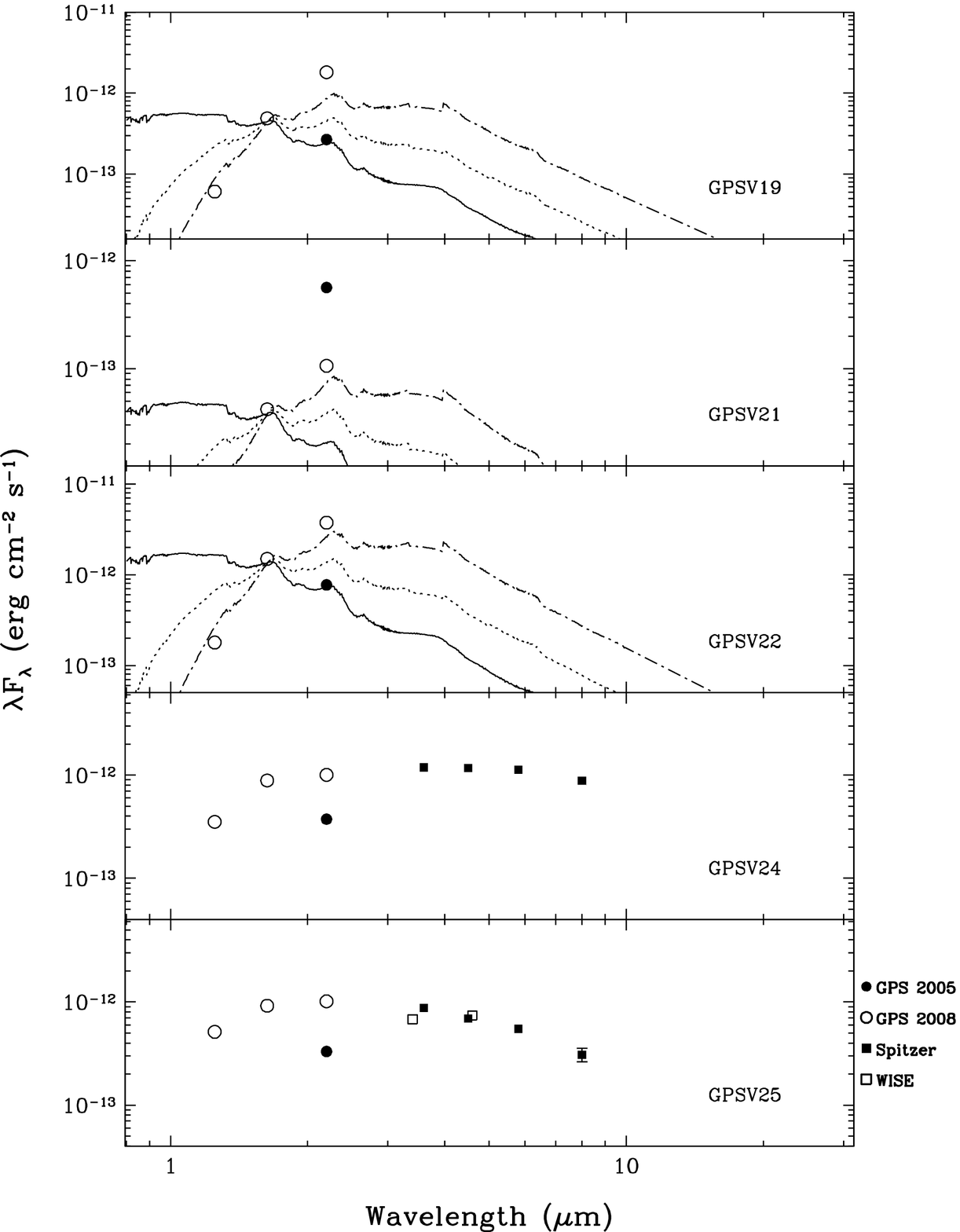}
\includegraphics[width=1\columnwidth]{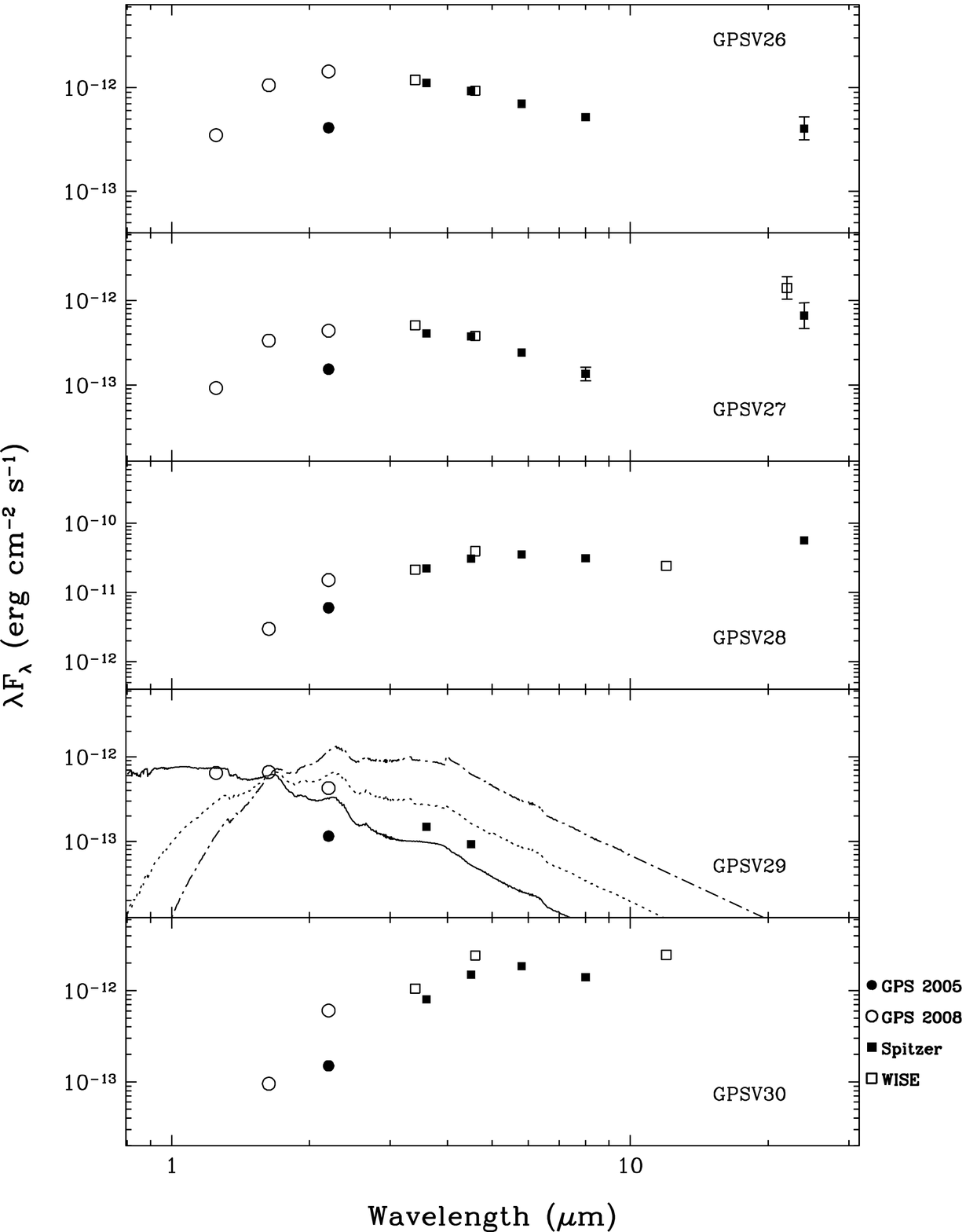}\\
\includegraphics[width=1\columnwidth]{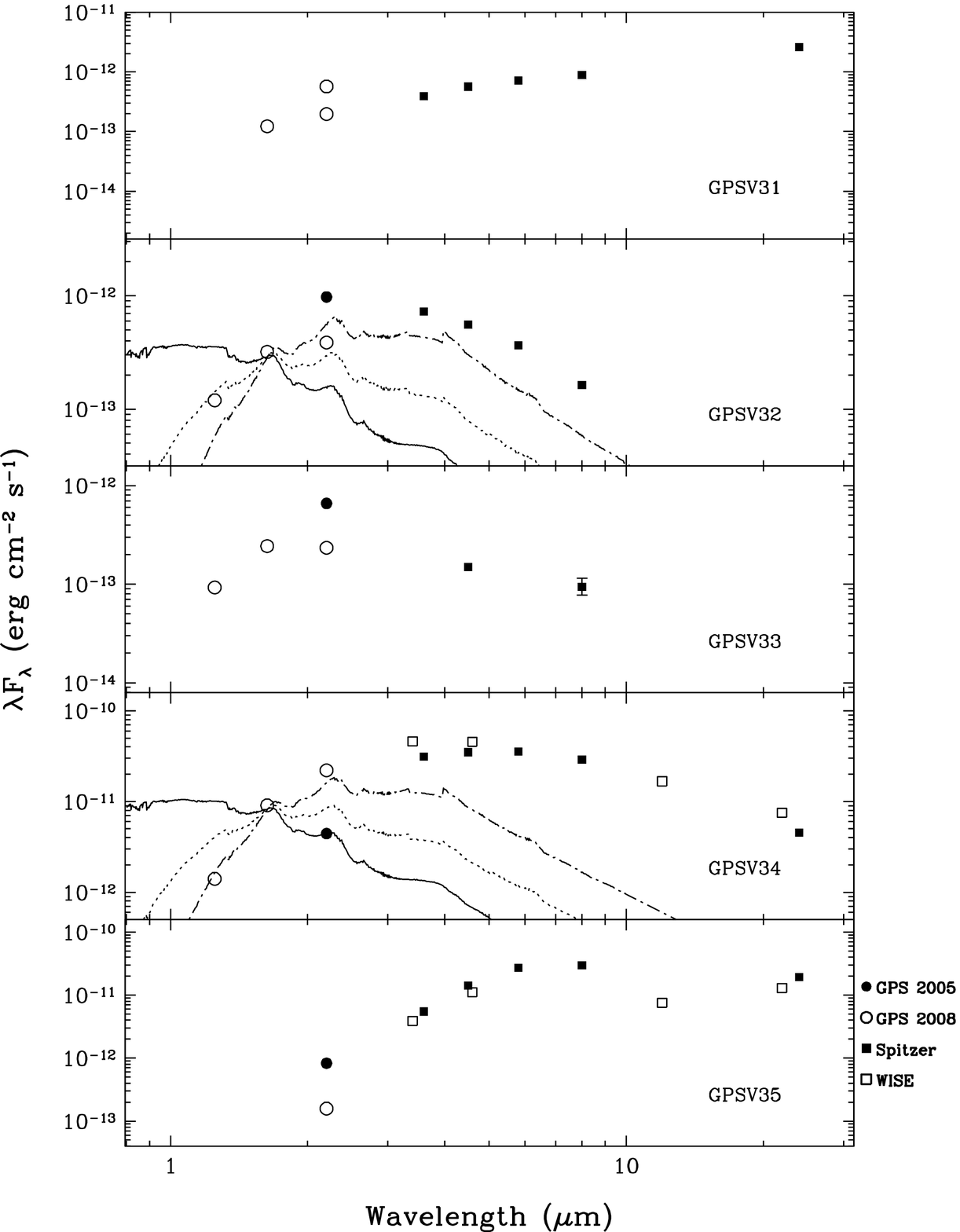}
\includegraphics[width=1\columnwidth]{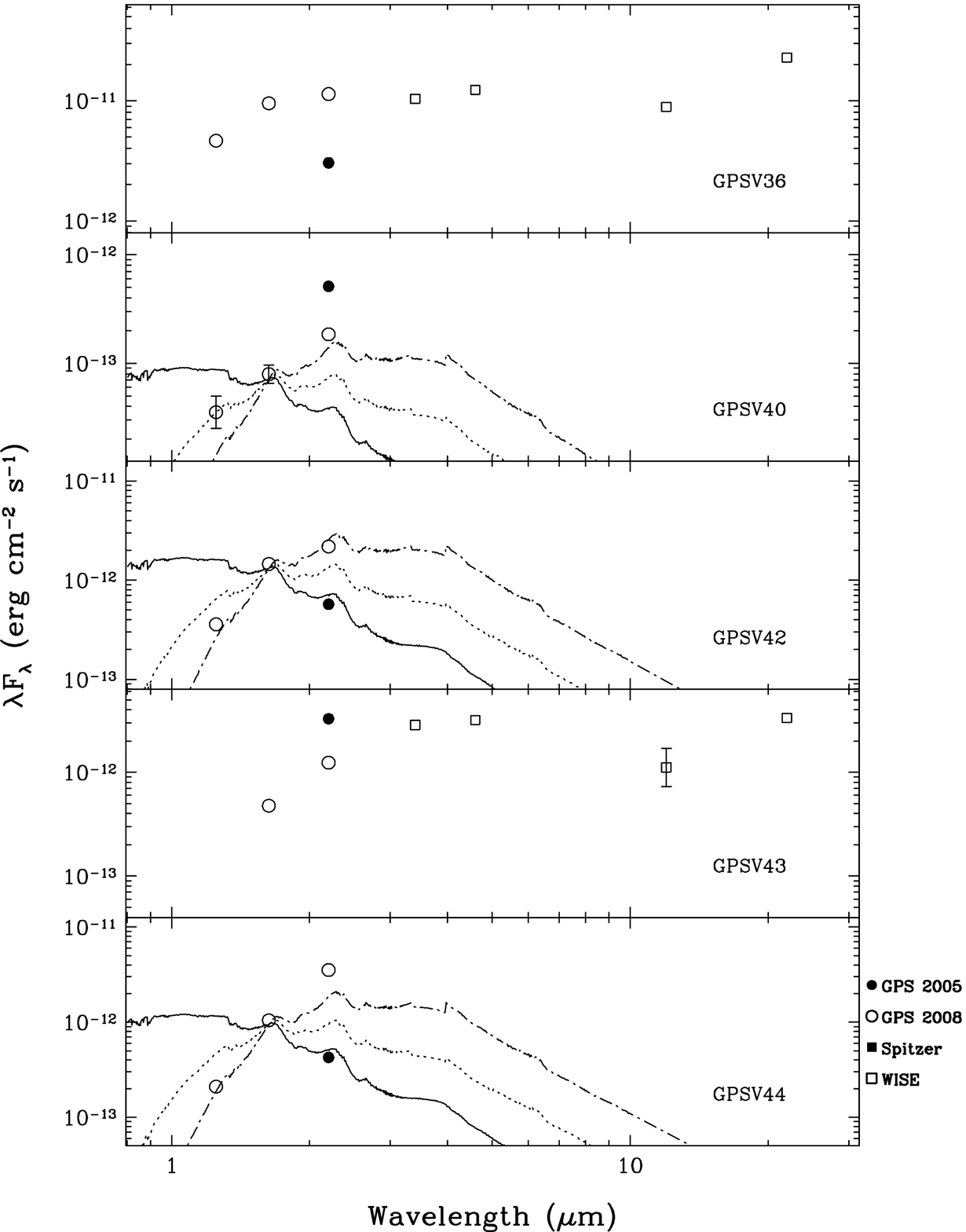}
\caption[]{Spectral energy distributions of DR7 objects in SFRs, including three objects with YSO characteristics which are not found within SFRs. Data for some of the objects are compared to a M5V (3240 K) \citeauthor{2004Castelli} model atmosphere ({\it solid line}) and the same model reddened to $A_{V}=10$~mag ({\it dotted line}) and $A_{V}=20$~mag ({\it dashed-dotted line}). Fluxes of the model where arbitrarily set to match the 2008-2009 $H$ observations of GPS. The models are reddened using the \citet{1989Cardelli} extinction law for wavelengths in the range $0.3 \mu$~m$ < \lambda < 3.3 \mu$~m. Extinction towards the wavelengths of the {\it Spitzer} bands is derived following \citet{2009Chapman}.} 
\label{app:ysos_seds1}
\end{center}
\end{figure*}

The spectral energy distributions for the 17 objects from DR7 associated with SFRs (section \ref{sec:conc_ext}) are presented in figure \ref{app:ysos_seds1}. In addition we present the SEDs for the three objects that have similar characteristics to YSOs in colour-colour plots but are not located near a known star formation region. These are GPSV36, 40 and 43. The SEDs are constructed using the 2 epoch UKIDSS GPS DR7 information (2005,2008) and public data, when available, from the Spitzer legacy survey, Cygnus-X (2007-2008 observations), and/or WISE (2010 observations). Some of the objects that lack mid-infrared data, or for which an infrared excess is not clear, are compared to an M5V (3240 K) \citeauthor{2004Castelli} model atmosphere, and the same model reddened to $A_V$=10 and 
20~mag. $A_V$=20 is near the upper end of the possible interstellar extinction column values for any of 
the variables (determined in $\S$2.2 using the local CMD for the field around each source).

The fact that the near- and mid-infrared observations are not contemporaneous does not allow to clearly classify some of the objects as YSOs. This is observed in GPSV32 where if we only take into account the first epoch of GPS and the Spitzer data, its SED resembles that of an M-dwarf with $A_V=20$~mag of extinction. This changes if we take the second GPS epoch and Spitzer data, where the SED would point to an infrared excess. The same issue arises in GPSV29.

It is also difficult to classify as YSOs objects in our sample without mid-infrared data. In every case, these objects could point to either infrared excess (not being observed) or correspond to highly extinct M-dwarfs. As an example we compare the object GPSV34 (a spectroscopically confirmed YSO, Contreras Pe\~{n}a et al., in prep.) with the same \citeauthor{2004Castelli} model atmospheres in figure \ref{app:ysos_seds1}. We see that the near infrared data from the second GPS epoch is well fitted by a M-dwarf with $A_V=20$~mag of extinction. However, the mid-infrared data allows us to confirm its probable YSO nature. Two of the objects with GPS data only are confirmed via its spectra in the analysis to be presented in an upcoming study (Contreras Pe\~{n}a et al., in prep.), these are GPSV19 and GPSV22.   

Objects GPSV36 and GPSV43 have no association with a star forming region (section \ref{sec:conc_ext}). However, their SEDs show a clear infrared excess that suggests a YSO classification. 

Even though we cannot confirm every object in SFR as a YSO, the SEDs shown here and the spectral characteristics for a sample of them (Contreras Pe\~{n}a et al., in prep.) point to such classification for the majority of them. This supports our conclusion that finding high amplitude variables close to areas of star formation is sufficient 
evidence that most of these stars are YSOs.

\label{lastpage}
\end{document}